\documentclass[11pt,a4paper]{article}
\pdfoutput=1	
\usepackage{multirow}
\usepackage{multicol}
\usepackage{alpha}
\usepackage{macros}
\usepackage{graphicx}
\usepackage{amsmath}
\usepackage{amssymb}
\usepackage{sidecap}
\usepackage{wrapfig}
\usepackage{comment}
\usepackage{color}

\newcommand{\C}{{\cal C}}
\newcommand{\Chat}{\widehat{\cal C}}
\newcommand{\Ct}{\ensuremath{\mathcal C^{\B\to\K} } }
\newcommand{\B}{{\ensuremath{\textnormal{B}_{\textnormal{s}}}}}
\newcommand{\Bs}{{\ensuremath{\textnormal{B}_{\textnormal{s}}}}}
\newcommand{\CB}{\mathcal C^\B}
\newcommand{\ChatB}{\Chat^\B}
\providecommand{\CK}{\ensuremath{\mathcal C^\K}}
\providecommand{\ChatK}{\ensuremath{\Chat^\K}}
\providecommand{\K}{{\ensuremath{\textnormal{K}}}}
\newcommand{\tk}{{t_\K}}
\newcommand{\tb}{{t_{\Bs}}}
\DeclareMathOperator{\e}{e}

\newcommand{\cO}{{\cal O}}

\newcommand{\kin}{\mathrm{kin}}
\newcommand{\spin}{\mathrm{spin}}

\newcommand{\ord}[1]{\mathcal{O}\left(#1\right)}

\newcommand{\Ri}{\mathcal{R}^{\,\textrm{I}}}
\newcommand{\Rii}{\mathcal{R}^{\,\textrm{II}}}
\newcommand{\Riii}{\mathcal{R}^{\,\textrm{III}}}

\newcommand{\Mi}{\mathcal{M}^{\textrm{I}}}
\newcommand{\Si}{\mathcal{S}^{\textrm{I}}}

\newcommand{\eq}[1]{Eq.~(\ref{#1})}
\newcommand{\fig}[1]{Fig.~(\ref{#1})}
\newcommand{\EK}[1]{E_{\rm K}^{(#1)}}
\newcommand{\EB}[1]{E_{\rm \Bs}^{(#1)}}
\newcommand{\tfit}[2]{t_{\rm #1}^{\rm #2}}  

\newcommand{\rmk}{\mathrm{k}}

\newcommand{\kn}{\textnormal{k}}

\newcommand{\Cxx}[2]{{\C}^{#1,#2}}
\newcommand{\Cxxx}[3]{{\C}^{#1,#2}_{#3}}
\newcommand{\Cxxxx}[4]{{\C}^{#1,#2}_{#3,#4}}
\newcommand{\Ex}[1]{E^{#1}}
\newcommand{\Exx}[2]{E^{#1}_{#2}}
\newcommand{\Axxx}[3]{{\rm A}^{#1,#2}_{#3}}
\newcommand{\Axxxx}[4]{{\rm A}^{#1,#2}_{#3,#4}}

\newcommand{\drixxx}[3]{\rho^{\textnormal{I},{#1}}_{#2,#3}}

\newcommand{\Rix}[1]{\mathcal{R}^{\,\textrm{I}}_{\mu, #1}}
\newcommand{\Rixx}[2]{\mathcal{R}^{\,\textrm{I},{#1}}_{\mu, #2}}

\newcommand{\vecp}{{\bf p}}

\newcommand{\vecx}{{\bf x}}

\newcommand{\rmd}{\mathrm{d}}
\newcommand{\Tra}{\mathrm{Tr}}

\def\psibar{\overline{\psi}}

\def\bec{\begin{center}}
\def\eec{\end{center}}
\def\beq{\begin{equation}}
\def\eeq{\end{equation}}
\def\bes{\begin{eqnarray}}
\def\ees{\end{eqnarray}}

\newcommand{\veca}[1]{{\bf #1}}

\begin{document}
\bibliographystyle{plain}

\preprintno
{
DESY~19-030\\
[3em]
}

\title{Extraction of bare Form Factors for 
$\mathrm B_\mathrm s \to \mathrm K \ell \nu$ Decays in non-perturbative HQET}
\collaboration{\includegraphics[width=2.8cm]{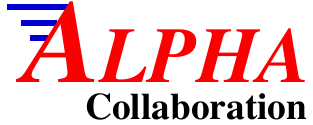}}

\author[desy]{Felix~Bahr}
\author[desy,hu]{Debasish~Banerjee}
\author[dresden]{Fabio~Bernardoni}
\author[desy]{Mateusz Koren}
\author[desy]{Hubert~Simma}
\author[desy,hu]{Rainer~Sommer}

\address[desy]{John von Neumann Institute for Computing (NIC), DESY, Platanenallee~6, D-15738~Zeuthen, Germany}
\address[hu]{Institut~f\"ur~Physik, Humboldt-Universit\"at~zu~Berlin,
             Newtonstr.~15, 12489~Berlin, Germany}
\address[dresden]{Medizinische Fakult\"at, Carl Gustav Carus, TU Dresden, Fetscherstra\ss e~74, D-01307~Dresden, Germany}
\begin{abstract}
We discuss the extraction of the ground state 
$\langle\K(\veca{p})|V_\mu(0)|\B(\veca{0})\rangle$
matrix elements from Euclidean lattice correlation functions.
The emphasis is on the elimination of excited state contributions.
Two typical gauge-field ensembles with lattice spacings $0.075, \; 0.05$~fm and pion masses $330,\;270$~MeV are used from the O($a$)-improved CLS 2-flavour simulations and the final state momentum is $|\veca{p}|=0.5\,\GeV$.
The b-quark is treated in HQET including the $1/m_\mathrm{b}$ corrections. 
Fits to two-point and three-point correlation functions and suitable
ratios including summed ratios are used, yielding consistent results 
with precision of around 2\%
which is {\em not} limited by the $1/m_\mathrm{b}$ corrections but by the dominating static form factors.
Excited state contributions are under reasonable control but are the 
bottleneck towards precision. We do not yet include a specific 
investigation of multi-hadron contaminations, a gap in the literature
which ought to be filled soon. 
\end{abstract}

\maketitle
\tableofcontents
\section{Introduction }
\label{sec:intro}

Decays of b-quarks in the form $\mathrm B \to \pi  \ell \nu$ and $\mathrm B_\mathrm s \to \mathrm K \ell \nu$ are very relevant in 
constraining the Standard Model of particle physics and hence also in the search 
for deviations from it. Lattice QCD is the method of choice for predicting 
the necessary form factors \cite{Lattice:2015tia
,Flynn:2015mha
,Bouchard:2014ypa
,Colquhoun:2015mfa
,Colquhoun:2017gfi
,Gelzer:2017edb
,Monahan:2018lzv
,Lattice:2017vqf
,Bazavov:2019aom
}.  While  
computations are progresssing significantly, there are still few groups carrying them out. Crosschecks by independent (lattice) methods are largely absent.  
In Ref.~\cite{Bahr:2016ayy} we have started an investigation which takes significantly different avenues in almost every choice that can be made. 
We use (improved) Wilson fermions,
treat the b-quark in non-perturbative HQET and 
compute three-point functions for all time separations 
rather than fixing its total time-span (called $\tau=\tk + \tb$ later) 
to one or few values. 
We opt for non-perturbative HQET because a complete and practical method is known 
to non-perturbatively renormalize the theory and match it to QCD \cite{
Heitger:2003nj
,Blossier:2010jk
,Blossier:2012qu
,DellaMorte:2013ega
,Hesse:2012hb
}. When completed, this will provide an independent
crosscheck.

Once these choices are made,  one needs to carry out the basic steps 
as written in Ref.~\cite{Bahr:2016ayy}:
\begin{itemize}\setlength{\itemsep}{-1mm}
\item[a)] obtain the {\em ground state} matrix elements 
  $\langle \K | V^\mu(0) | \Bs \rangle$ that mediate the transition, 
\item[b)] renormalize the currents (and thus matrix elements)
    and, if an effective theory is used, relate them to QCD (``matching''),
\item[c)] take the continuum limit of the matrix elements,
\item[d)] extrapolate to the quark masses realized in Nature,
\item[e)] map out the $q^2$ dependence.
\end{itemize}
In the previous work we saw that the continuum limit
is smooth in the static approximation, which is expected to yield 
the by far dominating part of the full result. In this paper we discuss the step a) in detail, including the $1/m$ terms. 
It is not at all an easy enterprise. As discussed at length below, 
the issue is that ground state matrix elements are obtained at large
Euclidean time separations. At time separations of around $\frac12$~fm 
--- a typical QCD length scale --- the low-lying states start dominating
the two- and three-point functions. Unfortunately the statistical quality
of the Monte Carlo estimates is typically deteriorating fast once 2~fm in total 
(or 1~fm for $K$ and $\Bs$ each) are reached. It is thus a matter of
the details of methods and QCD dynamics whether there is a window to
determine the desired form factors at the percent level.  

In other words, control over the extraction of the bare form factors is crucial in a proper
determination of the decay rates. In this publication we thus explain our methods and 
their limitations in detail.

\section{Setup for the non-perturbative evaluation}
\label{sec:setup}
\subsection{Form factors from Euclidean correlation functions}
\label{sec:general}

\begin{figure}[ht]
\centering
\includegraphics[width=5cm]{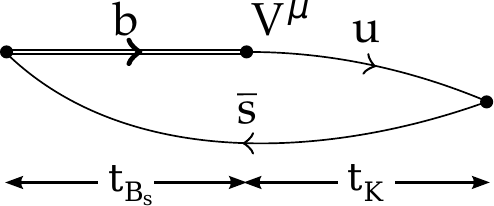}
\caption{The correlation function used to 
extract the $\Bs\to\K$ decay matrix element.}
\label{fig:3ptfct}
\end{figure}

To leading order in the weak interactions, the  
$\B \to \K \ell \nu$ decay rate 
depends on two form factors, 
$h_\parallel(E_\K)$ and $h_\perp(E_\K)$. 
which are related to the commonly used 
\begin{eqnarray}
 f_+(q^2) & = & \displaystyle\frac{1}{\sqrt{2m_\Bs}} \, \left[ h_\parallel + (m_\Bs-E_\K)\, h_\perp \right]\,,\\
 f_0(q^2) & = & \displaystyle\frac{\sqrt{2m_\Bs}}{m_\Bs^2-m_\K^2} \, \Bigl[ 
           (m_\Bs-E_\K) \, h_\parallel + \left(E_\K^2-m_\K^2\right) \, h_\perp \Bigr]\,.
\end{eqnarray}
The kinematics is explained in Ref.~\cite{Bahr:2016ayy}, and in
the rest-frame of the 
$\B$, we have
\begin{eqnarray}
(2m_{\B})^{-1/2}\langle\K(\veca{p}_\K)|\hat{V}^0_\mathrm{M}(0)|\B(\veca{0})\rangle &=&
h_\parallel(E_\K)\,,\\
(2m_{\B})^{-1/2}\langle\K(\veca{p}_\K)|\hat{V}^k_\mathrm{M}(0)|\B(\veca{0})\rangle &=& p^k_\K
h_\perp(E_\K)\,,
\end{eqnarray}
where $(E_\K,\veca{p}_\K)$ is the four-momentum of the final state
and $|\B(\veca{0})\rangle$ is the initial state with the $\B$ at rest,
$\veca{p_\B}=\veca{0}$. The above hadron-to-hadron matrix elements
are in Minkowski space, as indicated by the subscript M. They can be 
obtained from the Euclidean three-point function (fig. \ref{fig:3ptfct}), 
where the spatial volume is a $L^3$ torus,
\begin{align}
\begin{split}
  \C^{\Bs\to\K}_\mu(\tk,\tb;\veca{p}_\K) =&
  \frac{a^9}{L^3}\sum_{\veca{x}_\mathrm{f},\veca{x}_\mathrm{v},\veca{x}_\mathrm{i}}
  e^{-i\veca{p}_\K\,(\veca{x}_\mathrm{f}-\veca{x}_\mathrm{v})} \times 
  \langle \overline{\cO}_{\rm us}(x_\mathrm{f})
  V_\mu(x_\mathrm{v}) \cO_{\rm bs}(x_\mathrm{i}) \rangle,   \label{eqn:3pt} \\
  & \tk = (x_\mathrm{f}-x_\mathrm{v})_0>0\,,\quad
    \tb  = (x_\mathrm{v}-x_\mathrm{i})_0>0\,,
\end{split}
\end{align}
with suitable interpolating fields $\cO_{q q'},\overline{\cO}_{q q'}$ (see \eqref{eq:cObs},\eqref{eq:cOsu}) and 
\begin{equation}
  V_\mu(x) = \psibar_{\rm u}(x)\gamma_\mu\psi_{\rm b}(x)
  \label{eqn:Vmu}\,.
\end{equation}
Removing the overlaps 
$\langle \B(\veca{0})|\cO_{\rm bs}|0\rangle$,
$\langle 0|\overline{\cO}_{\rm us}|\K(\veca{p}_\K)\rangle$ (amputation), requires also the 
two-point functions 
\begin{align}
  \CK(t;\veca{p}_\K) &= \frac{a^6}{L^3} \sum_{\veca{x}_\mathrm{f},\veca{x}_\mathrm{i}} 
  e^{-i\veca{p}_\K\,(\veca{x}_\mathrm{f}-\veca{x}_\mathrm{i})}\langle 
  \overline{\cO}_{\rm us}(x_\mathrm{f})\cO_{\rm us}(x_\mathrm{i}) \rangle,\quad t = (x_\mathrm{f}-x_\mathrm{i})_0\\
  \CB(t) &=  \frac{a^6}{L^3} \sum_{\veca{x}_\mathrm{f},\veca{x}_\mathrm{i}} 
  \langle \overline{\cO}_{\rm bs}(x_\mathrm{f})\cO_{\rm bs}(x_\mathrm{i})\rangle
  ,\quad t = (x_\mathrm{f}-x_\mathrm{i})_0\,.
  \label{eqn:2pt1}
\end{align}
The explicit relation between the Euclidean correlation functions
and the desired matrix elements is (see e.g. \cite{Best:1997qp})
\bes
  \langle\K(\veca{p})|\hat{V}_\mathrm{M}^0(0)|\B(\veca{0})\rangle 
  &=& \lim_{t\to\infty}
  {\C^{\Bs\to\K}_0(t,t;\veca{p})
   \over
   [\CK(2t;\veca{p}) \CB(2t) ]^{1/2} }\,,
   \nonumber\\[-1ex] \label{eqn:me1} \\[-1ex] \nonumber
  \langle\K(\veca{p})|\hat{V}_\mathrm{M}^k(0)|\B(\veca{0})\rangle 
  &=& i \,\lim_{t\to\infty}
  {\C^{\Bs\to\K}_k(t,t;\veca{p})
   \over
   [\CK(2t;\veca{p}) \CB(2t) ]^{1/2} }\,.   
\ees
A few comments are in order
\begin{itemize}
    \item[-]We chose to define the correlation functions on 
    a lattice. Replacing $a^3\sum_\veca{x} \to \int\rmd^3\veca{x}$
    yields the continuum expressions.
    \item[-] The derivation of \eq{eqn:me1} (on a lattice) is based 
    just on the existence of a transfer matrix with standard 
    properties and on interpolating fields which are local in time; 
    non-locality (smearing) in space is allowed.
    \item[-] Of course, on a finite lattice the limit 
    $t\to\infty$ has to be replaced by a proper procedure.
    Controlling it is the main topic of this paper. As a first,
    straightforward, procedure we note that it would be sufficient to have 
    $t$ (as well as the spatial box length $L$) large compared
    to the typical QCD scales $\frac12\fm, \; 1/m_\pi$. 
    \item[-] In order to obtain the physical form factors,
    the current $V_\mu$ has to be 
    properly renormalized and the continuum limit has to be taken.
\end{itemize}
So far we have spoken of standard relativistic (lattice) QCD. 
The non-perturbative HQET expressions including $\ord{1/m}$ terms are obtained by the following
generic expansion of correlation 
functions~\cite{Heitger:2003nj,Sommer:2010ic,Blossier:2010mk}
\bes
 \log(\C^{\Bs\to\K}_\mu) = 
 \log Z^\mathrm{HQET}_{V_\mu} +  \log([\C^{\Bs\to\K}_\mu]_\mathrm{stat}) 
 + \sum_k \omega_k \rho_\mu^k + \delta_\cO
 + \ord{1/m_\mathrm{b}^2}\,, 
 \label{eqn:corrfcthqet}              
\ees
where $[\C^{\Bs\to\K}]_\mathrm{stat}$ is the correlation function in the static approximation and 
\bes
  \rho_\mu^k = {[\C^{\Bs\to\K}_\mu]_k \over [\C^{\Bs\to\K}_\mu]_\mathrm{stat}}\,.
  \label{e:rhomuk}
\ees
The "stat" terms refer to lowest order (static) HQET, while
$k \in\{\rm kin,spin\}$ are the terms due to the 
$1/m$ corrections to the action, and the remaining terms in the sum over $k$
are $1/m$ corrections to the currents, see Sect.~\ref{sec:currins}.
In $\delta_\cO$, we collect all terms which arise from the renormalization and $1/m$ 
corrections to the interpolating fields.
Their form is irrelevant since they cancel 
in the ratios of interest \eq{eqn:me1}.
Apart from $\delta_\cO$, all divergences strictly cancel within \eq{eqn:corrfcthqet} 
iff the parameters $\log Z^\mathrm{HQET}_{V_\mu}$ and $\omega_k$
are determined non-perturbatively. The expansion of
 $\log(\C^{\Bs})$ reads the same apart from the current-terms 
which are absent. Finally, functions of correlation functions
such as the ratio \eq{eqn:me1} are trivially obtained from the 
expansion of the correlation functions.

For completeness we remind the reader of the rule of the game:
the expansion is ``derived'' by considering the parameters 
$\omega_k$ as $\ord{1/m}$ and expanding in $1/m$,
despite the fact that individual pieces may be highly divergent.
The logarithm is taken in \eq{eqn:corrfcthqet} since then
all expressions are automatically linear in the parameters
and higher order terms in the parameters (and thus $1/m$) are directly avoided.
Those would otherwise have to be dropped to preserve renormalizability.
\vfill

\subsection{Lattice Setup}
\label{sect:lattice}

\newcommand{\mlight}{m_{\rm l}}
\def\Nsm{I}

\begin{table}[t!]
\begin{center}
\begin{tabular}{cccccc}
\toprule
ensemble & $\beta$ & $L/a$ & $a$ [fm] &  $m_\pi$ [MeV] & $m_\pi L$ \\
\midrule
  A5  & $5.2$ & $32$ & $0.0749(8)$ & $330$  & $4.0$ \\ 
  O7  & $5.5$ & $64$ & $0.0483(4)$ & $270$  & $4.2$ \\ 
\bottomrule  
\end{tabular}
\caption{Overview of the subset of $N_{\rm f} = 2 $ CLS ensembles used. They have $T=2\times L$. Lattice
spacings, $a$, are taken from Ref.~\cite{strangemass} and Ref.~\cite{lottini1}. The
pion mass is denoted by $m_\pi$, and we have $m_\pi L \geq 4$ 
such that finite-volume effects are sufficiently (exponentially) suppressed at the
level of accuracy we are aiming for. 
}
\label{tab:ens}
\end{center}
\end{table}

\begin{table}[t!]
\begin{center}
\begin{tabular}{ccccccc}
\toprule
ensemble & $\kappa_s$ & $N_{\rm cfg} $ & $\Nsm^\K$ & $(\Nsm^\Bs_1, \Nsm^\Bs_2, \Nsm^\Bs_3)$ & $\veca{n}$ & $\vartheta$ \\ 
\midrule
A5 & $0.13535(2)$ & $1000$ & 35 & $(30, 60, 155)$  & $(1,0,0)^T$ & $+0.2141$ \\
O7 & $0.13627(1)$ & $490$  & 75 & $(68, 135, 338)$ & $(1,1,0)^T$ & $-0.3593$ \\
\bottomrule
\end{tabular}
\caption{Details of the measurements. 
The hopping parameter of the valence strange quark, $\kappa_s$, has
been determined as described in the text. 
The number of configurations on which we computed the correlations functions is
denoted by $N_{\rm cfg}$.
The number of smearing iterations applied to the light quark(s) in the $\K$ and $\Bs$ meson
is given by $\Nsm^\K$ and $\Nsm^\Bs_r$, respectively.
The integer vector $\veca{n}$ and the scalar twist angle $\vartheta$
determine the momentum of the $\K$ meson 
by setting $\theta^\mathrm{s} = \veca{n} \vartheta$ in \eqref{eqn:pK},
i.e. $\veca{p}_\K = (2\pi - \vartheta) \veca{n}/L$.
}
\label{tab:meas}
\end{center}
\end{table}

For the lattice computation of the correlation functions 
we use gauge field configurations generated by the 
Coordinated Lattice Simulations (CLS) effort
\cite{Fritzsch:2012wq} with two degenerate flavors of 
O$(a)$-improved 
Wilson fermions~\cite{Jansen:1998mx} and Wilson gauge action\cite{Wilson:1974sk}. 
In the following, we present only results from the two representative ensembles, 
labeled A5 and O7, see \tab{tab:ens}. The lattice spacings were determined from the Kaon 
decay constant $f_\K$ at physical pion mass in Ref.~\cite{Fritzsch:2012wq} and updated 
in Ref.~\cite{Lottini:2014zha}. They do not depend on the 
quark mass, i.e. we use a mass-independent renormalisation scheme.

The natural choice for the mass of the valence up quark is equal to
the mass of the degenerate up and down sea quarks, $\mlight$, and
thus fixed by the gauge field ensembles.
For the mass of the spectator (valence) strange quark we are free, 
however, to choose any smooth function $\mstrange(\mlight)$ passing
through the physical point. 
We follow strategy 1 of Ref.~\cite{Fritzsch:2012wq} and define $\mstrange(\mlight)$ 
by fixing the squared ratio between the Kaon mass and the Kaon decay constant to 
its physical value $R^{\rm phys}_\mathrm{K} \equiv (m^{\rm phys}_{\rm K}/f^{\rm phys}_\K)^2$ 
with $m^{\rm phys}_{\rm K} = 494.2\,\MeV$ and $f^{\rm phys}_\K= 155\,\MeV$. 
We expect that this will lead to a flat extrapolation to the physical 
value of $\mlight$, the ``physical point''. 
At the simulated (unphysical) values of $\mlight$, however,
the resulting Kaon mass is $m_\K(\mlight) = f_\K(\mlight)\cdot (R^{\rm phys}_\mathrm{K})^{1/2}$
and only approximately equal to its physical value (see e.g. Fig.~\ref{fig:kaon_disp}).

In the static correlation functions on the r.h.s. of Eq.~\eqref{eqn:corrfcthqet},
the heavy quark field $\psi_\mathrm{b}$ is replaced by a static quark 
$\psi_\mathrm{h} = \gamma_0\psi_\mathrm{h}$ with the HYP2 discretizations 
\cite{DellaMorte:2003mn,DellaMorte:2005yc}. 
At NLO also the $k \in  \{\rm kin,spin\}$ correlation functions are needed, see Appendix~\ref{s:appNLO}. 
They are static correlation functions with an extra insertion of $\cO^\mathrm{kin}$ or $\cO^\mathrm{spin}$
from the HQET action \cite{Heitger:2003nj}.
Finally there are the extra $1/m$ terms in the current $V_\mu$,
see Sec.~\ref{sec:currins} and Appendix~\ref{s:appNLO}.
We implement  O($a$) improvement perturbatively by writing the vector current as
\begin{subequations}
\begin{align}
 V_0^\textnormal{stat} &=  \overline \psi_\textnormal u \gamma_0          \psi_\textnormal h + ac_{\textnormal V_0}(g_0)
 \overline \psi _\textnormal l \sum_l \overleftarrow\nabla^\textnormal S_l \gamma_l          \psi_\textnormal h, \\
V_k^\textnormal{stat} &=  \overline \psi_\textnormal u \gamma_k          \psi_\textnormal h - ac_{\textnormal V_k}(g_0) 
\overline \psi _\textnormal l \sum_l \overleftarrow\nabla^\textnormal S_l \gamma_l \gamma_k \psi_\textnormal h\,.
\end{align}
\label{deltav}
\end{subequations}
with \cite{Grimbach:2008uy} 
\begin{equation}
  c_{\textnormal V_k} = 0.0518 g_0^2  \,,\quad
                 c_{\textnormal V_0}  = 0.0380 g_0^2\,
\end{equation}
when we evaluate the static matrix elements, but we use 
$c_{\textnormal V_k}=c_{\textnormal V_0} = 0$ 
when $V_\mu^\stat$ enters ratios at order $1/m$
such as \eq{e:rhomuk}. 

For the interpolating field of the $\Bs$ meson we use three different quark bilinears ($r = 1,2,3$)
\begin{subequations}
\begin{align}
   \cO_{\mathrm{bs},r}(x) & \equiv  
   \psibar_{\mathrm{h}}(x)  \gamma_5 \psi^\mathrm{sm}_{\mathrm{s},r}(x)\,,
   \\
   \overline{\cO}_{\mathrm{bs},r}(x) & \equiv  
   \psibar^\mathrm{sm}_{\mathrm{s},r}(x)  \gamma_0 \gamma_5 \gamma_0 \psi_{\mathrm{h}}(x)\,.
\end{align}
\label{eq:cObs}
\end{subequations}
The smeared light quark fields are 
constructed by $\Nsm^\Bs_r$ iterations of Gaussian smearing \cite{Alexandrou:1990dq,Gusken:1989ad}
\begin{equation}
    \psi^\mathrm{sm}_{\mathrm{s},r}= (1 - \kappa_\mathrm{G}\, a^2 \Delta )^{\Nsm^\Bs_r} \psi_s \, .
\end{equation}
where the gauge links in the covariant Laplacian $\Delta$ 
are defined by 3 iterations of (spatial) APE smearing \cite{Albanese:1987ds}
and $\kappa_\mathrm{G} = 0.1$.

The interpolating field for the Kaon, 
\begin{subequations}
\begin{align}
   \cO_{\mathrm{su}}(x_\mathrm{f}) & \equiv 
   \psibar^\mathrm{sm}_\mathrm{s}(x_\mathrm{f})  \gamma_5 \psi^\mathrm{sm}_\mathrm{u}(x_\mathrm{f})\,,
   \\
   \overline{\cO}_{\mathrm{su}}(x_\mathrm{f}) & \equiv 
   \psibar^\mathrm{sm}_\mathrm{u}(x_\mathrm{f}) \gamma_0 \gamma_5 \gamma_0 \psi^\mathrm{sm}_\mathrm{s}(x_\mathrm{f})\,,
\end{align}
\label{eq:cOsu}
\end{subequations}
is constructed from two smeared quark fields, $\psibar^\mathrm{sm}_{\mathrm{s}}$ and $\psi^\mathrm{sm}_{\mathrm{u}}$, 
each with $\Nsm^\K$ iterations of Gaussian smearing. 

Integrating out the Grassmann-valued quark fields yields
the correlation functions in terms of traces of products
of quark propagators and other factors such as smearing operators.
In order to make full use of translation invariance, we evaluate the traces by a stochastic estimator
which we represent by a single random $U(1)$ vector with support on time slice $(x_\mathrm{f})_0$. 
All values of 
$(x_\mathrm{f})_0$ are averaged over~(``full time dilution'').
More details can be found in Appendix \ref{sec:cf}.

The momentum transfer $q^2$ is computed from the continuum relation
\begin{equation}
    q^2 = (m^{\rm phys}_\Bs)^2 + (m^{\rm phys}_\K)^2 - 2 m^{\rm phys}_\Bs \sqrt{(m^{\rm phys}_\K)^2 + \veca{p}_\K^2}
\end{equation}
with $\veca{p}_\K$ in the rest frame of the $\Bs$ and physical values of the masses 
(using $m^{\rm phys}_\Bs = 5366.77\, \MeV$). 
In order to have $q^2$ fixed to the same value on all ensembles, we need to 
adjust $\veca{p}_\K$ accordingly. This is achieved by using flavor-twisted boundary
conditions for the (valence) $\mathrm{s}$ and $\mathrm{b}$ quarks, i.e.
instead of periodic boundary conditions, we impose 
\begin{subequations}
\begin{align}
\psi_\mathrm{s}(x_0,\vecx + \hat{\veca{k}}L) & = \, e^{i\theta_k^\mathrm{s}}\, \psi_\mathrm{s}(x_0,\vecx)\,,
\\
\psi_\mathrm{b}(x_0,\vecx +\hat{\veca{k}}L) & = \, e^{i\theta_k^\mathrm{b}}\, \psi_\mathrm{b}(x_0,\vecx)\,,
\end{align}
\label{eqn:theta}
\end{subequations}
for the unit vectors $\hat{\veca{k}}$ in $k$-direction. 
In this way, the momentum of the $\K$ meson becomes 
\begin{equation}
    (p_\K)_k = (2\pi n_k - \theta^\mathrm{s}_k)/L\,,
    \label{eqn:pK}
\end{equation}
while the $\Bs$ meson is kept at rest ($\veca{p}_\Bs \sim \theta^\mathrm{b}-\theta^\mathrm{s} = 0$) 
by choosing $\theta^\mathrm{b} = \theta^\mathrm{s}$. 
We fix
the value of $\vert\veca{p}_\K\vert = 0.535\,\GeV$, and hence $q^2 = 21.23\,\GeV^2$.
\footnote{
  Note that small changes of the input parameters lead to 
  $$
    \delta q^2 \approx 
    10\, \GeV \cdot \delta m_B 
    +  5\, GeV \cdot\delta m_K 
    +  2\,GeV^2 \cdot \delta \vartheta 
    +  5\,GeV^2 \cdot \displaystyle\frac{\delta a(\beta)}{a(\beta)}\,.
  $$
The particular value for $\veca{p}_\K$ was chosen 
on the ``N6'' lattice which we do not discuss here, but 
which will
be used in the physics analysis. On N6 we have $L/a=48$ 
and $\beta=5.5$ such that 
 $\veca{n} = (1,0,0)^T$ and $\theta^\mathrm{s} = \veca{0}$ lead to $\vert\veca{p}_\K\vert = 0.535\,\GeV$.
}

\section{Static correlation functions}
\label{sec:cf_stat}
For $\tk$, $\tb \gg a$, where any possible violations of
positivity are exponentially 
damped\footnote{In the pure Wilson regularization and with static quarks,
positivity of the lattice transfer matrix is exact.
With the O$(a)$-improvement term 
this may be different, but universality means that non-positive 
contributions may appear only at distances close to the lattice
spacing.
For a concrete example of such contributions we refer to~Ref.~\cite{Luscher:1984xn}.}, 
and assuming infinite ${\rm T}$ for the moment, we can decompose
the Euclidean correlation functions as (in the following we keep the $\veca{p}$
dependence of $\CK$ and $\Ct$ implicit):
\begin{align}
  \CK(\tk) =& \sum_{m=0}^{\infty}(\kappa^{(m)})^2e^{-E_\K^{(m)}
  \tk} \cong\sum_{m=0}^{N_\K-1}(\kappa^{(m)})^2e^{-E_\K^{(m)}
  \tk},\label{eq:c2ll}\\
  \CB_{rs}(\tb) =&
  \sum_{n=0}^{\infty}\beta_r^{(n)}\beta_s^{(n)}e^{-E_{\B}^{(n)}
  \tb} \cong\sum_{n=0}^{N_\B-1}\beta_r^{(n)}\beta_s^{(n)}e^{-E_{\B}^{(n)}
  \tb},\label{eq:c2hl}\\
  \begin{split}
  \C^{\Bs\to\K}_{\mu,r}(\tk,\tb)
  =&\sum_{n=0}^{\infty}\sum_{m=0}^{\infty}\kappa^{(m)}\varphi^{(m,n)}_\mu\beta_r^{(n)}
  e^{-E_\K^{(m)} t_\K} e^{-E_\Bs^{(n)} t_\Bs}\cong\\
  &\sum_{n=0}^{N_\K-1}\sum_{m=0}^{N_\B-1}\kappa^{(m)}\varphi^{(m,n)}_\mu\beta_r^{(n)}
  e^{-E_\K^{(m)} t_\K} e^{-E_\Bs^{(n)} t_\Bs},
  \end{split}
  \label{eq:c3}
\end{align}
where the indices $r,s$ label the smearing levels used for the $\B$ meson, while
the indices $m,n$ label the Kaon and $\B$ meson energy levels respectively. A rather small number of states
leads to a good precision if the time separations $t_\K, \,t_\Bs$ are sufficiently large. In our fits 
we will use $N_\K=1$ and $N_\B\leq3$.

By comparing the Euclidean representations of the correlation functions with
\eq{eqn:me1} we see that the bare form factors of interest are given by
$\varphi^{(0,0)}_\mu$ from \eq{eq:c3}, e.g.
\begin{align}
   \langle\K(\veca{p})|\hat{V}_{\rm M}^0(0)|\B(\veca{0})\rangle =&~ \varphi^{(0,0)}_0. 
   \label{eq:ffmel}
\end{align}
In the following we describe two methods to determine them:
\begin{enumerate}
  \item{By directly fitting\footnote{All the fits in this paper are uncorrelated,
  i.e.\ we use a diagonal covariance matrix.}} the correlation functions in a window of time
  extents where one can reliably limit the influence of excited states,
  statistical noise and other systematic effects (such as the finite-$T$
  extent of the available periodic lattices). This will be the focus of Section
  \ref{sec:fits_stat}.
  \item{By forming suitable ratios of correlation functions (like the one in
\eq{eqn:me1}) where at large enough $t$ the dependence on all or most of the
additional parameters cancels out. This will be the topic of Section
\ref{sec:rat_stat}.}
\end{enumerate}

As we shall see, for both of these methods it is beneficial to first obtain a
good determination of the two-point function parameters, as well as to determine
regions in the three-point functions where one is safe from the finite-$T$
effects. These prerequisite steps are described in the following subsections.
First we describe our methods to extract $\kappa^{(0)}$ and $E_\K^{(0)}$ from \eq{eq:c2ll}.

\subsection{Kaon correlation functions}
\label{sec:ck}
While \eq{eq:c2ll} is true in the limit of infinite $T$, in
practice we have to take into account that $T$ is finite.
With our periodic boundary conditions in time, the ground-state contribution reads
\begin{equation}
\CK(t) \cong (\kappa^{(0)})^2 (e^{-\EK{0}t}+e^{-\EK{0}(T-t)}).
\label{eq:c2ll_nk1}
\end{equation}

Since the signal-to-noise problem in the Kaon sector is
very mild, we average $\CK(t)$ and $\CK(T-t)$ and always fit until the middle of the lattice, $\tfit{K2}{max}=T/2$.
For the start of the fit range, we follow
Refs.~\cite{strangemass,Bahr2015Form} and use the (rather conservative) criterion
that $t^{\K2}_{\rm min}$ is the first $t$ for which the excited-state
contribution, estimated by a two-state fit, is less than 25 \% of the statistical
error of the effective mass 
$E_{\K,{\rm eff}}(t) = - \frac{1}{a}\log[\CK(t+a)/\CK(t)]$. In practice, we obtain
$\tfit{K2}{min}\approx1.3-1.4\,\textrm{fm}$.
In this range, we fit the correlator to \eq{eq:c2ll_nk1} to get the ground-state
energy and amplitude. Note that the gap between the ground state and excited
states in the Kaon sector is large, approximately 
$1\,\GeV \approx 5\,\fm^{-1}$.

\begin{figure}[tbp!]
\begin{center}
\includegraphics[width=12cm]{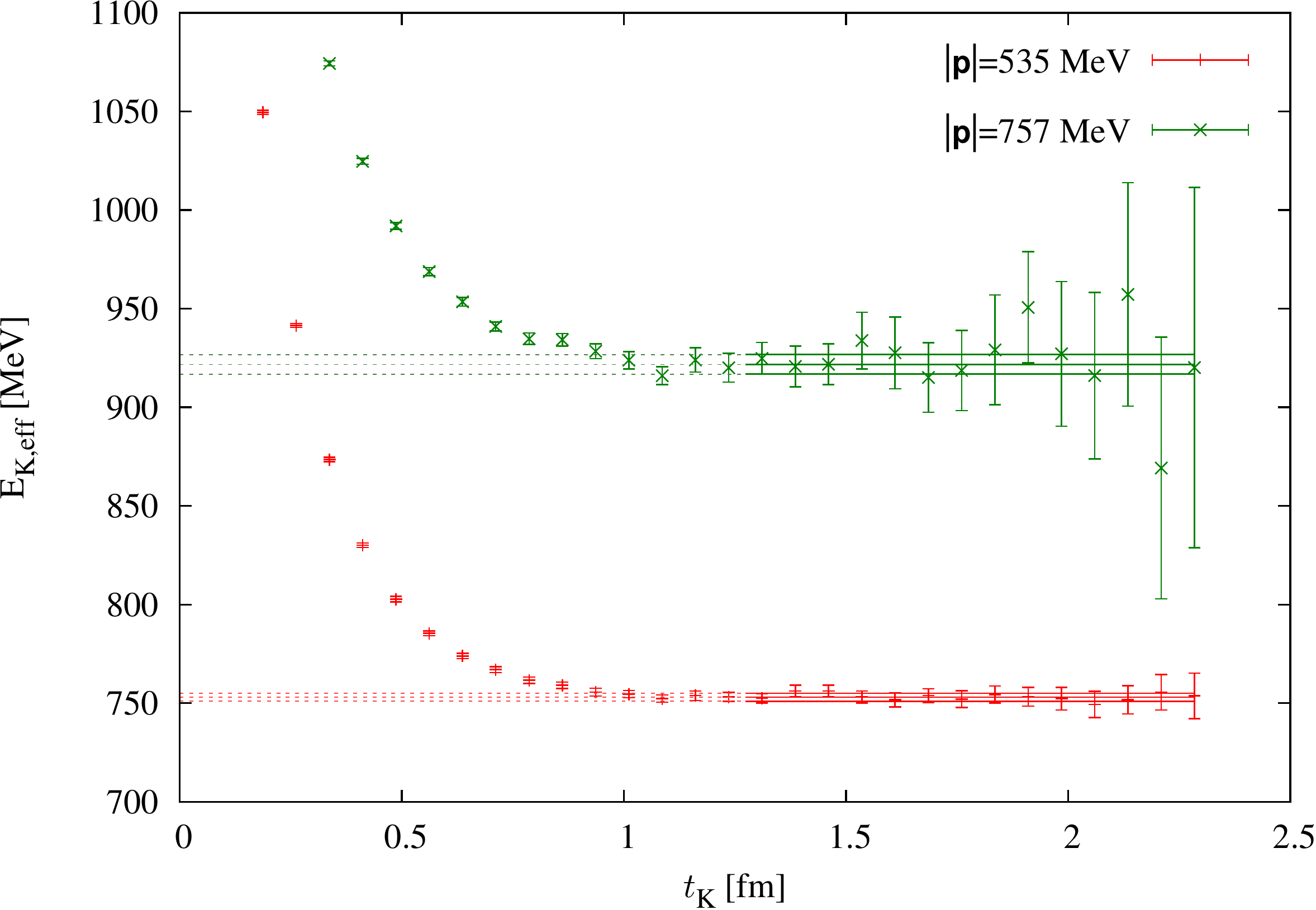}
\caption{Effective masses, $E_{\rm K, eff}(t_\K) = -\partial_t \log\C^{\K}(t_\K)$ for ensemble A5 and two values
of momentum, together with the fitted energies. In this graph as well as all 
following ones we use MeV and fm units just for illustration. We thus convert 
from lattice units to physical ones {\em without} taking into
account the error in the lattice spacings. The lattice parameters for the Kaon with
momentum $\vert\veca{p}\vert= 535$ MeV are given in Table \ref{tab:meas}. For the Kaon at
$\vert\veca{p}\vert= 757$ MeV, we have used $\veca{n} = (1,0,0)$ and $\vartheta = -0.9263$.}
\label{fig:kaon_meff}
\end{center}
\end{figure}

\begin{figure}[tbp!]
\begin{center}
\includegraphics[width=12cm]{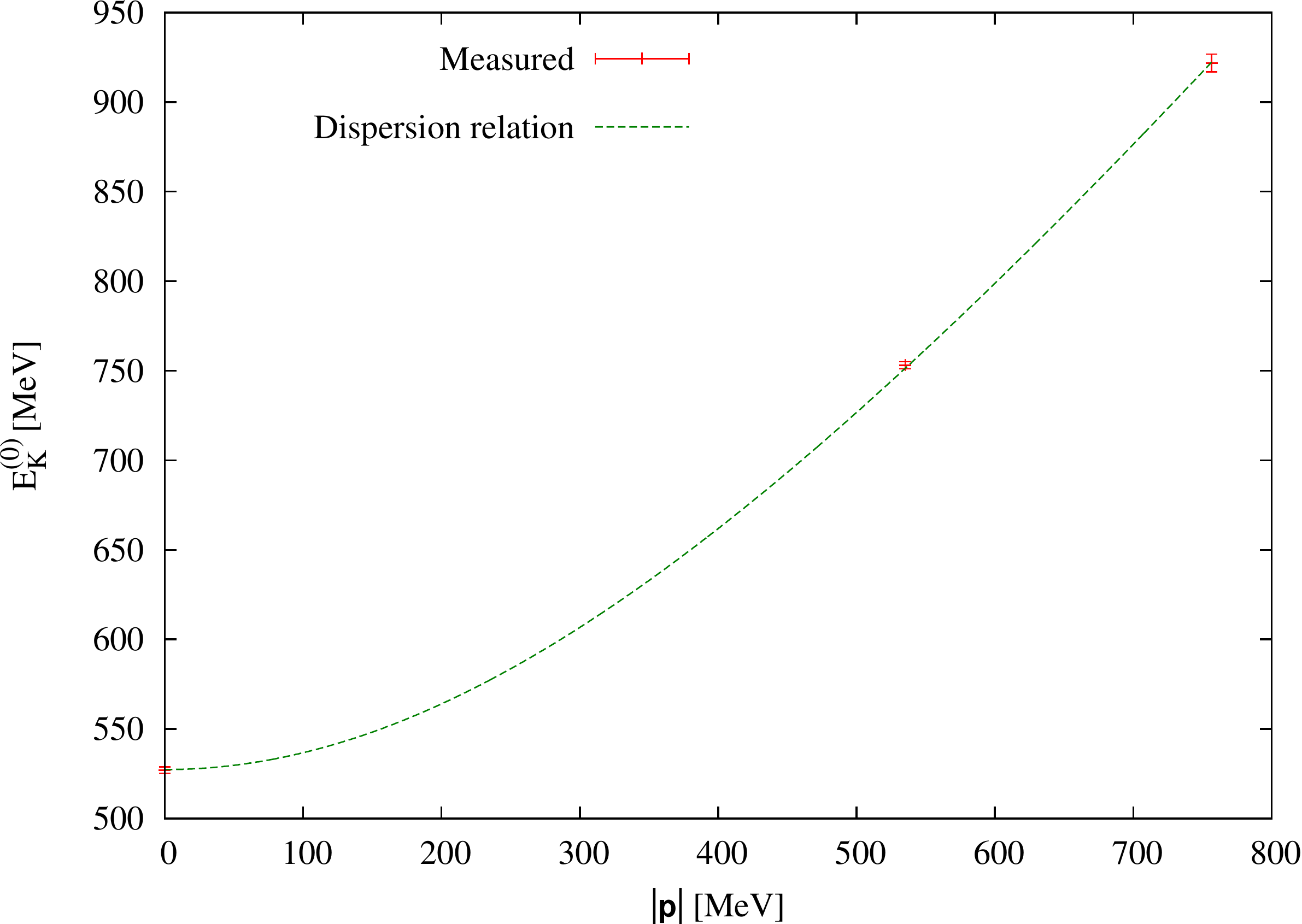}
\caption{Kaon energies at three values of momentum for ensemble A5. The lattice parameters are indicated
in the caption of Fig \ref{fig:kaon_meff}. The dashed curve shows the continuum dispersion relation evaluated 
with the rest mass from Ref.~\cite{strangemass} (it is not a fit).}
\label{fig:kaon_disp}
\end{center}
\end{figure}

To check the behaviour of the signal-to-noise ratio and the
dispersion relation, we calculated the Kaon two-point function 
at two values of momentum.
The effective mass plots for both momenta are presented together with the corresponding fits in
Fig.~\ref{fig:kaon_meff}. At higher momentum the signal-to-noise
problem becomes significantly more severe. The dispersion relation, also
including a measurement for $\veca{p}=0$, is presented in
Fig.~\ref{fig:kaon_disp}. The continuum dispersion relation describes the data
very well even at the coarsest lattice spacing, showing that the cutoff
effects in $E_\K/m_\K$ are below our statistical precision.

\subsection{$\B$ correlation functions}
\label{sec:cb}
For the two-point heavy-light ($\B$) correlator, we have three different
smearings (in addition to the unsmeared correlator, which is not used due to its
large contamination from highly excited states). With the diagonal and the
off-diagonal terms, we have six independent correlators in a symmetric $\CB_{rs}$ matrix.
The determination of the parameters of the $\B$ correlation functions is divided
into several steps, which are described in the following subsections.

\subsubsection{Energies from the GEVP}
\label{sec:gevp}
\begin{figure}[tbp!]
\begin{center}
\makebox[\textwidth][c]{
\includegraphics[height=7cm]{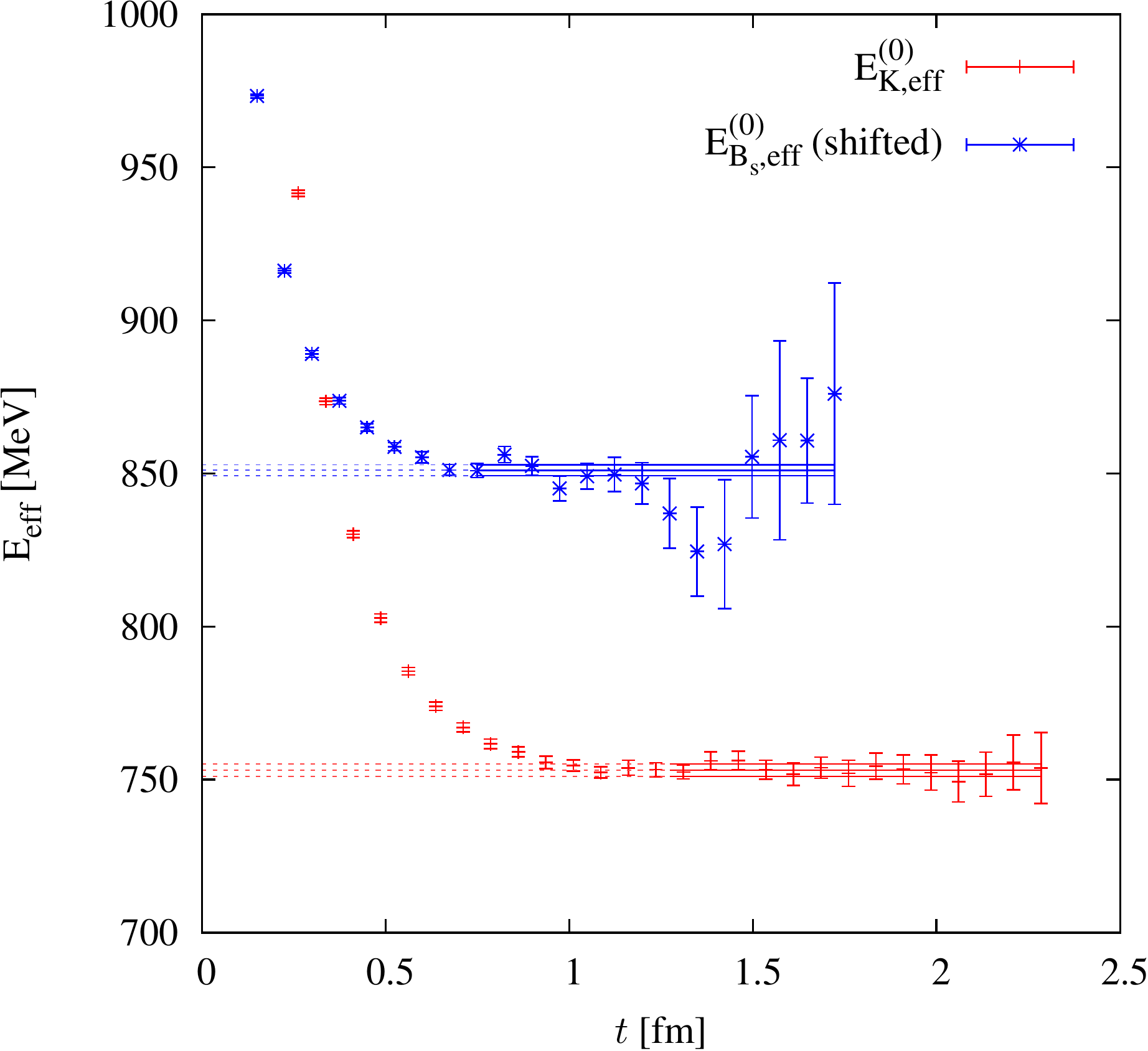}
\hspace{0.2cm}
\includegraphics[height=7cm]{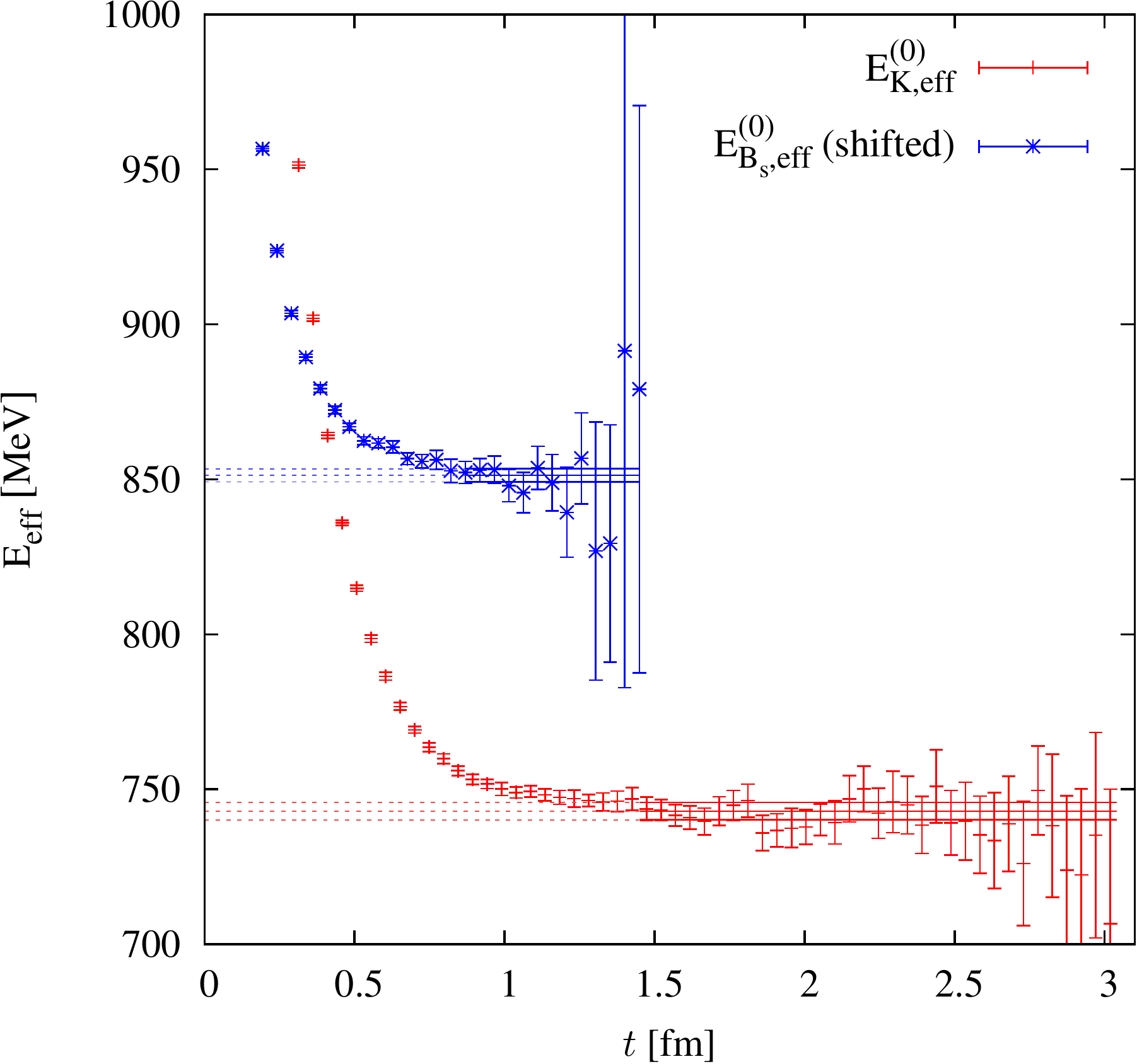}
}
\caption{Dependence of the $\K$ and $\B$ ground-state energies on $t$ for
ensemble A5 (left) and O7 (right). $E_{\Bs,{\rm eff}}^{(0)}$ 
 is shifted
vertically for presentation. Dashed (and full) horizontal lines show the plateau fit 
values (and ranges).}
\label{fig:E0_gevp}
\end{center}
\end{figure}

We first determine the energies by solving the generalized eigenvalue problem (GEVP) on $\CB_{rs}$
\cite{gevplw}. We use all $N_{\B}=3$ interpolating fields. The GEVP is defined as
\begin{equation}
\CB(t)v^{(n)}(t,t_0) = \lambda^{(n)}(t,t_0)\, \CB(t_0) \, v^{(n)}(t,t_0)\,, \quad n=1,\ldots, N_\B\,,
\end{equation}
and yields approximations to the lowest $N_{\B}$ energy 
levels via
\begin{equation}
E_{\Bs,{\rm eff}}^{(n)}(t) = -\tilde{\partial}_t \log\big(\lambda^{(n)}(t,t_0))\big) = \frac1{2a} \log\big(\lambda^{(n)}(t-a,t_0)/\lambda^{(n)}(t+a,t_0)\big).
\end{equation}
We use $t_0=\lceil t/2\rceil$ since this asymptotically accelerates the 
convergence (with $t$) to the energies $E_\Bs^{(n)}$ \cite{gevp}.
For each $n$ separately we then find the GEVP estimate of the energy 
$E_{\Bs}^{(n)}$ by a plateau fit (a weighted average) of $E_{\Bs,{\rm eff}}^{(n)}(t)$. The
upper end, $t_\textrm{max}$, of the plateau is chosen with a noise criterion: only 
$E_{\Bs,{\rm eff}}^{(n)}(t)$ with a relative error less than 20\% (50\% for the excited
states) are included in the average. The start of the plateau, $\tmin$, is chosen by requiring
$
r(\tmin) =
\frac{|E^{(n)}_\textnormal{plat}(\tmin)-E^{(n)}_\textnormal{plat}(\tmin-\delta
t)|} {\sqrt{\sigma^2(\tmin)+\sigma^2(\tmin-\delta t)}} \leq 3,
\label{eq:gevp_plat}
$
where $E^{(n)}_\textnormal{plat}(\tmin)$ is the result of the plateau fit
starting at $\tmin$ and $\delta t \approx 2/(E_\Bs^{(N)}-E_\Bs^{(0)})$, which we
approximate by choosing $\delta t = 0.3$ fm.
For the ground state this criterion coincides with that of Ref.~\cite{bqmass},
ensuring that the statistical error dominates over the systematic one. Examples
of $E_{\Bs,{\rm eff}}^{(0)}(t)$ are presented in Fig.~\ref{fig:E0_gevp}
(together with $E_{\K,{\rm eff}}(t)$ for comparison). Note that a good
estimate of the ground-state energy will be important in
Sec.~\ref{sec:rat_stat}.

For the excited states one should in principle use a higher value of $\delta t
\approx 2/(E_\Bs^{(N)}-E_\Bs^{(n)})$ but this is not possible with the current precision
of the data, as one ends up in regions dominated by noise. However, as opposed
to the ground state, for the excited-state energies we only need reasonable
rough estimates, as any residual systematic error will be eliminated in the
subsequent fitting steps as described in Sec.~\ref{sec:fits_stat}.

\subsubsection{Amplitudes}
Having obtained the energies, we determine the amplitudes $\beta_r^{(n)}$. We
start with the diagonal elements of $\C^{\Bs}$,
\begin{equation}
\C_{rr}^\Bs(t) \cong \sum_{n=0}^{N_\Bs-1} (\beta_r^{(n)})^2 e^{-\EB{n}t}\,
\end{equation}
and estimate the squared amplitudes through a linear fit. These
values are used as initial guess for a non-linear fit for the amplitudes $\beta^{(n)}_r$ to all the
correlators, including the off-diagonal ones. In this way also
the relative signs of the amplitudes are obtained.

Then, we try the full fit for both amplitudes and energies whenever
possible. On certain ensembles the full fit becomes unstable with $N_\Bs=3$,
in this case we fall back to the fit with fixed energies.

The upper end of the fit range, $\tfit{B2}{max}$, is common for all the six
correlators and is determined by the relative noise criterion, i.e. as the
last point at which the relative error is smaller than 2.5\% for all of the
$\CB_{rs}$. For our ensembles this gives $2.1-2.4\;\textrm{fm}$. Also the 
lower end of the fit range, $\tfit{B2}{min}$, is the same for all $\CB_{rs}$.
It is a tunable parameter that is common for the determination of the amplitudes 
and for the combined fit (cf. Sec.~\ref{sec:fits_stat}).

\subsection{Three-point static correlation functions}
\label{sec:3pt_CF}

\begin{figure}[h]
\centering
\includegraphics[width=5.8cm]{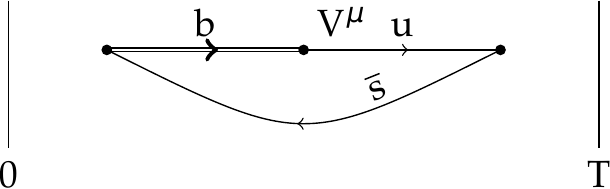}
\hspace{0.6cm}
\includegraphics[width=5.8cm]{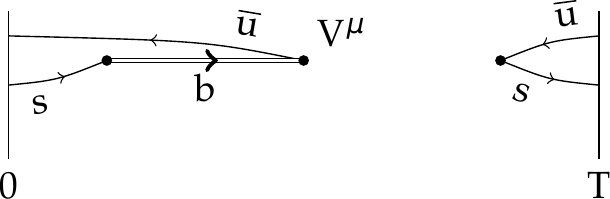}
\caption{Contributions to the three-point function corresponding to the 
$\Bs\to\K$ decay (left), and a wrap-around Kaon (right). Figures taken from Ref.~\cite{Bahr2015Form}.}
\label{fig:wrap}
\end{figure}

Due to the finite time extent of the lattice, the three-point function also receives
contributions from particles propagating (``wrapping'') around the torus in time
direction and is the sum of the contributions from the two
diagrams shown in Fig.~\ref{fig:wrap}. At large enough time separations, the
contributions of the corresponding lowest states can be written as

\begin{equation}
\begin{aligned}
\C^{\Bs\to\K}_{\mu,r}(t_\K,t_\Bs) \cong &\kappa^{(0)}\varphi^{(0,0)}_\mu
\beta_r^{(0)} e^{-\EB{0}t_\Bs}e^{-\EK{0}t_\K}+\\
&\kappa^{(0)}\xi_{\mu,r} e^{-E_{{\rm B}^*}t_\Bs} e^{-\EK{0}(T-t_\Bs-t_\K)},
\end{aligned}
\label{eq:wrap_full}
\end{equation}
where $\xi_{\mu,r}=\langle0|V_\mu|{\rm B}^*\rangle\langle{\rm
B}^*|P_{\rm hl}|\K\rangle$ contains unknown matrix elements of the state
${\rm B}^*$, the lightest heavy-light state contributing to the ``wrapper'' diagram.
At static order the energy $E_{{\rm B}^*}$ is equal to $\EB{0}$.

Instead of introducing extra fit parameters $\xi_{\mu,r}$ (and possibly others, if
excited states need to be included), we restrict ourselves to time separations 
where the wrapper contributions are negligible.

\begin{figure}[tbp!]
\begin{center}
\includegraphics[width=11cm]{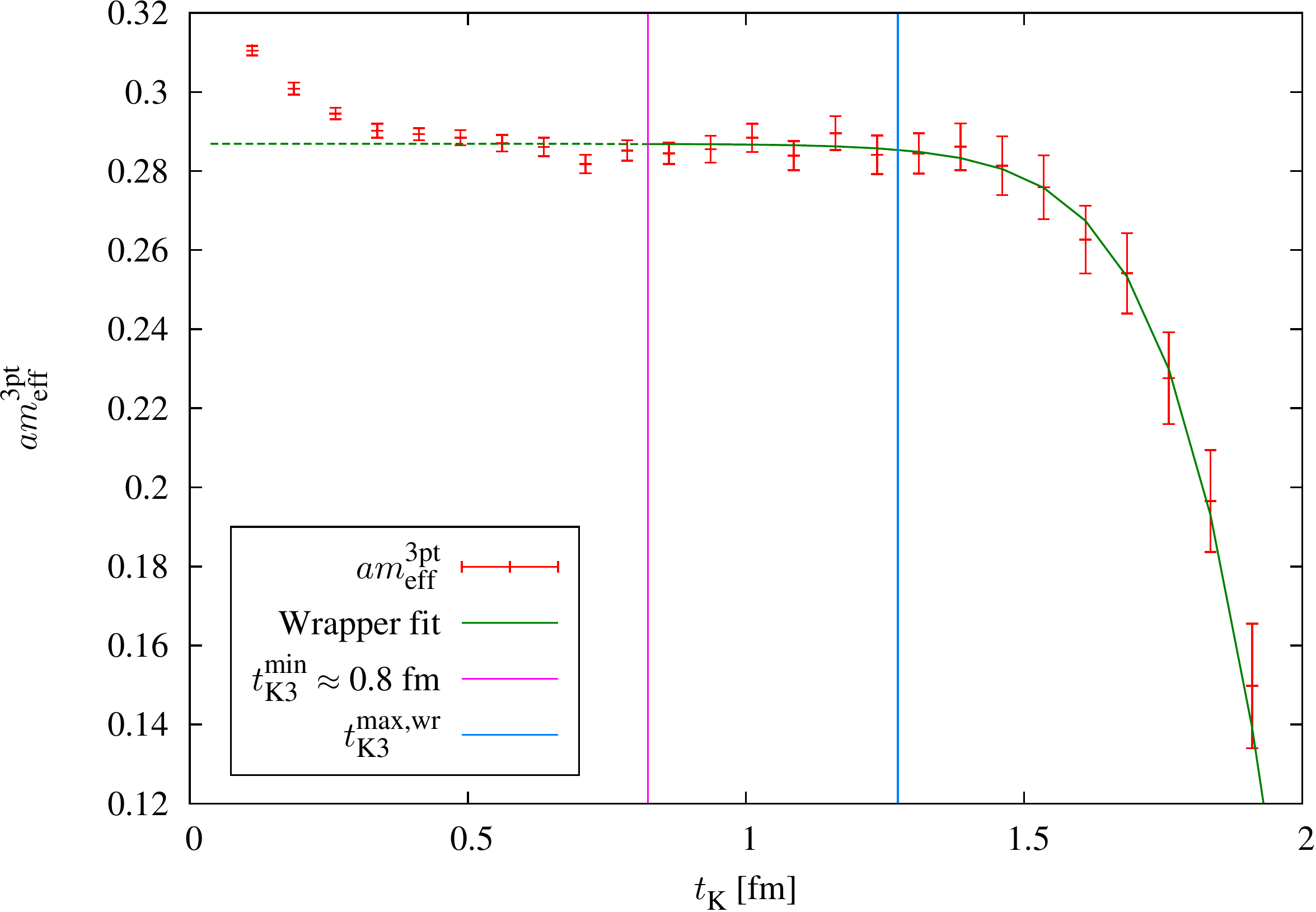}
\caption{The effective mass $am^{\rm 3pt}_{\rm eff}(\tk) =
\log(C^{\Bs\to\K}_{1,3}(\tk,t_\Bs)/(C^{\Bs\to\K}_{1,3}(\tk+a,t_\Bs))$ with $t_\Bs = 8a \approx 0.6\,\fm$ for
ensemble A5, together with the fitted function. The start of the finite-$T$ fit
and $\tfit{K3}{max,wr}$ are also shown.}
\label{fig:tk_prof}
\end{center}
\end{figure}

For this purpose, at every fixed $t_\Bs$ we fit the three-point function
to the form
\begin{equation}
\C^{\Bs\to\K}_{\mu,r}(t_\K) \cong
B_{\mu,r}e^{-E t_\K}+C_{\mu,r}e^{+E t_\K},
\end{equation}
with $B_{\mu,r}$ and $C_{\mu,r}$ being linear fit parameters (which one can express in terms
of the amplitudes and matrix elements from \eq{eq:wrap_full}) and $E$
being a non-linear fit parameter (although one could in principle set it to
$E_\K$ extracted from the two-point function). The fit is done in a region
where ground-state dominance in the Kaon sector is expected (starting at
approximately $\tfit{K3}{min}=0.8$ fm).
Fig.~\ref{fig:tk_prof} shows an example of such a fit at fixed $t_\Bs$.

For every $t_\Bs$, $\mu$, and $r$ separately, we then find $\tfit{K3}{max,wr}$ as
the last $t_\K$ that fulfills the condition:
\begin{equation}
\frac{C_{\mu,r}e^{+E t_\K}}{B_{\mu,r}e^{-E
t_\K}+C_{\mu,r}e^{+E t_\K}}<c_{\rm
wr}\,\frac{\delta\C^{\Bs\to\K}_{\mu,r}(t_\K,t_\Bs)} {\C^{\Bs\to\K}_{\mu,r}(t_\K,t_\Bs)} \,,
\label{eq:wrapper}
\end{equation}
with $c_{\rm wr}=0.25$. 

\begin{figure}[tp!] \centering \makebox[\textwidth][c]{
\includegraphics[height=8.25cm]{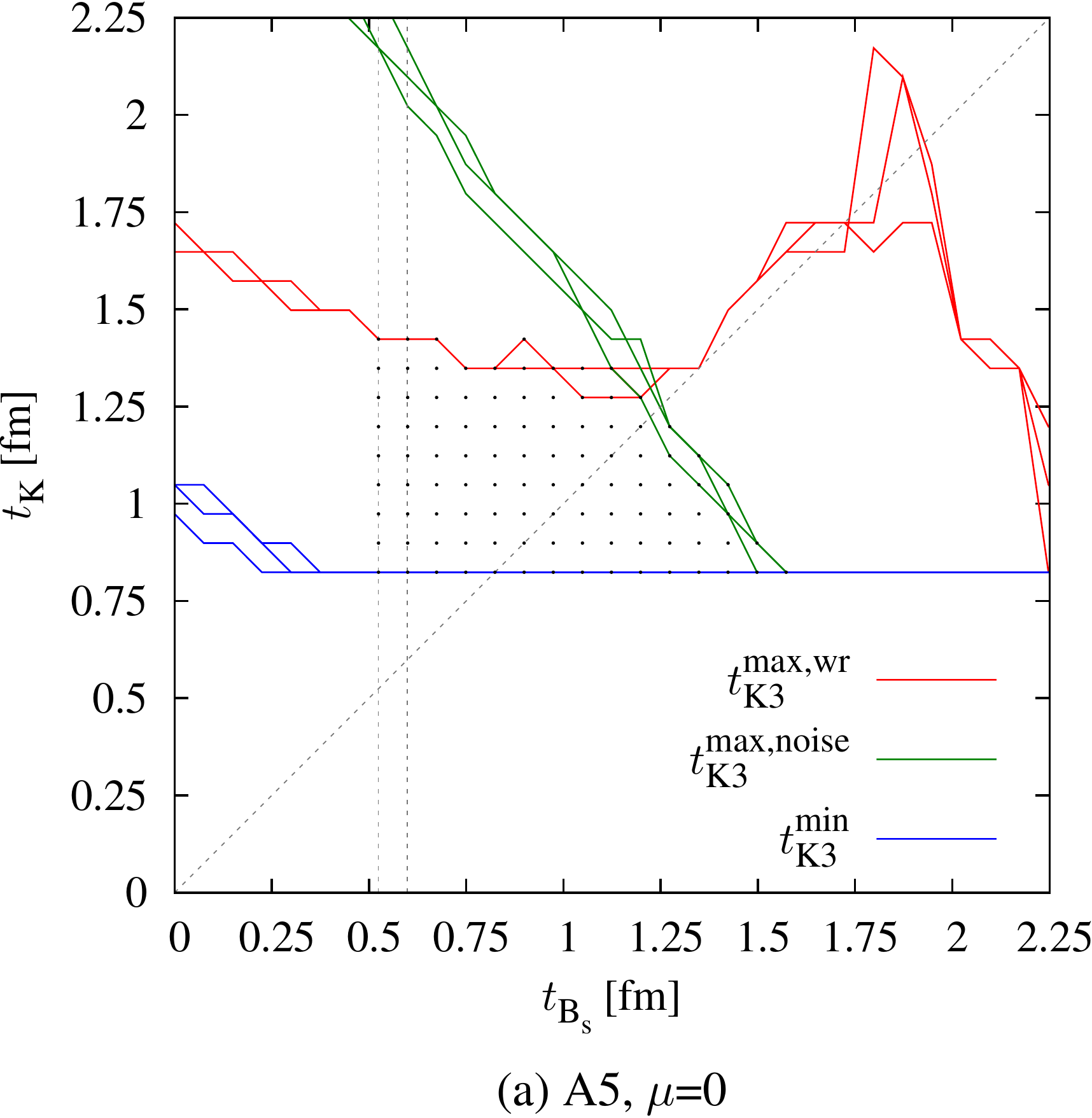}
\hspace{0.5cm}
\includegraphics[height=8.25cm]{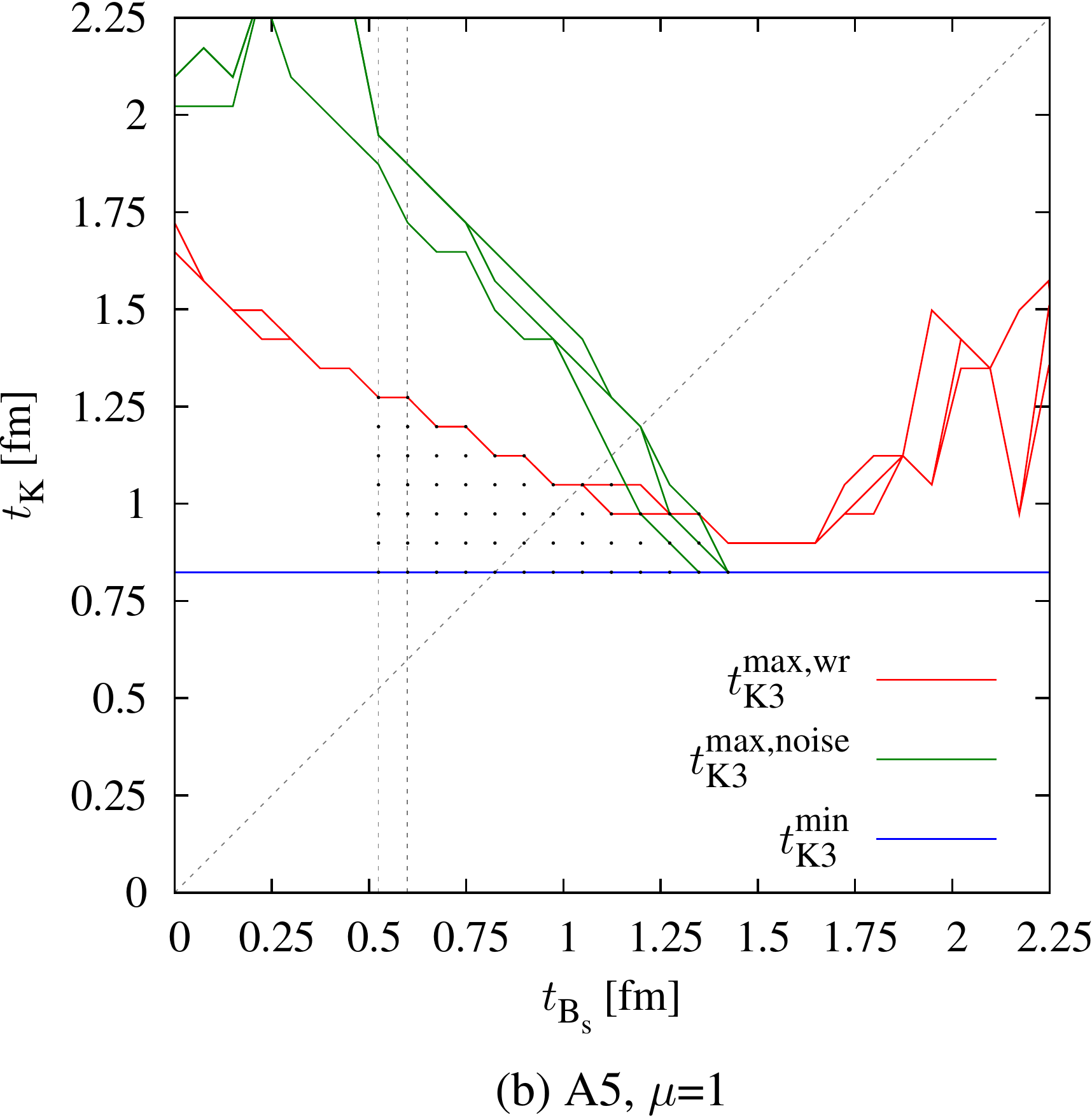}}\\[15pt]
\makebox[\textwidth][c]{
\includegraphics[height=8.25cm]{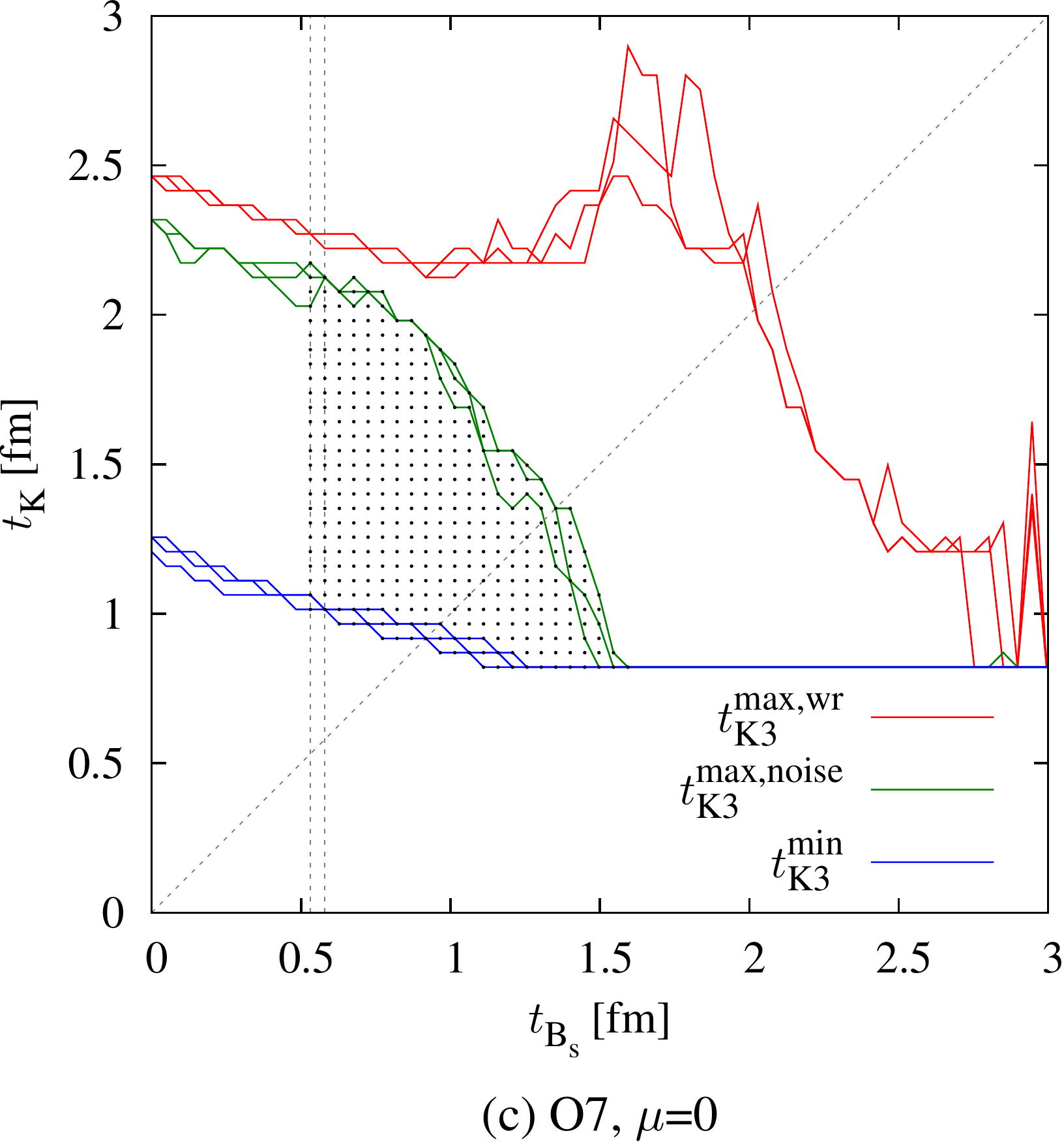}
\hspace{0.5cm}
\includegraphics[height=8.25cm]{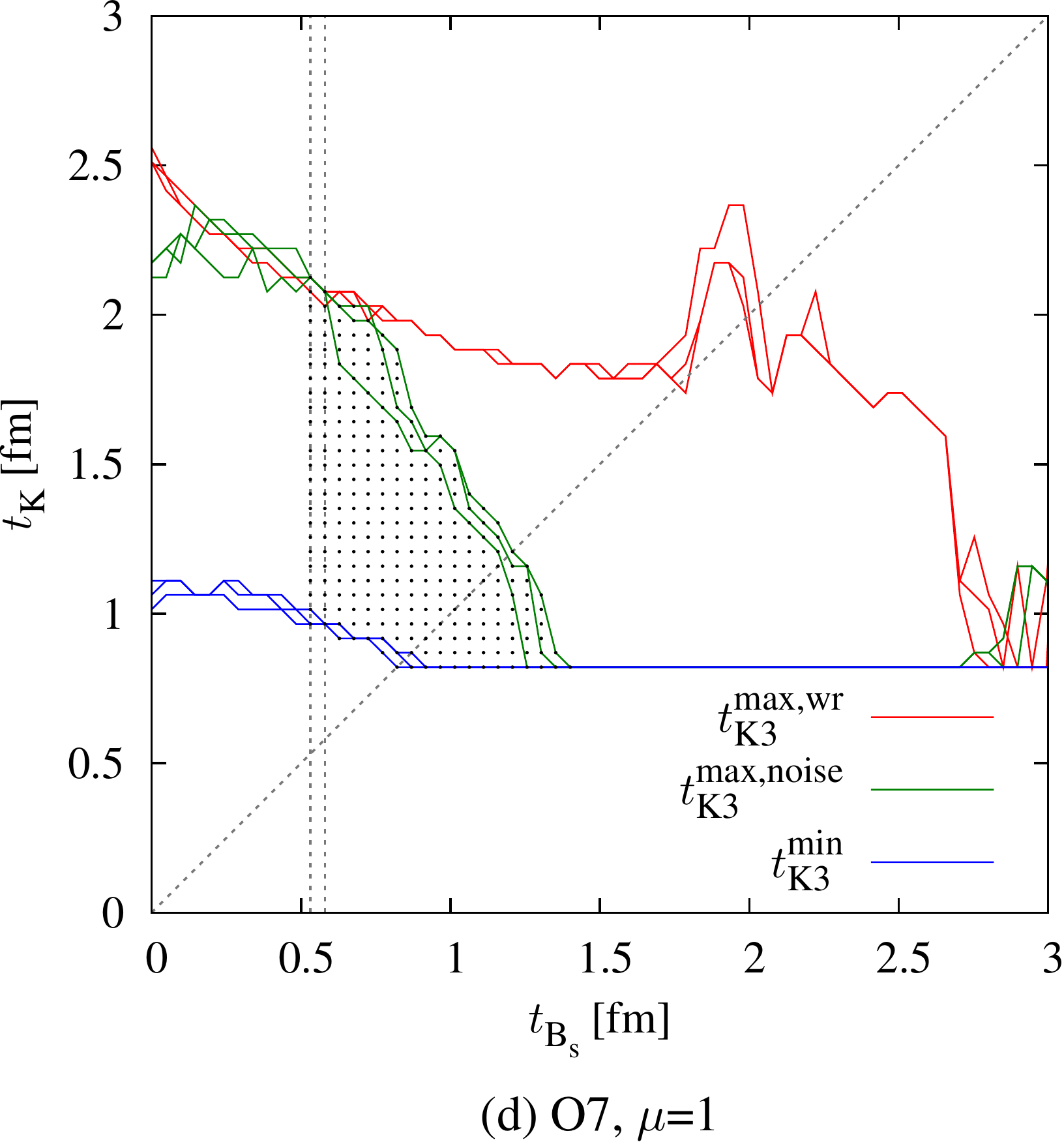}}\\[3pt]
\caption{Time ranges $\tfit{K3}{max,wr}$, $\tfit{K3}{max,noise}$, and
$\tfit{K3}{min}$ for lattices A5 and O7. 
The different lines with same color show the values for the three smearings.
The vertical dashed lines are the two choices of $\tfit{B3}{min}$ for
the combined fit of Sec.~\ref{sec:fits_stat} and the dots correspond to
the data points included in the fit.
To guide the eye, we also show the line $\tk=\tb$ which is relevant for the ratios in Sec.~\ref{sec:rat_stat}. 
}
\label{fig:tk_min_max}
\end{figure}

The times $\tfit{K3,\mu}{max,wr}(t_\Bs)$ are shown by the red lines 
in Fig.~\ref{fig:tk_min_max} for the two representative ensembles A5 and O7 
(together with other constraints for the fits as discussed in Sect. \ref{sec:fits_stat}). 
We observe that the wrapper criterion yields considerably different $\tfit{K3}{max,wr}$
for $\mu=0$ and $\mu=1$ (with lower $\tfit{K3}{max,wr}$ for $\mu=1$). On the
other hand, the different smearing levels give very similar limits, but are
still all treated separately.

\section{Matrix elements at static order from a combined fit}
\label{sec:fits_stat}
From the two-point function fits, we have obtained (estimates of) the energies 
and amplitudes, $\kappa^{(m)}$, $\beta_r^{(n)}$. Taking them as input
values, we can now fit \eq{eq:c3} for each $\mu=0,1$ with 
$\varphi^{(m,n)}_\mu$, i.e. the form factors, as $N_\K\times N_\Bs$ linear parameters.
These could in principle serve as the final result. However, for better
stability, accuracy, and control of the systematic errors, we take all these
results only as initial conditions for a combined simultaneous fit to the two-point and
the three-point functions. In this way, we can use the information on the
energies and amplitudes also from the three-point functions, which contain many
data points. To estimate the errors we use \cite{Wolff:2003sm,Schaefer:2010hu} 
but with the derivatives with respect to the fit parameters calculated analytically.

The numerical results presented in this section are for the fit with $N_\K=1$ 
and $N_\Bs=3$, corresponding to a total of 20 fit parameters\footnote{$N_\K+N_\Bs$ energies,
$N_\K^2+N_\Bs^2$ amplitudes, and $2 \times N_\K\times N_\Bs$ form factors}. We find that
this choice gives the best stability with respect to the changes of the fit ranges,
and therefore we use it to extract the values of $\varphi^{(0)}_{\mu}$. An
alternative fit with $N_\K=1$, $N_\Bs=2$ is discussed in Appendix
\ref{sec:fit12_stat}. The fits with both $N_\K > 1$ and $N_\Bs > 1$
can easily become very unstable without some form of prior assumptions about the
values of the excited-state form factors.

To keep statistical and systematic errors under control, 
only a small and carefully chosen subset of all the
available time separations $0 \le t_\K + t_\Bs < T$
of the three-point function in the $t_\Bs$-$t_\K$-plane can be used in the fit.

Moreover, it is important to have a criterion for good determination of the 
ground-state form factors by the combined fit. However, it is not easy to find 
a simple strict criterion, like the one from Sec.~\ref{sec:ck} for $E_\K$,
and we monitor several criteria.

Details of both issues, the choice of the fit ranges and the quality of the fit,
are discussed in the following.

\subsection{Fit ranges}
\label{sec:fit_ranges}

For the three-point functions we select the data points to be included
in the combined fit by the constraints 
$\tfit{B3}{min} \le t_\Bs $ and $\tfit{K3}{min}(t_\Bs) \le t_\K \le \tfit{K3}{max}(t_\Bs)$.

The function $\tfit{K3}{max}(t_\Bs)$ is chosen, for each $\mu$ and smearing level
separately\footnote{
   In \cite{Bahr2015Form,Bahr:2016ayy} we used one and the same value of
   $\tfit{K3,\mu}{max}$ for all $t_\Bs$, corresponding roughly to the one that
   produced a rectangle of largest area in the plane of the allowed points (for
   both values of $\mu$ separately).
}, as 
\begin{equation}
\tfit{K3}{max}(t_\Bs) = \min\big\{\tfit{K3}{max,wr},\,\tfit{K3}{max,noise}\big\}\,,
\end{equation}
were $\tfit{K3}{max,wr}(t_\Bs)$ is given by the wrapper criterion of \eqref{eq:wrapper}
and $\tfit{K3}{max,noise}(t_\Bs)$ is determined from a relative noise criterion,
i.e. as the largest  $t_\K$ which fulfills
\begin{equation}
\frac{\delta\C^{\Bs\to\K}_{\mu,r}(t_\K,t_\Bs)}{\C^{\Bs\to\K}_{\mu,r}(t_\K,t_\Bs)}<c_{\rm
noise}\,,
\end{equation}
with $c_{\rm noise}=2.5\%$. The corresponding curves are shown in Fig.~\ref{fig:tk_min_max}
and illustrate that A5 is an example of a wrapper-limited lattice (due to a small physical 
time extent). On the other hand, O7 is noise limited and by reducing statistical errors 
one may hope to improve the determination of the form factors.

Also the function $\tfit{K3}{min}(t_\Bs)$ is selected by an automatic 
criterion\footnote{In Ref.~\cite{Bahr:2016ayy} $\tfit{K3}{min}$ was a manually
tunable parameter of the fit proceduce, chosen in common for all smearings and $\mu$.
}
analogous to the one described in Sec.~\ref{sec:ck}: for each $t_\Bs$, $\mu$, and smearing, we do a 
two-exponential fit to $t_\K\in[0.4\;{\rm fm}, \tfit{K3}{max}(t_\Bs)]$ (with the 
energies fixed to the ones extracted from the two-point function). We then find 
the minimum time $\tfit{K3*}{min}(t_\Bs)$ at which the excited-state contribution 
is smaller than 25\% of the statistical error. The final value is set to
\begin{equation}
\tfit{K3}{min}(t_\Bs) = \max\big\{0.8\;{\rm fm},\,\tfit{K3*}{min}(t_\Bs)\big\}
\end{equation}
to avoid values getting too small in the regions with large noise, where
the two-state fit does not work well. The resulting $\tfit{K3}{min}$ 
is shown by the blue curves in Fig.~\ref{fig:tk_min_max}.

The remaining tunable parameters of the fit are $\tfit{B2}{min}$ and $\tfit{B3}{min}$.
In Ref.~\cite{Bahr:2016ayy} we used $\tfit{B2}{min} \approx 0.45$ \fm, which seems 
sufficient when $N_{\B}=3$. With better
smearings and improved analysis methods, we can use slightly lower values, with
two choices: $\tfit{B2}{min}\approx 0.32\,\fm$ (``aggressive'') and $0.38\,\fm$ (``conservative'').

Concerning $\tfit{B3}{min}$ it was observed in Ref.~\cite{Bahr:2016ayy} that the fit 
performs better
if one takes a positive $\tfit{B\Delta}{min} \equiv \tfit{B3}{min} - \tfit{B2}{min}$.
This can be understood by noting that the amplitudes in the two-point function are 
proportional to $|\beta_i^{(n)}|^2$, while in the three-point functions they are only
proportional to $\beta_i^{(n)}$. Thus, the suppression of excited states is
expected to be stronger in the two-point functions if $\beta_i^{(0)} > \beta_i^{(1)}$.
We find that this is in fact the case, except for the lowest
smearing.\footnote{For the ensembles we analysed, typical values are:
$\beta_3^{(1)}/\beta_3^{(0)} \approx 0.25-0.4$, $\beta_2^{(1)}/\beta_2^{(0)}
\approx 0.67$, $\beta_1^{(1)}/\beta_1^{(0)} \approx 1$.} 

With increasing time separation $t_\Bs$ the suppression of the first excited state 
is proportional to $e^{-(E_\Bs^{(1)}-E_\Bs^{(0)})t}$, which approximately amounts to 
0.67 at $t=0.2\;\fm$, 0.4 at $t=0.4\;\fm$, and 0.25 at $t=0.5\;\fm$. 
Although it is not immediately obvious which value one should take, 
this justifies a non-zero value of $\tfit{B\Delta}{min}$.
We keep $\tfit{B\Delta}{min} \approx 0.2\,\fm$, which gives a
good compromise in suppressing the higher excited-state contributions and
keeping the number of data points large enough for good fit stability.

\subsection{Fit quality and results}
\label{sec:fit_qual}

The combined fit is uncorrelated, i.e.\ we use a diagonal covariance matrix.
For such a big matrix the full covariance matrix is very badly determined and conditioned (in
fact, if the number of data points is bigger than the number of measurements,
it is not possible to invert \cite{Michael:1993yj,Michael:1994sz}).

As a way to monitor the quality of the uncorrelated fit, we use $\chi^2_{\rm
exp}$, the expectation value of $\chi^2$ given normally distributed
data with the measured covariance matrix~\cite{chiexpectBB}.
It can be estimated efficiently including 
autocorrelations \cite{chiexpectBS}, but
the uncertainty on its determination can be large and
it is not easy to exactly quantify what is a good fit.
In general we consider fits with $\chi^2 \lesssim 2 \chi^2_{\rm exp}$ as
acceptable.

Another way of monitoring the fit quality is to look at the contributions to the
$\chi^2$ coming from different correlation functions. In particular, an
unusually large contribution from $\CB$ should be monitored as it can indicate
that $\tfit{B2}{min}$ was chosen too small (even though the overall $\chi^2$ 
looks acceptable because of the many points in the three-point functions).

Apart from monitoring the $\chi^2$, we analyze the stability of the extracted fit 
values and errors with respect to changes of $\tfit{B2}{min}$ and $\tfit{B3}{min}$. 

The stability plots are always organized in the following way: the data is
divided in groups with different $\tfit{B2}{min}$, which is also plotted on the
$x$-axis in the middle of each group. Inside every group $\tfit{B\Delta}{min}$
varies from 0 to approx.\ 0.5 fm. Additionally, we highlight in different colors
the values for selected $\tfit{B2/3}{min}$, which are described in the
following.

\begin{figure}[p!] \centering
\makebox[\textwidth][c]{
\includegraphics[height=7cm]{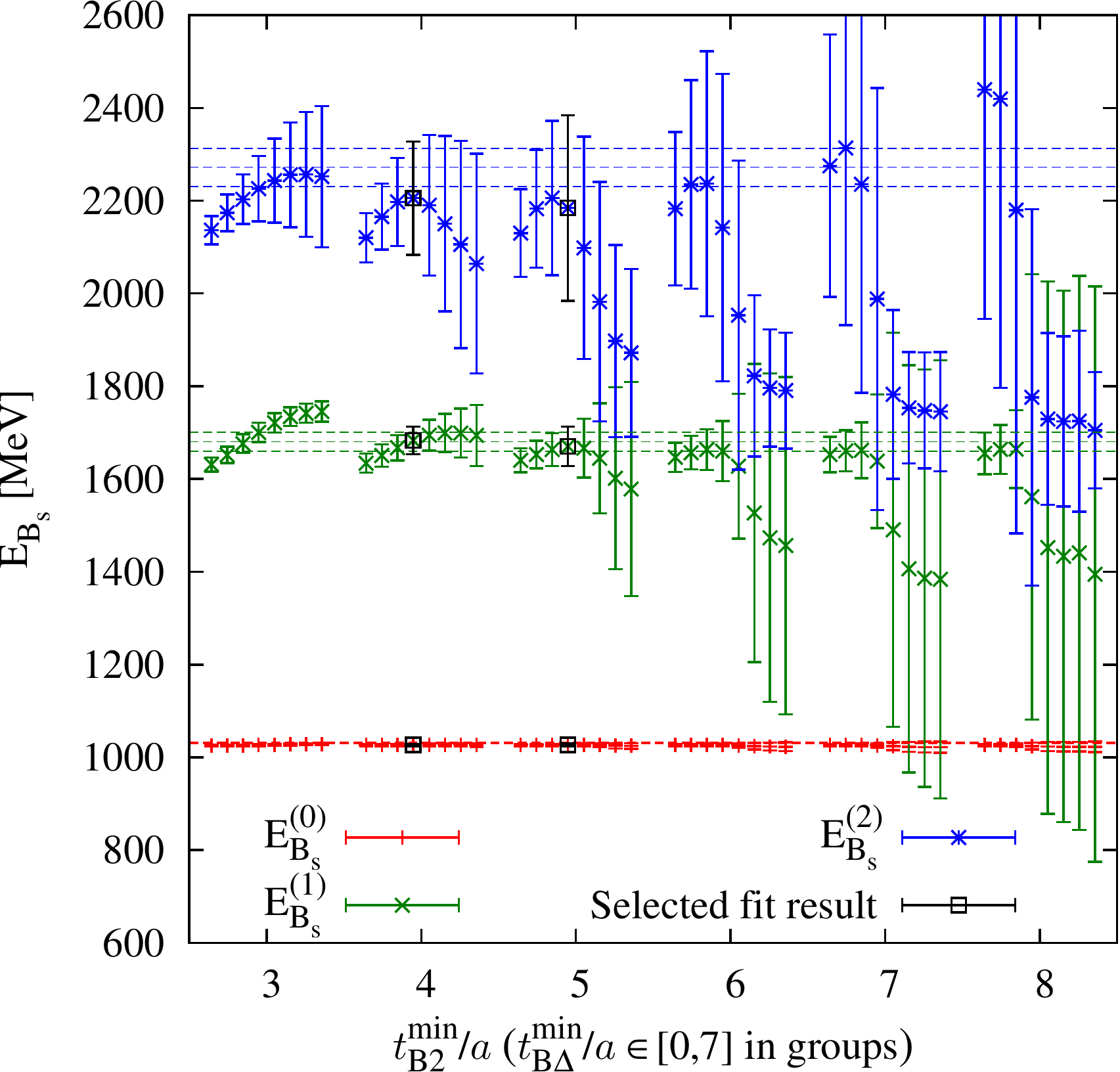}
\hspace{0.2cm}
\includegraphics[height=7cm]{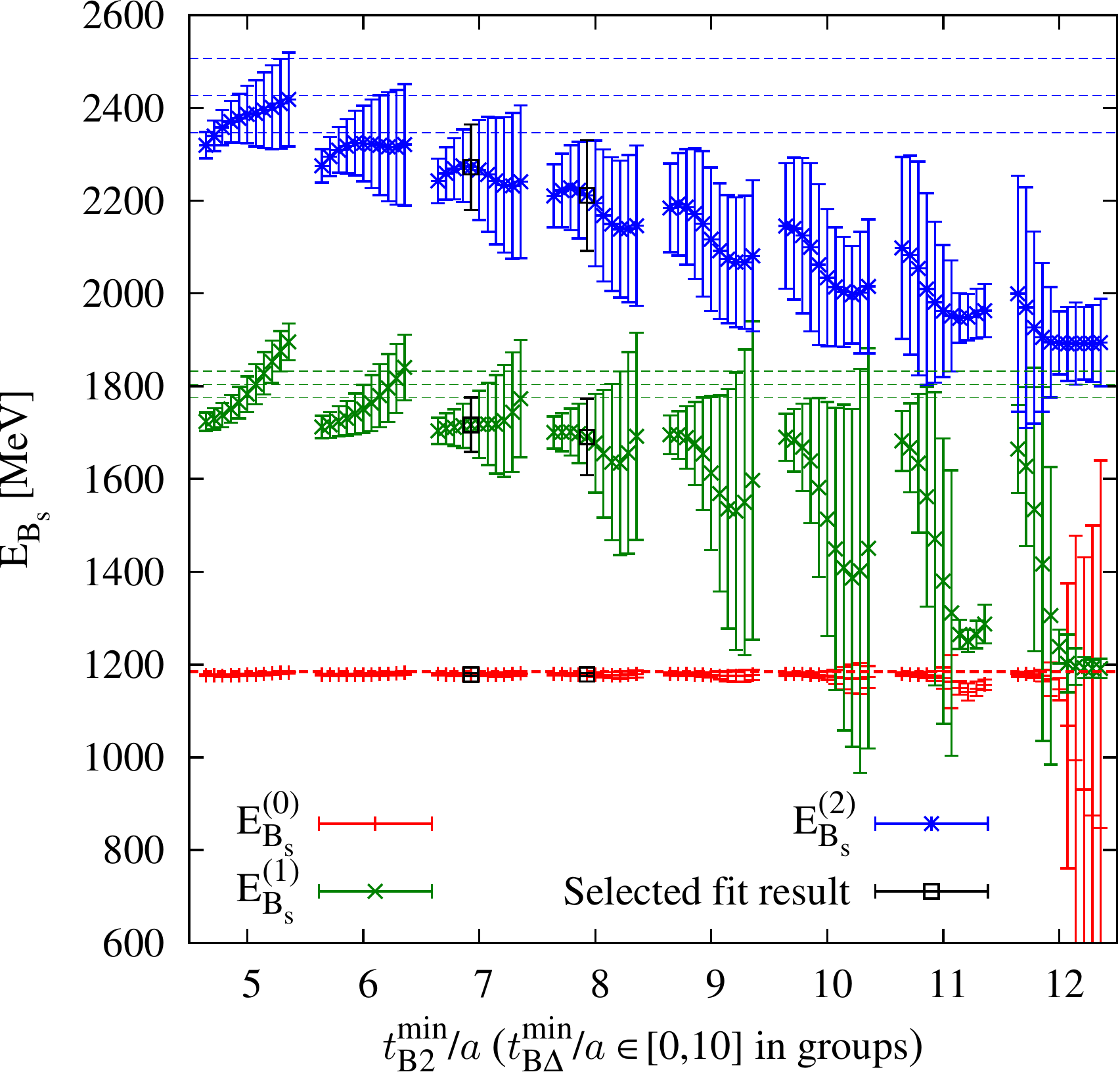} }
\caption{Stability plots for all three $\B$ energies on ensemble A5
(left) and O7 (right). The selected $\tfit{B2/3}{min}$ values are highlighted
with a black color, while the dashed lines are values from the GEVP, cf.\
Sec.~\ref{sec:gevp}. The behaviour at large $\tfit{B2}{min}$ and
$\tfit{B3}{min}$ can be traced to the fact that the fit can no longer resolve
three separate states. The selected fit values are discussed in Sec.~\ref{sec:fit_ranges}.
A zoom into the behaviour of $\EB{0}$ is shown in Fig.~\ref{fig:stab_1x3_eb0}.}
\label{fig:stab_1x3_all_eb}
\end{figure}

\begin{figure}[p!] \centering
\makebox[\textwidth][c]{
\includegraphics[height=7cm]{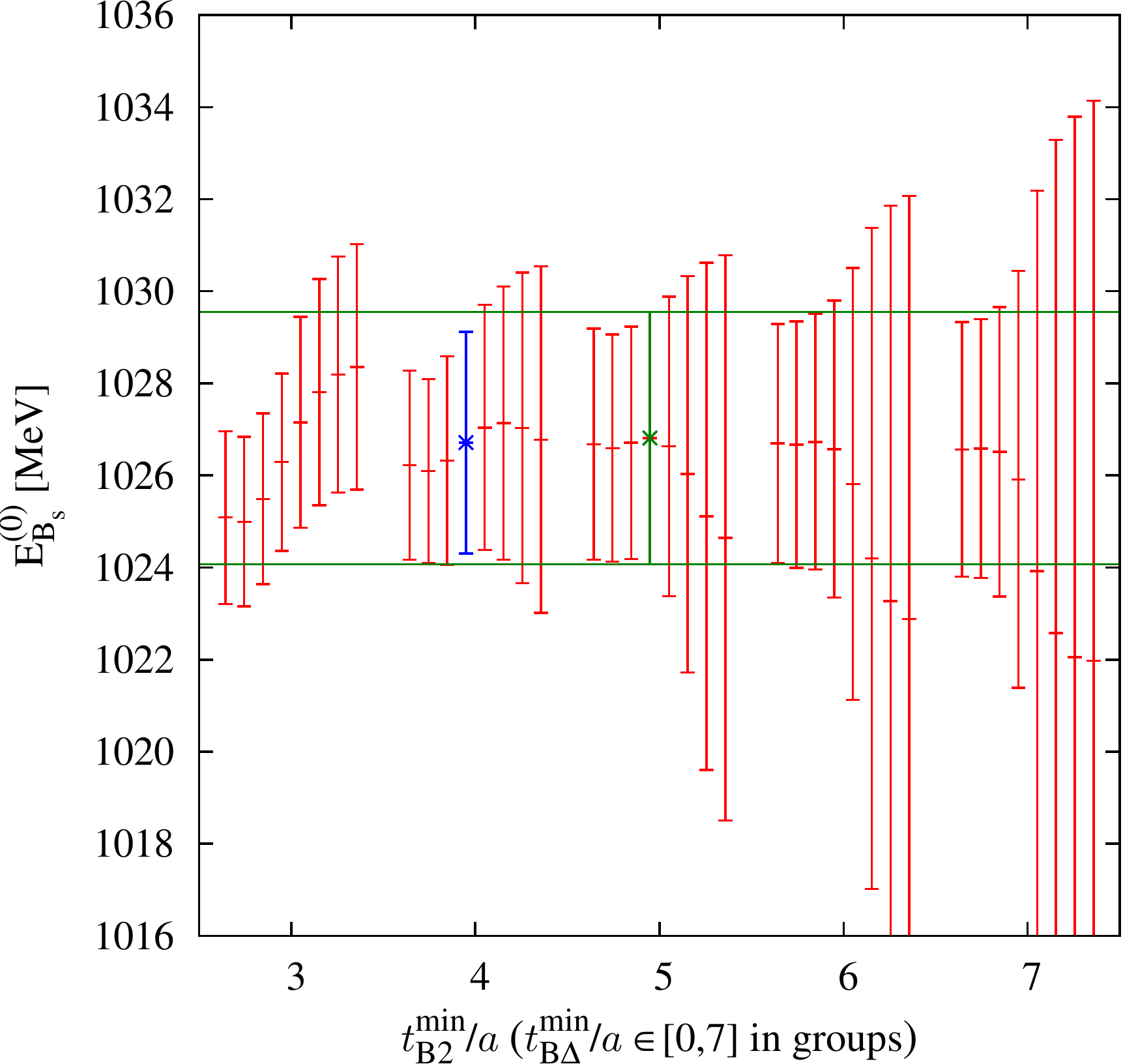}
\hspace{0.3cm}
\includegraphics[height=7cm]{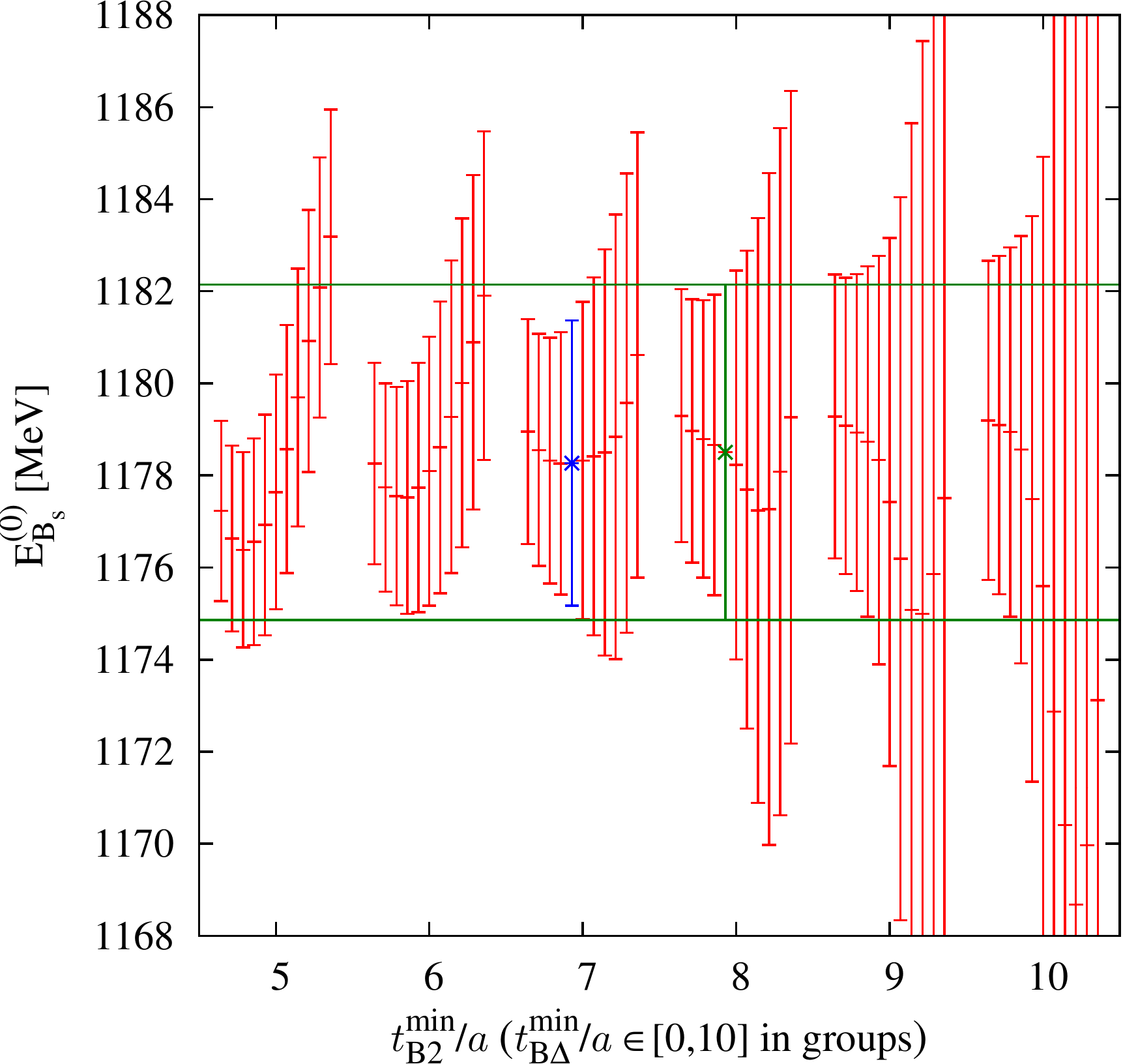}}\\[9pt]
\caption{Stability plots for $\EB{0}$ for ensemble A5 (left) and O7 (right).}
\label{fig:stab_1x3_eb0}
\end{figure}

\begin{figure}[p!] \centering
\makebox[\textwidth][c]{
\includegraphics[height=9.25cm]{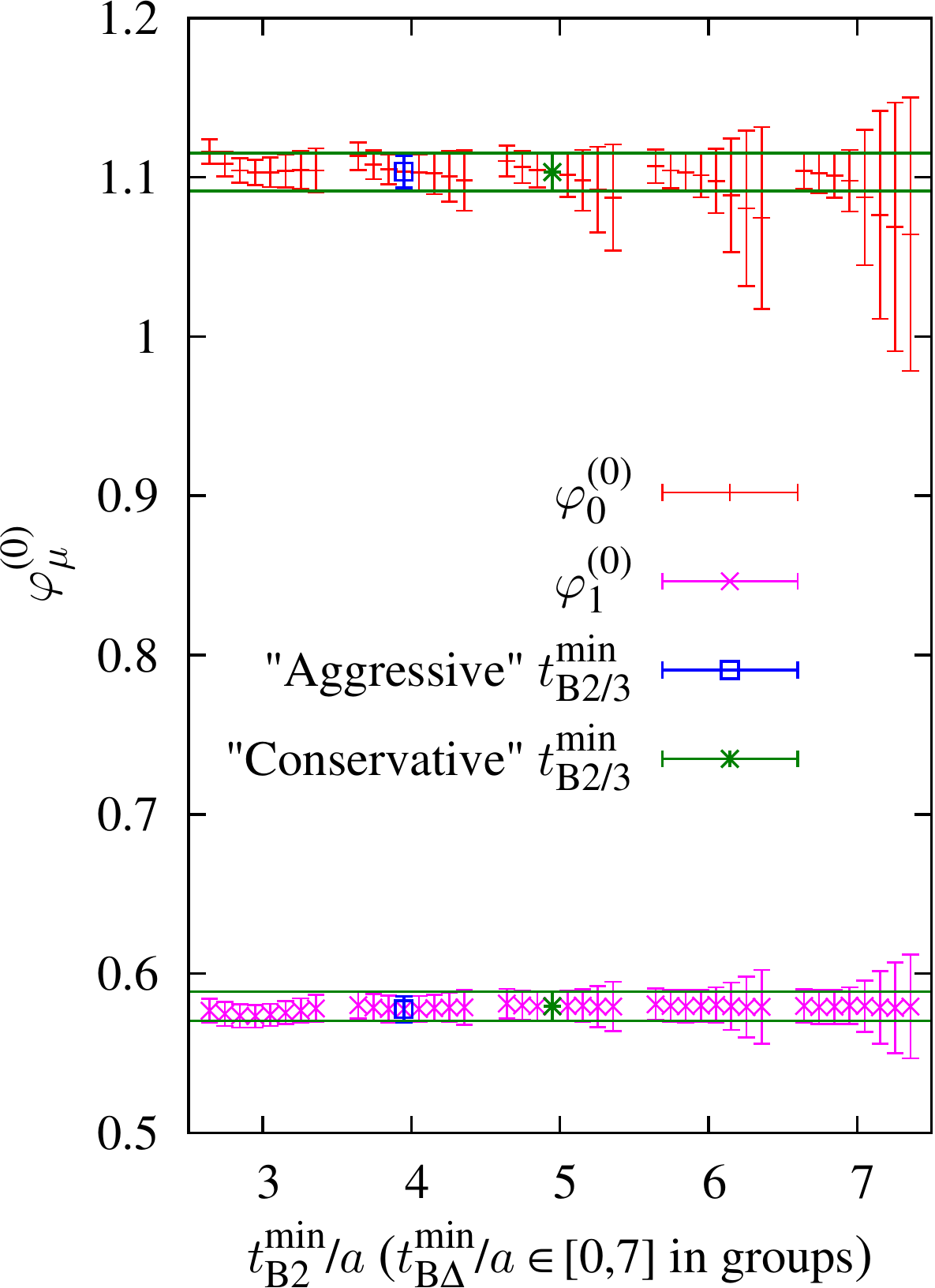}
\hspace{0.2cm}
\includegraphics[height=9.25cm]{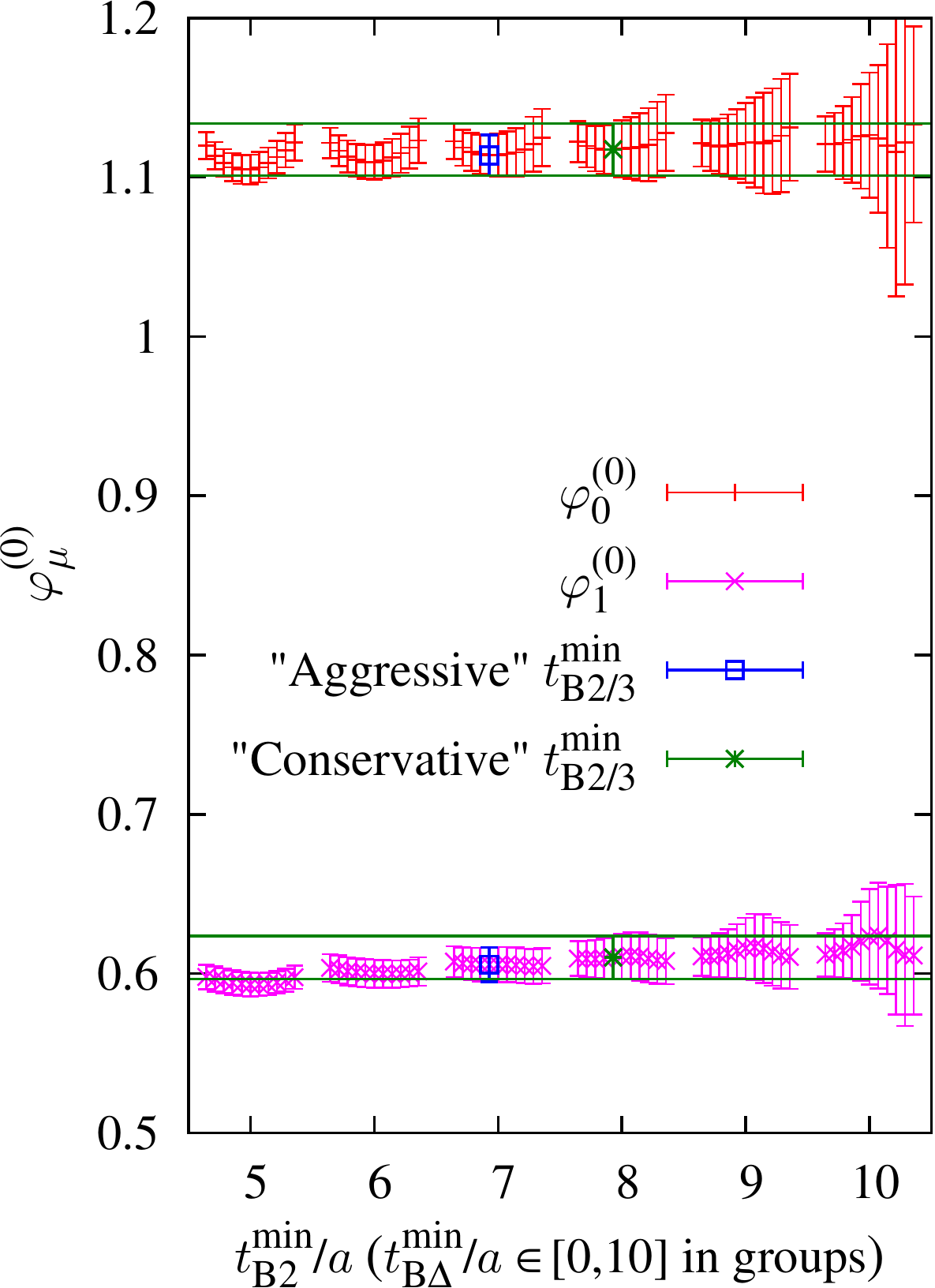}}
\caption{Stability plots of the ground-state form factors for ensemble A5 (left)
and O7 (right). The results are rather stable with respect to changes of
$\tfit{B2/3}{min}$, although errors grow quickly.}
\label{fig:stab_1x3_phi0}
\end{figure}

\begin{figure}[tp!] \centering
\makebox[\textwidth][c]{
\includegraphics[height=7cm]{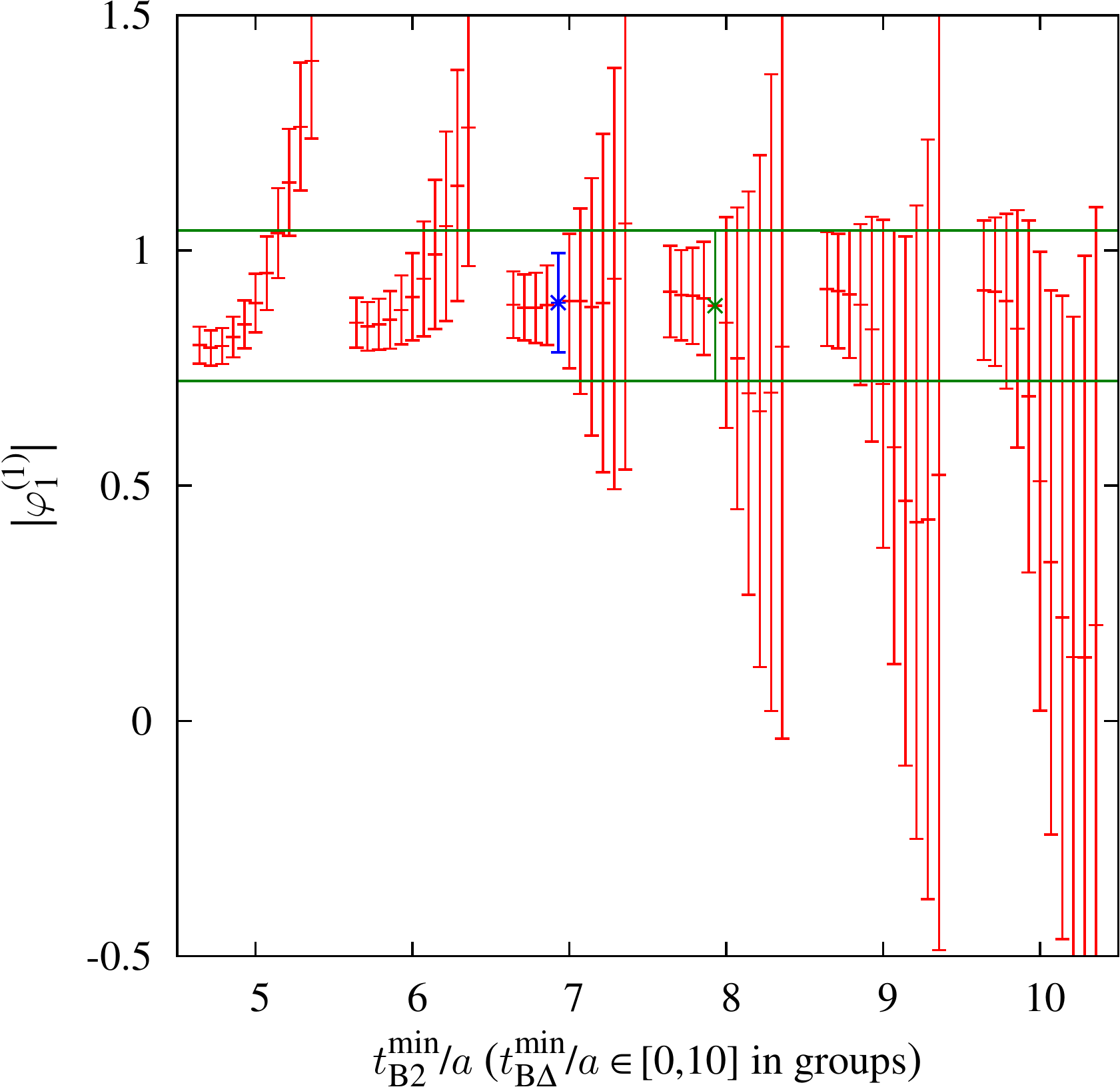}
\hspace{0.3cm}
\includegraphics[height=7cm]{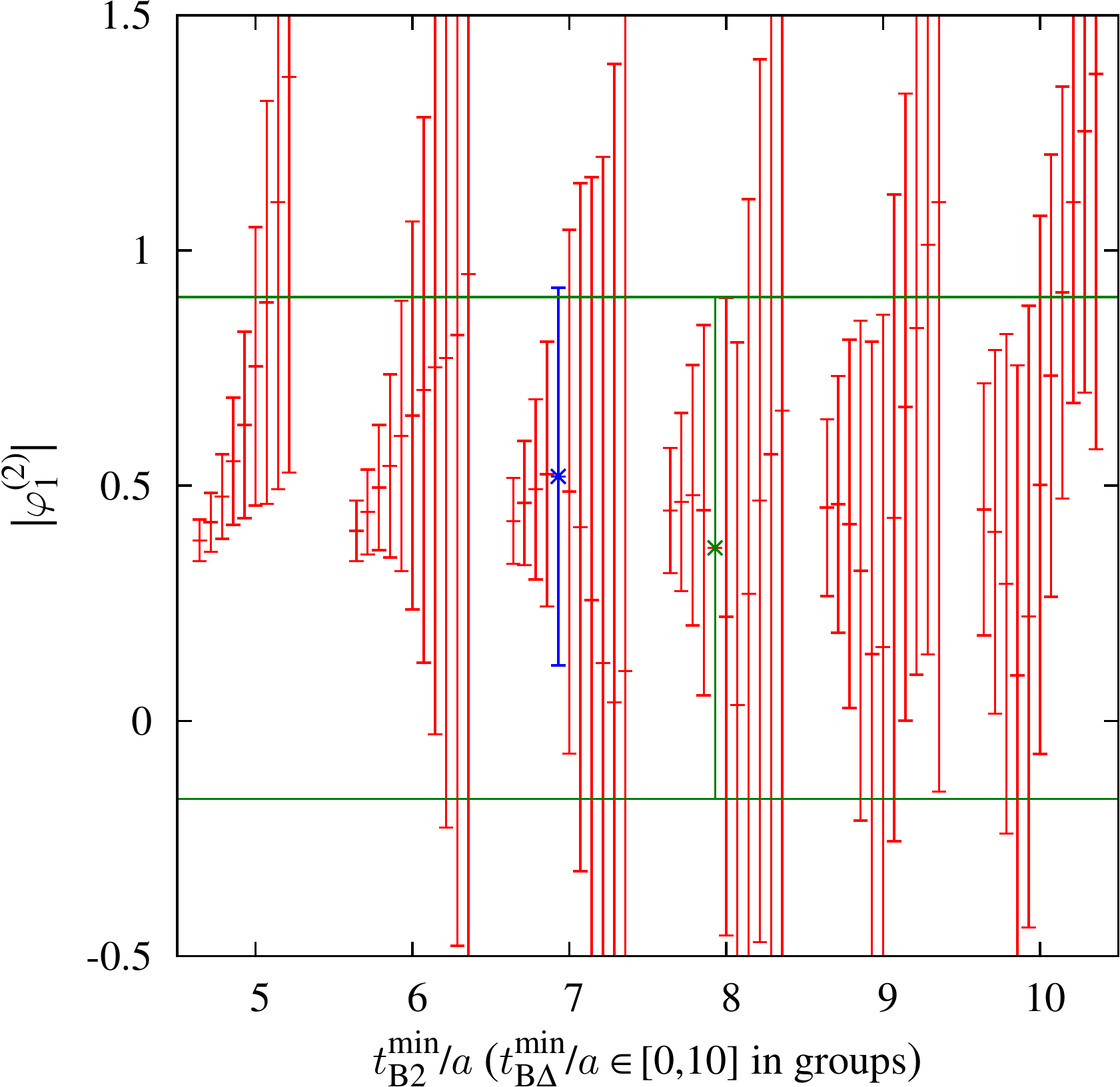}}
\caption{Excited-state form factors at $\mu=1$ for ensemble
O7.}
\label{fig:stab_1x3_phix}
\end{figure}

In choosing $\tfit{B2}{min}$ and $\tfit{B3}{min}$ one must find a good window
between too small values, where the fit is plagued by contamination from the
higher excited states (this can be seen by high $\chi^2$ as well as by lack of
stability of the extracted fit parameters with respect to small changes in
$\tfit{B2/3}{min}$) and too high values, where it can no longer resolve the
three $\Bs$ states. The latter is demonstrated in Fig.~\ref{fig:stab_1x3_all_eb}. At
large $\tfit{B2/3}{min}$ we see that for both ensembles at least one of the
energies is no longer resolved. Therefore, we refrain from showing the largest 
values of $\tfit{B2}{min}$ in Figs.~\ref{fig:stab_1x3_eb0} -- \ref{fig:stab_1x3_phix}.

In general, the ground-state energies (cf.~Fig.~\ref{fig:stab_1x3_eb0}) and
the ground-state form factors (cf.~Fig.~\ref{fig:stab_1x3_phi0}) are
reasonably stable with respect to changes of $\tfit{B2/3}{min}$ (within quickly
growing errors).
We also show representative examples for the excited-state
form factors in Fig.~\ref{fig:stab_1x3_phix}. Their precision,
especially of $\varphi^{(2)}_{\mu}$, is much worse than of their ground-state
counterparts.

\begin{table}[tbp!]
\begin{center}
\begin{tabular}{cccccc}
\toprule
id & $\mu$ & \multicolumn{2}{c}{Fit}  &
Ratio  & Summed ratio \\
 &  & $\tfit{B2}{min}\approx 0.32\,\fm$   & $\tfit{B2}{min}\approx 0.38\,\fm$  &
$\Rii$ & $\Mi$ \\
\midrule
A5 & 0 & 1.103(12) & 1.104(10) & 1.129(18) & 1.097(17) \\
A5 & 1 & 0.579(9)  & 0.578(8)  & 0.579(13)(05) & 0.587(11) \\
O7 & 0 & 1.118(16) & 1.114(13) & 1.113(19) & 1.109(13) \\
O7 & 1 & 0.610(13) & 0.606(11) & 0.602(15) & 0.601(10) \\
\bottomrule
\end{tabular}
\caption{Results for the ground-state form factors $\varphi_\mu^{(0,0)}$. The second error in the
result for $\Rii_{\mu=1}$ for ensemble A5 is the systematic error due to
the wrapper effect. The last column shows
the results of the linear fit to $\Si_{\mu,3}(\tau)$, see the shaded band in \protect\fig{fig:summed}.}
\label{tab:res_stat}
\end{center}
\end{table}
 
The values for the ground-state form factors extracted with both choices
of  $\tfit{B2}{min}$ are collected in Table \ref{tab:res_stat}, together
with the results from ratio methods described in the next section.

\section{Matrix elements at static order from ratios}
\label{sec:rat_stat}
\subsection{Ordinary ratios}

An alternative to extracting the ground-state matrix elements from fitting the
correlation functions directly is to construct an appropriate ratio of the
correlation functions such that the dependence on all or most of the other
parameters cancels in the limit of large time.

One example of such a ratio was actually used to \textit{define} the form
factor, cf.~\eq{eqn:me1}. In the following it will be beneficial to generalize
this definition to the case where $\tk\neq\tb$ in the three-point function
($\tau\equiv\tk+\tb$): 
\begin{equation}
\Ri_{\mu,r}(\tk,\tb) = 
  \frac{\C^{\Bs\to\K}_{\mu,r}(\tk,\tb)}{\big[\C^\K(\tau)\C^\Bs_{rr}(\tau)\big]^{1/2}}
  \exp\big\{(\tilde{E}^{(0)}_\Bs(\tau)-\tilde{E}^{(0)}_\K(\tau))\tfrac{\tb-\tk}{2}\big\}\,.
\label{eqn:r1}
\end{equation} 
In the limit $\tk,\tb\to\infty$, this ratio converges to the desired bare form factor 
$\varphi^{(0,0)}_\mu$ if additional parameters $\tilde{E}_\K(\tau),\tilde{E}^{(0)}_\Bs(\tau)$ 
satisfy $\tilde{E}(\tau)=E + \ord{\exp(-\Delta E \tau)}$. Possible choices will be discussed below.

In Sec.~\ref{sec:sum_stat} we will see the advantage of this definition, while 
its obvious disadvantage is that in the denominator one needs the
two-point correlation functions at time $\tau=\tk+\tb$. This means that e.g.
for a simple estimate of a plateau in the three point function at
$\tk\approx\tb\approx 1\,\fm$ one needs the two-point correlation functions at a
large time separation $\tau\approx 2\,\fm$, which is particularly problematic in
the case of $\C^\Bs$ where the signal-to-noise problem is much more severe than for
$\CK$.

For lattices with short time extent, like A5, $\tau$ may
come close to the middle of the lattice and we need to take into account the  
influence of the second term in \eq{eq:c2ll_nk1}. We do that by multiplying the
ratio by an additional factor \mbox{$\big(1 + \e^{E_\K (T - 4t)}\big)^{1/2}$}
which cancels the unwanted contribution up to excited states
which are negligible in this region.

In addition, one can consider another ratio
\begin{equation}
\Rii_{\mu,r}(\tk,\tb) = 
  \frac{\C^{\Bs\to\K}_{\mu,r}(\tk,\tb)}{\big[\C^\K(\tk)\C^\Bs_{rr}(\tb)\big]^{1/2}}
  \exp\big\{\tilde{E}^{(0)}_\Bs(\tau)\tfrac{\tb}{2}
  + \tilde{E}^{(0)}_\K(\tau)\tfrac{\tk}{2}\big\}
 \label{eqn:r2}
\end{equation} 
which has a more favourable signal-to-noise behaviour of the denominator 
but requires energy estimates even for the choice $\tk=\tb$.

A similar ratio is
\begin{equation}
  \Riii_{\mu,i}(\tk,\tb) =
    \frac{\C^{\Bs\to\K}_{\mu,r}(\tk,\tb)}{\mathcal{N}^\K\C^\K(\tk)\,
    \mathcal{N}^\Bs_r\C^\Bs_{rr}(\tb)},
\end{equation}
where the additional normalization factors $\mathcal{N}^\K$,
$\mathcal{N}_r^\Bs$ can be expressed in terms of the correlation function
parameters $\kappa^{(0)}$ and $\beta^{(0)}_r$ respectively.
In practice, our data for $\Riii$ shows very similar or slightly inferior
behaviour compared to $\Rii$, therefore we do not present the numerical results
for this ratio.

Let us now discuss the choice of the ground-state energy estimates. One can set them to the effective masses of the
corresponding two-point functions,
avoiding any fit procedure. 
This, however, results in large statistical
fluctuations in the large-time region. For our data, it is beneficial to
instead use the time-independent estimates $\EK{0}, \EB{0}$ extracted in
Sections \ref{sec:ck} and \ref{sec:cb} respectively. The plateaux in both 
cases were chosen in a conservative manner, and as a crosscheck we
calculated the ratios for different choices of the plateau ranges of both
energies. From that exercise we find that the systematic error associated with using the fitted
ground-state energies is negligible with respect to our statistical
uncertainties.

\begin{figure}[p!]
\makebox[\textwidth][c]{
\includegraphics[height=9.25cm]{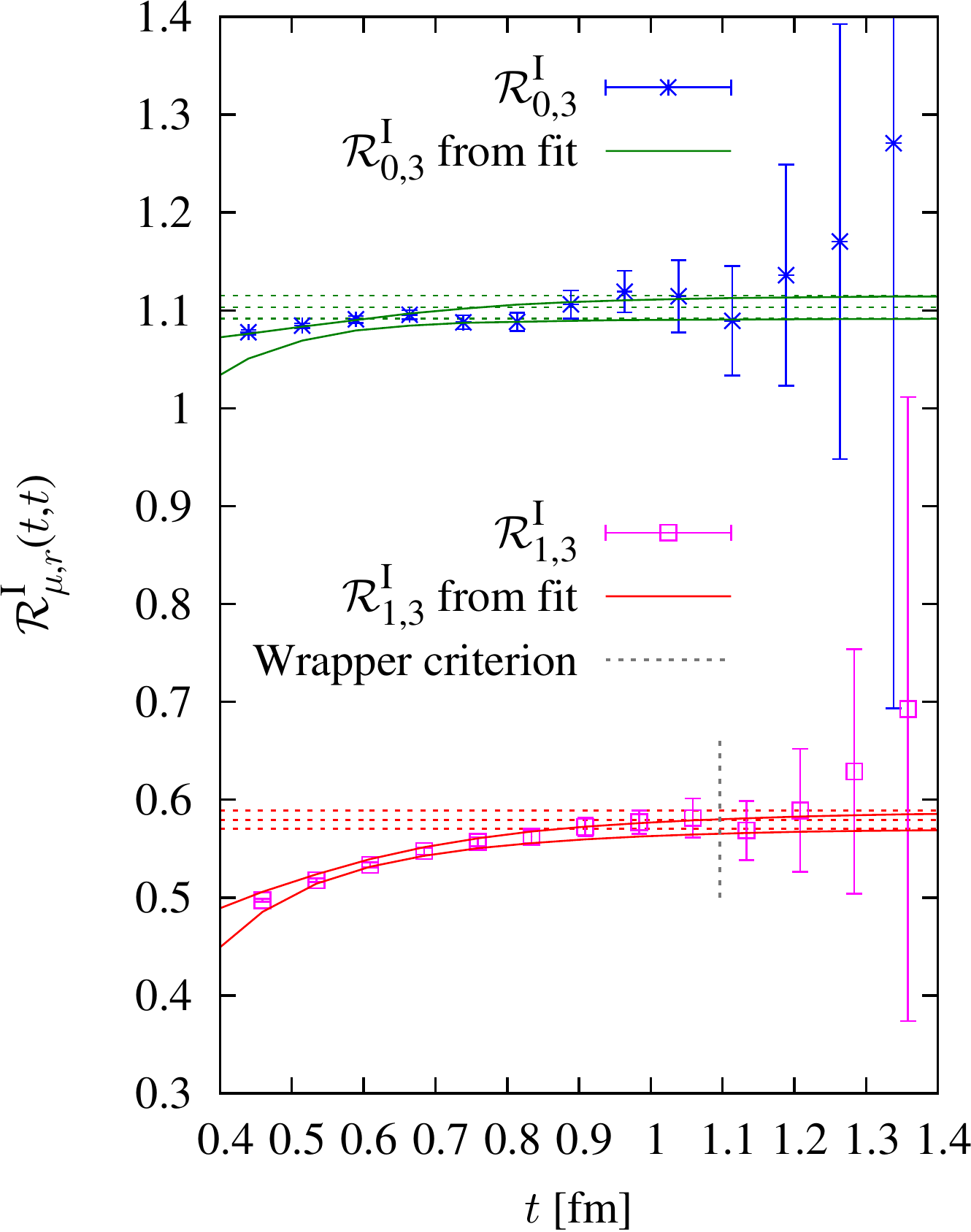}
\hspace{0.2cm}
\includegraphics[height=9.25cm]{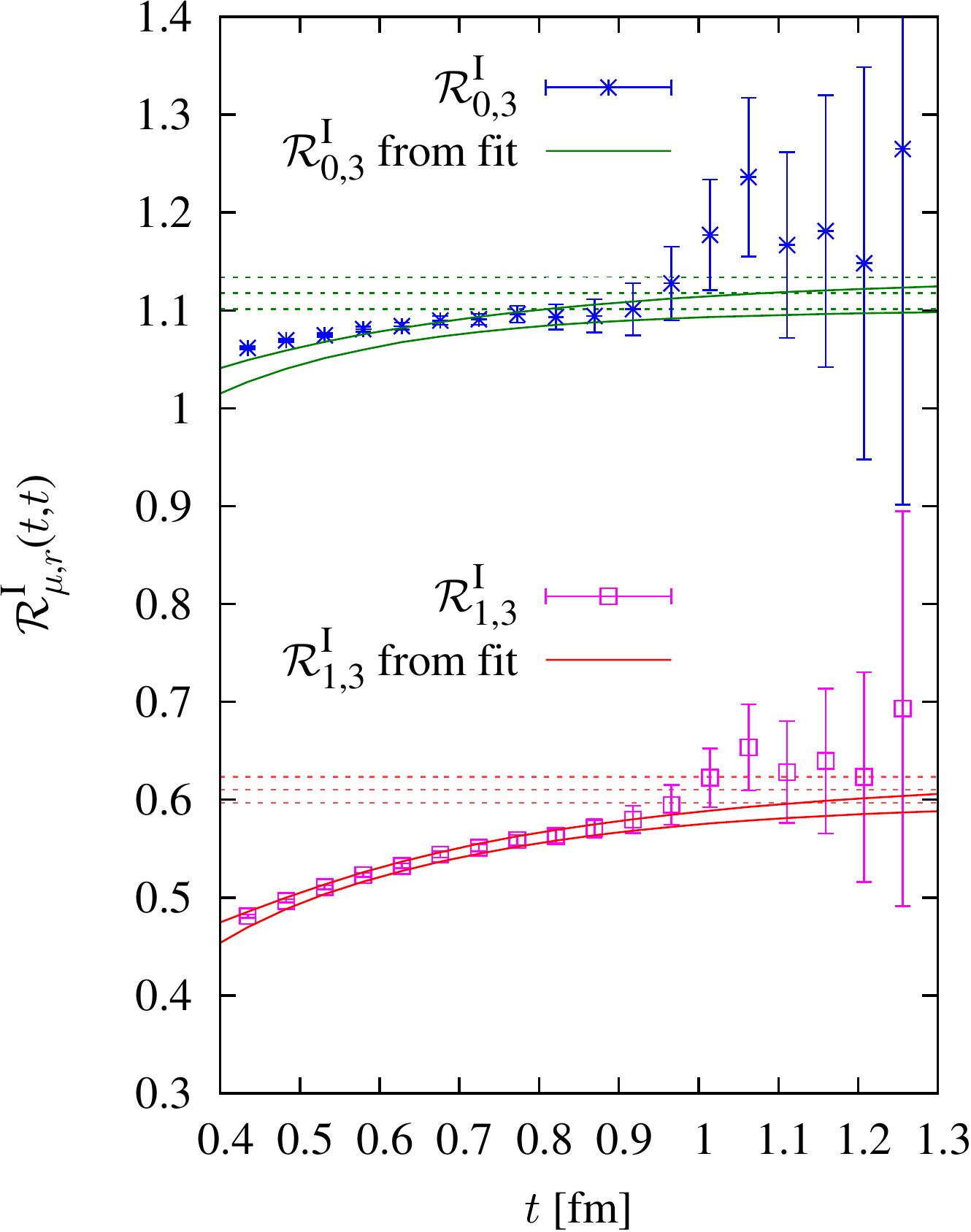}
}
\vspace{-0.75cm}
\caption{Overview of the ratios $\Ri$ with $\tk=\tb$ for ensembles A5 (left) and
O7 (right). Only the highest smearing, $r=3$, is presented. 
For comparison we also show the results from the combined fit in Sec.~\ref{sec:fits_stat}:
the dashed horizontal lines are the fitted values of $\varphi^{(0,0)}_\mu$ and the
curves ($1\sigma$ bands) are the respective ratios of correlation functions reconstructed from the 
fitted parameters.
}
\label{fig:r1_both_mu}
\end{figure}

\begin{figure}[p!]
\makebox[\textwidth][c]{
\includegraphics[height=9.25cm]{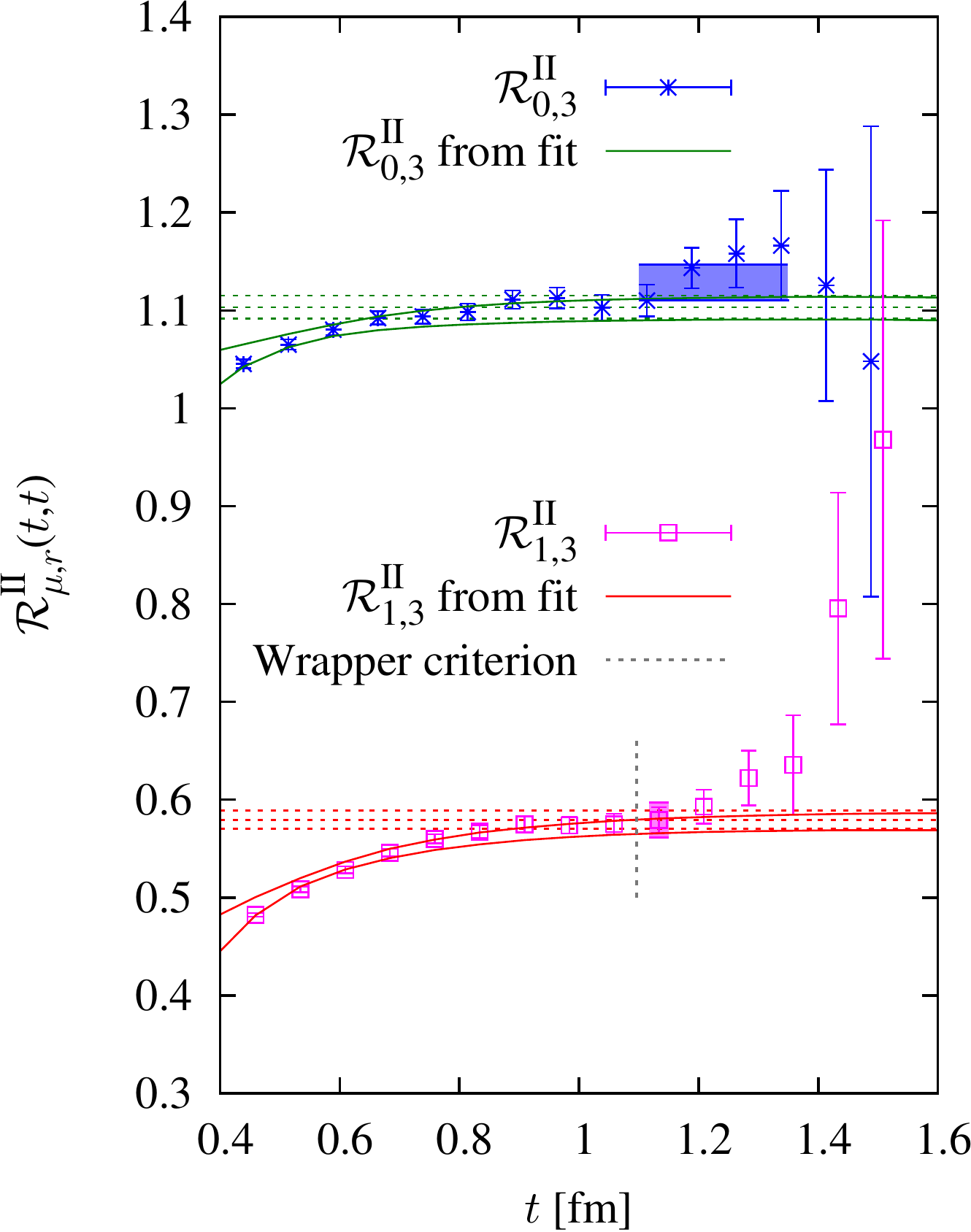}
\hspace{0.2cm}
\includegraphics[height=9.25cm]{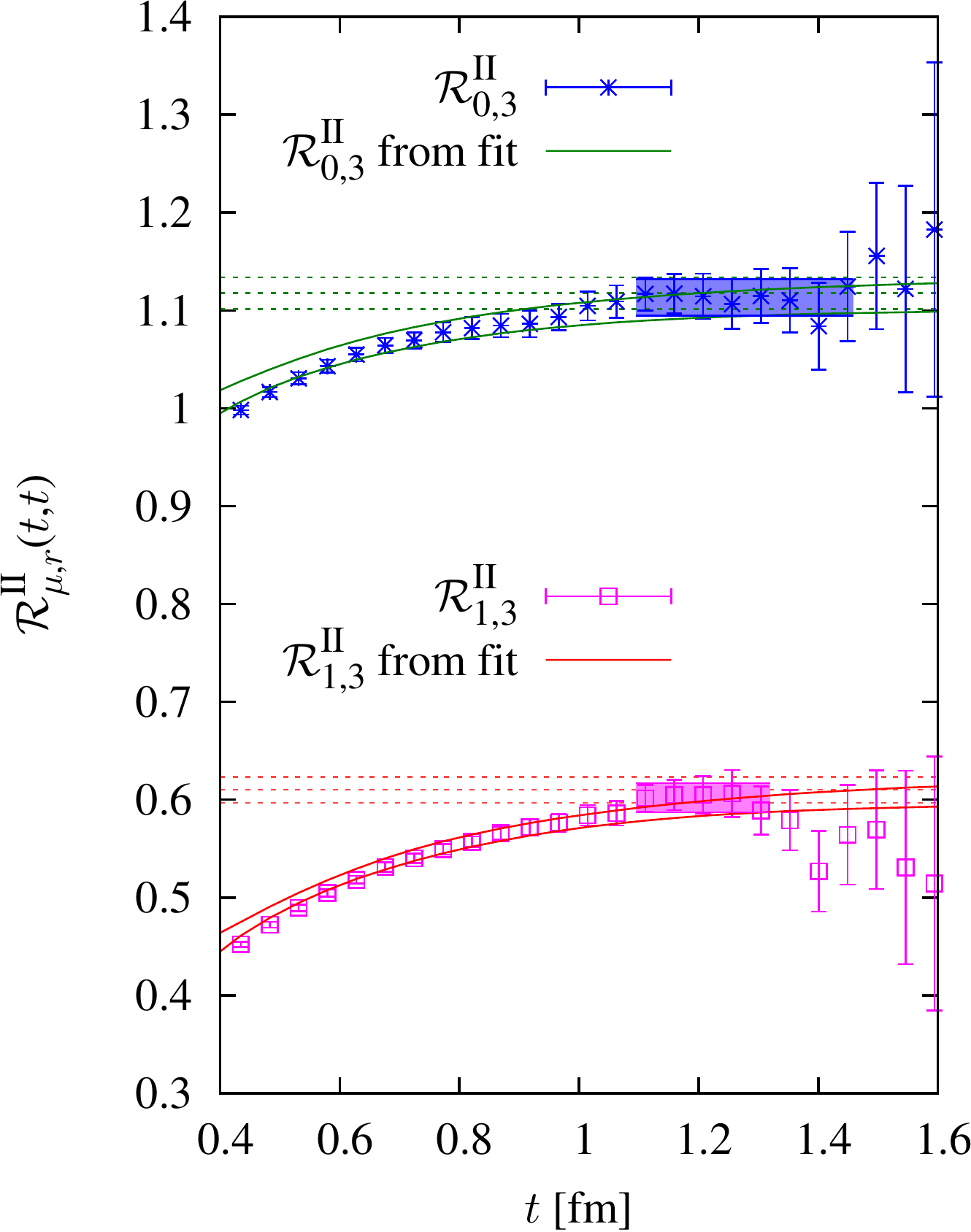}
}
\vspace{-0.75cm}
\caption{Overview of the ratios $\Rii$ with $\tk=\tb$ for ensemble A5 (left) and
O7 (right), with the fit curves analogous as in Fig.~\ref{fig:r1_both_mu}.
Selected plateaux are also shown. For $\mu=1$ on ensemble A5 the wrapper
contribution limits the available time range to $t \le \tfit{K3}{max,wr}$,
cf.~Fig.~\ref{fig:tk_min_max}.}
\label{fig:r2_both_mu}
\end{figure}

The resulting ratios $\Ri$ and $\Rii$ are presented in Figs.~\ref{fig:r1_both_mu} 
and \ref{fig:r2_both_mu}, respectively. In addition to the data we also plot the fit results from
Sec.~\ref{sec:fits_stat}, including both the fitted values of
$\varphi^{(0,0)}_\mu$ and the respective ratios of correlation functions
reconstructed by inserting the fit parameters ($\varphi^{(0,n)}_\mu$, $\kappa^{(0)}$, 
$\beta_r^{(n)}$, $E_\K^{(0)}$ and $E_\Bs^{(n)}$) into
Eqs.~(\ref{eq:c2ll})--(\ref{eq:c3}). There is a good agreement of the actual
ratios with their fitted counterparts at large values of $t$, where excited states 
neglected in the fits are irrelevant.

On ensemble A5, due to the short time extent of the lattice, the wrapper criterion \eq{eq:wrapper} 
becomes relevant. For $\mu=1$ it restricts the available times 
to approx.\ 1.1 fm, cf.~Fig.~\ref{fig:tk_min_max}.
At larger times we observe that the ratio $\Rii_1$ starts to grow rapidly,
while for $\Ri_1$ this is most likely masked by the large uncertainties. 
Furthermore, for $\mu=0$ the data is close to violating the wrapper criterion 
(see Fig.~\ref{fig:tk_min_max}) at the maximal time separations used.

On both lattices, A5 and O7, we see that at large times, $t\gtrsim 1$ fm, $\Rii$ is clearly superior
to $\Ri$ in terms of the signal to noise and has a comparable precision to the
results of the combined fit. To use $\Rii$ for the extraction of $\varphi^{(0,0)}_\mu$, 
we need to select a suitable plateau: looking at the bands from the
fit which give us an estimate of the excited-state contamination from the $\Bs$
sector, we start the plateau at $t^{\rm min}=1.1$ fm and fit until the loss of
precision of the signal below 5\% or until we hit the wrapper
criterion. On A5 for $\mu=1$ there is no valid plateau, because the wrapper
is hit before $t^{\rm min}$. In this case, we quote
the first data point above 1.1 fm as the final result and add to it a systematic error associated
with the wrapper contribution estimated by the fit.

\subsection{Summed ratios}
\label{sec:sum_stat}

\begin{figure}[tbp!]
\begin{center}
\vspace{0.25cm}
\makebox[\textwidth][c]{
\includegraphics[height=9.25cm]{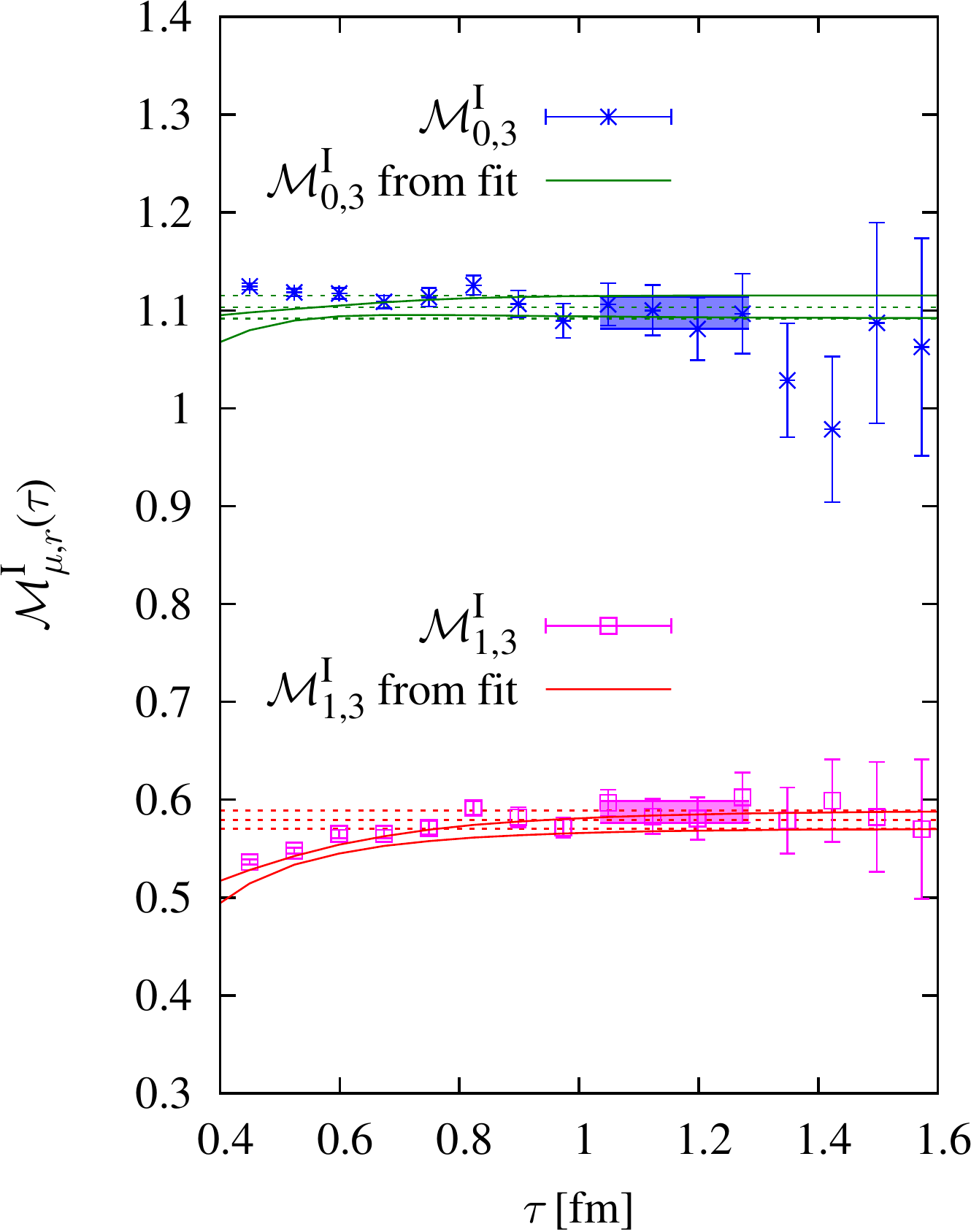}
\hspace{0.2cm}
\includegraphics[height=9.25cm]{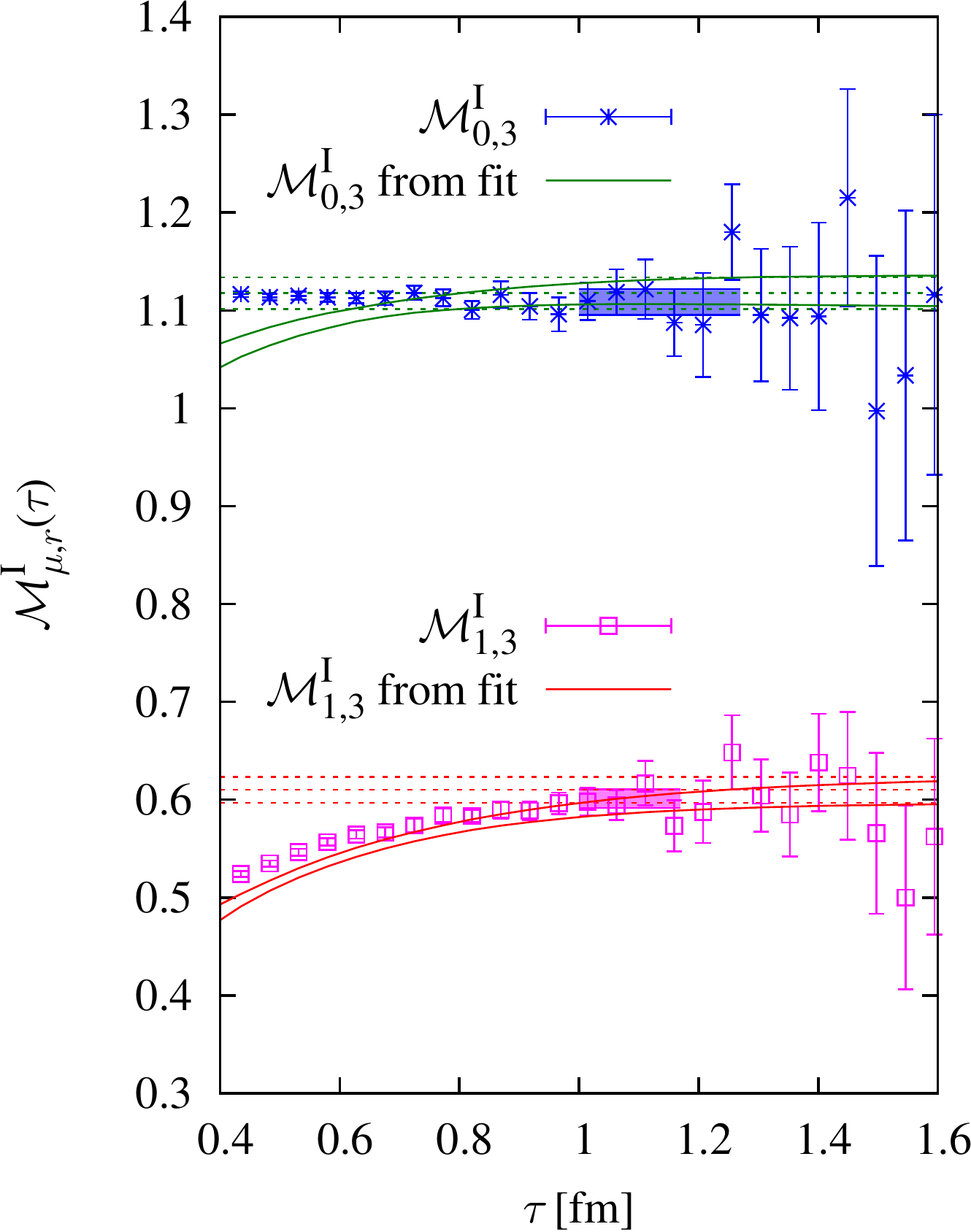}
}
\vspace{-1cm}
\end{center}
\caption{Overview of the $\Mi$ for ensemble A5 (left) and 
O7 (right), together with the fit curves analogous as in
Figs.~\ref{fig:r1_both_mu} and \ref{fig:r2_both_mu}. 
The shaded bands represents $c_2$ of fits $\Si_{\mu,r}(\tau) = c_1+c_2 \tau$ to the sum
 in the indicated
range of $\tau$, chosen such that wrapper
effects are negligible.}
\label{fig:summed}
\end{figure}

A way to get improved convergence to the ground state is to sum\footnote{We write this in the 
form used in our analysis, but note that often it is advantageous and natural to sum 
the position of the operator that causes the transition 
over all spacetime instead \cite{Maiani:1987by,Bulava:2011yz}.} 
the ratio \cite{Maiani:1987by,Capitani:2010sg,Bulava:2011yz} $\Ri$ and
determine the matrix element from
\begin{equation}
\Mi_{\mu,r}(\tau)=\partial_\tau\,
\Si_{\mu,r}(\tau)=\partial_\tau\,
a\sum_{\tb=0}^{\tau}\Ri_{\mu,r}(\tau-\tb,\tb)\, .
\label{eq:summed}
\end{equation}
The asymptotic excited-state contaminations are then $\mathcal{O}(\tau\Delta
{\rm e}^{-\tau\Delta})$, where $\Delta=\min(\EK{1}\!\!-\EK{0},
\EB{1}\!\!-\EB{0})$, as opposed to $\mathcal{O}({\rm e}^{-\tau\Delta/2})$ in
ordinary ratios \cite{Capitani:2010sg,Bulava:2011yz}.\footnote{One could in principle attempt to sum the other ratios
as well. However, the improvement in convergence is only proven to work (at
least asymptotically) for $\Ri$. Empirically we see that it does not work
for $\Rii$ and $\Riii$.} The accelerated convergence of the 
summed ratios can be a decisive advantage in the case
of ensembles with a limited extent in time such as A5:
contaminations by wrappers are less relevant.
Example results are shown in Fig.~\ref{fig:summed}. We observe 
that the convergence is indeed improved. 

In fact, especially for $\mu=0$ the plateaux seem to start {\em very} early. This
early onset of the plateaux is discussed in Appendix \ref{sec:sumrat_conv} where
we show that it is caused by an accidental 
cancellation of the excited-state contributions
in the $\Bs$ and $\K$ sector and therefore should be treated with caution.

Note, however, that the ratios reconstructed from the fits of Sec.~\ref{sec:fits_stat}
only include the influence of the excited states in the $\Bs$ sector.  
As in the previous subsection, these fit bands can be used to
estimate the start of the plateaux. We choose $\tau^{\rm min}=1$ fm and
$\tau^{\rm max}$  by the 5\% relative noise criterion.\footnote{One could
also choose 10\%, the plateaux would be a bit longer but that does not help in
reducing the statistical error.}
In this range we extract the form factor from the slope of the linear fit
directly to $\Si(\tau)$. Note that on both lattices we are not affected by 
the wrapper criterion as opposed to ordinary ratios.

The results from the different methods are gathered in Table \ref{tab:res_stat}
and are in good agreement. One slight difference is seen in
$\Rii_0$ on ensemble A5. However, in this case if we start the
plateau one data point earlier (which, judging from Fig.~\ref{fig:r2_both_mu},
seems legitimate) the result goes down to 1.117(15) which removes most of the
difference. In general the results from the fits and the summed ratio have
similar precision, in which case we prefer to quote the result from the summed
ratio as the final results, as the latter method is simpler. 

\subsection{GEVP method}
\label{s:gevp}

We have also implemented the GEVP method of Ref.~\cite{Bulava:2011yz}, both the ratio one, eq.~(2.16) 
of \cite{Bulava:2011yz}, and the summed one, eq.~(3.7) of \cite{Bulava:2011yz}. In both cases we 
specialized to using the GEVP only in the $\B$ channel. The results are very similar to the ones of the previous two
sections (with the largest wave function),
but with errors which are a little bit larger. The fact that we 
do not see a significant improvement might be due to our 
interpolating fields, which likely do not distinguish between 
single hadron and (excited) multi-hadron states. 
The GEVP is then not able to significantly reduce the
multi-hadron contributions.
We comment further on this issue in the conclusions.

\section{Matrix elements at $1/m$ order}
\label{sec:1m}

The HQET expansion of B-meson observables becomes a precision tool 
only when $1/m$ terms are included. We now discuss the determination of
these crucial terms for the matrix elements. As in the static approximation
we have the option to perform fits to the (two-point and three-point) correlation functions, or to consider ratios or summed ratios. 
Fits have been discussed and applied in Ref.~\cite{Blossier:2010mk}
for the somewhat simpler case of the B-to-vacuum matrix element
$f_\B$. As there, in our present case, all parameters 
$\beta_s^{(n)},\,\Ex{(n)}$ get triplicated with static and $1/m$ 
pieces corresponding to kin and spin insertions. Furthermore pure exponentials turn into exponentials plus terms of the form
$\Exx{\kin}{n} t \exp(-\Exx{\stat}{n} t)$. This proliferation of terms
in addition to the $1/m$ corrections to the matrix elements themselves 
makes an analysis in terms of fits cumbersome and difficult. We therefore
concentrate on the analysis of ratios, which we already have seen to be just as good as fits in the static approximation. We will also see that there 
are some simplifications in the $1/m$ expansion of ratios which make them 
rather accessible.

As a preparation for expanding the ratios, we start with the underlying two- and three-point functions. First note that we choose
the arbitrary interpolating fields
$\cO_{\mathrm{bs},r}$, eq.~\eqref{eq:cObs}, not to contain
a $1/m$ piece. Therefore, the $1/m$ expansion of the two-point function reads
\begin{align}
 \log(\CB (t)) &= -  m_\textnormal{bare}t + \log(\Cxx{\B}{\stat} (t))  +
 \sum_{k \in \{\kin,\spin\}} \omega_{k} 
  \left( \frac{ \Cxx{\B}{k} (t)}{\Cxx{\B}{\stat}(t)} \right).
\end{align}
Only the (dimensionful) HQET parameters of the action,
$m_\textnormal{bare}$ as well as $\omega_\rmk \sim 1/m$,
appear. Here and everywhere below all O$(1/m^2)$ terms are dropped without notice. Note also that $m_\textnormal{bare}$ drops 
out in the expressions for matrix elements.
 
The energy $E^\Bs = \lim_{t\to\infty} -\partial_t \log(\CB (t)) $ is $1/m$-expanded as
$E^\Bs = m_\textnormal{bare} + \Ex{\stat} +  \omega_\kin \Ex{\kin} + \omega_\spin \Ex{\spin}$ with $\Ex{x}$ appearing in the 
large time behaviour of 
\cite{Bernardoni:2013xba} 
\begin{align}
&-\partial_t \log \Cxxx{\B}{\stat}{rr}(t) = \Exx{\stat}{} + \ord{e^{-\Delta \Ex{\stat}t}}, \\
&-\partial_t \frac{\Cxxx{\B}{\kn}{rr}(t)}{\Cxxx{\B}{\stat}{rr}(t)} = \Exx{\kn}{} + \ord{te^{-\Delta \Ex{\stat}t}}\,, \quad
k \in \{\kin,\spin\}.
\label{e:dtratio}
\end{align}
All $\Ex{x}$ refer to the ground state; the first excited state contribution
leads to the term with $\Delta \Ex{\stat} = \Exx{\stat}{2} - \Exx{\stat}{1}$.
Because one always first expands in $1/m$ and then takes a limit of large time, terms such as $\exp(-\Delta \Ex{\kin}t)$ do not appear.

In the numerical applications we will take $\partial_t$ to be the
forward derivative, $a \partial_t f(t) = f(t+a) - f(t)$. 
 On integrating \eq{e:dtratio}
, we get:
\begin{align}
 &\frac{\Cxxx{\B}{\kn}{rr}(t)}{\Cxxx{\B}{\stat}{rr}(t)} = \Axxx{\B}{\kn}{r} - \Exx{\kn}{} t + \ord{te^{-\Delta\Ex{\stat}t}}\,.
 \label{eq:A2pt}
\end{align}
 The integration constants $\Axxx{\B}{\kn}{r}$ do depend on the $\kin$ and $\spin$ insertions as well as the smearing level used. 

In complete analogy we have for the three-point functions ($t=t_\Bs = t_\K$): 
\begin{align}
 &\frac{\Cxxxx{\B\to\K}{k}{\mu}{r}(t,t)}{\Cxxxx{\B\to\K}{\stat}{\mu}{r}(t,t)} = \Axxxx{\B\to\K}{k}{\mu}{r} - 
 \Exx{k}{} t + \ord{te^{-\Delta \Ex{\stat}t}}.
 \label{eq:A3pt}
\end{align}

Now we turn to the the ratios 
and insert the $1/m$ expansion of the quantities which enter their definition
 in \eqref{eqn:r1} and \eqref{eqn:r2}.
To understand the structure, consider the expansion 
\begin{align}
\nonumber \Rix{r}(t) 
 &= \Rixx{\stat}{r}(t) \left[1 + \sum_k \omega_k \,\drixxx{k}{\mu}{r} (t) + \sum_j \omega_{\mu j} \,\drixxx{j}{\mu}{r} (t) \right]
\end{align}
with
\begin{align}
\drixxx{k}{\mu}{r} (t,t)&= \frac{\Cxxxx{\B\to\K}{k}{\mu}{r}(t,t)}{\Cxxxx{\B\to\K}{\stat}{\mu}{r}(t,t)} 
     - \frac12\frac{\Cxxx{\B}{k}{r}(2t)}{\Cxxx{\B}{\stat}{r}(2t)}\,, \quad k \in \{\kin,\spin\}\,,
\\
\drixxx{j}{\mu}{r} (t,t)&=     
     \frac{\Cxxxx{\B\to\K}{j}{\mu}{r}(t,t)}{\Cxxxx{\B\to\K}{j}{\mu}{r}(t,t)}\,.
     \label{eq:rhojmut}
\end{align}
The sums over $k$ run as indicated and for $j$ the range is
seen in \tab{tab:vec}.
It follows that the desired $1/m$ corrections to the ground state matrix elements are given by 
the large time limits,
\begin{align}
    \rho^{k}_{\mu} &= \lim_{t\to\infty} \drixxx{k}{\mu}{r} (t) 
    = \Axxxx{\B\to\K}{k}{\mu}{r} - \frac{1}{2} \Axxx{\B}{k}{r}\,.
    \\
    \rho^{j}_{\mu} &= \lim_{t\to\infty} \drixxx{j}{\mu}{r} (t) 
    \label{eq:rhojmu}\,,
\end{align}
These equalities are what we alluded to before as simplification in the 
$1/m$ expansion. In fact, since the ground state matrix elements in the last expression are just given by the $A$ terms, it should not come as a surprise that the same final
formulae for $\rho^{j}_{\mu}$ and $\rho^{k}_{\mu}$ are obtained if one starts from
ratios $\Rii$ or $\Riii$. 

The strategy to obtain the $1/m$ terms of the matrix elements is then to
extract $\Axxx{\B}{k}{r}$ and $\Axxxx{\B\to\K}{k}{\mu}{r}$
from fits to \eq{eq:A2pt} and \eq{eq:A3pt} and $\rho^{j}_{\mu}$
from \eq{eq:rhojmu}.

\begin{figure}[!tbh]
\begin{center}
\makebox[\textwidth][c]{
\includegraphics[height=7cm]{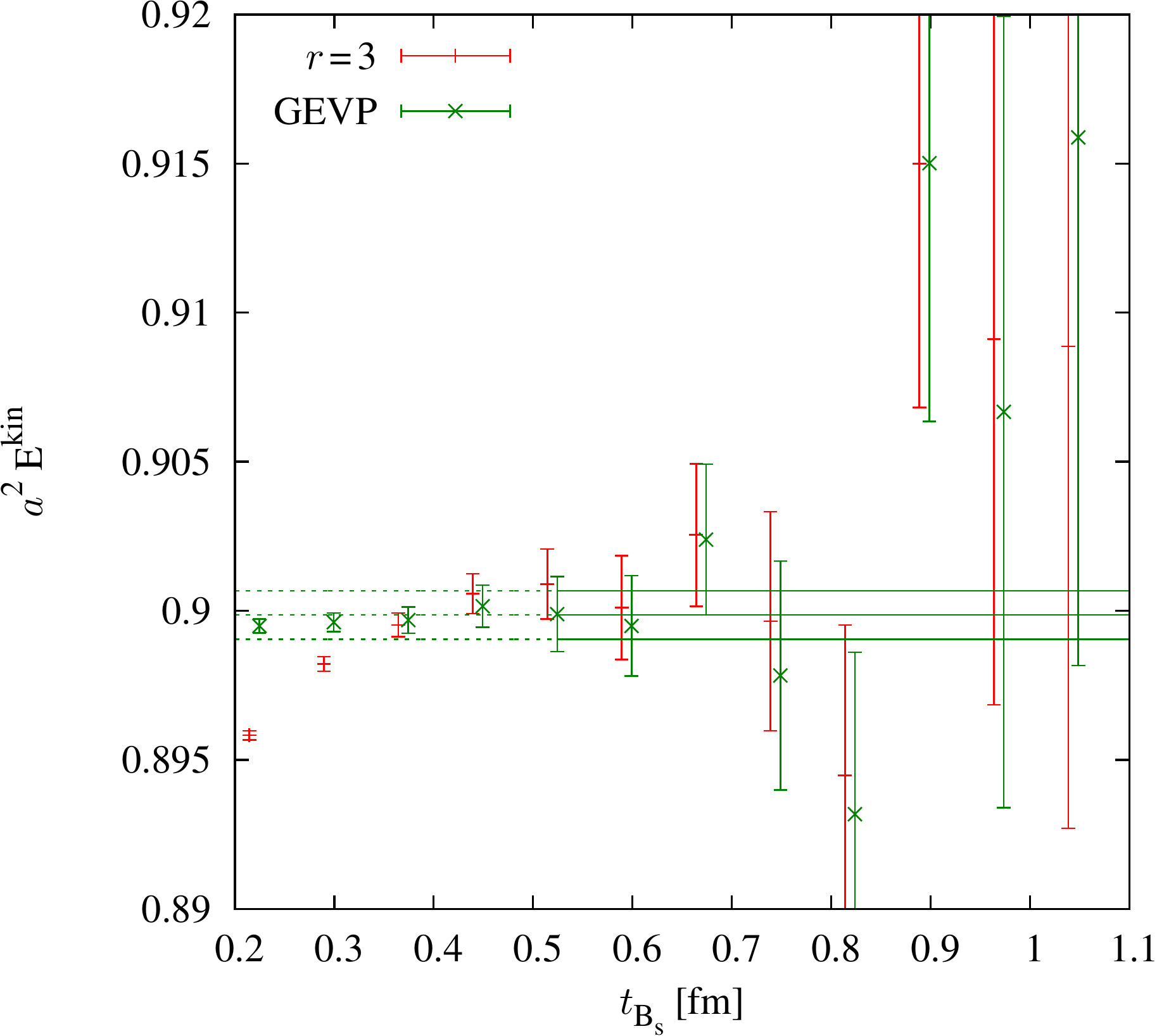}
\hspace{0.2cm}
\includegraphics[height=7cm]{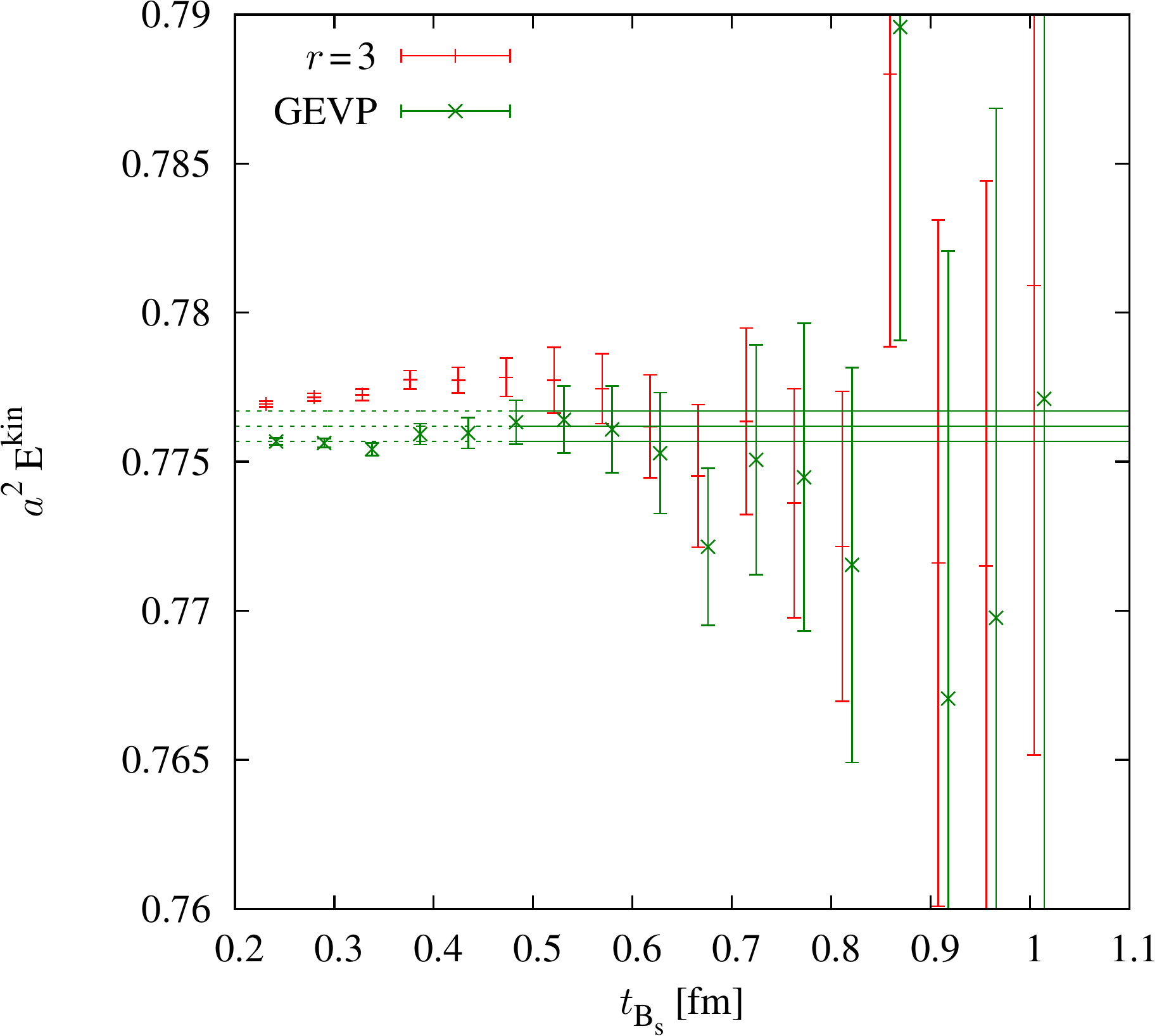}
}
\caption{$a^2 E_{\rm kin}$ for ensembles A5 (left) and O7 (right). The highest
smearing and the GEVP result are shown. The band shows the GEVP plateaux average.}
\label{fig:ekin}
\end{center}
\end{figure}

\begin{figure}[!tbh]
\begin{center}
\makebox[\textwidth][c]{
\includegraphics[height=7cm]{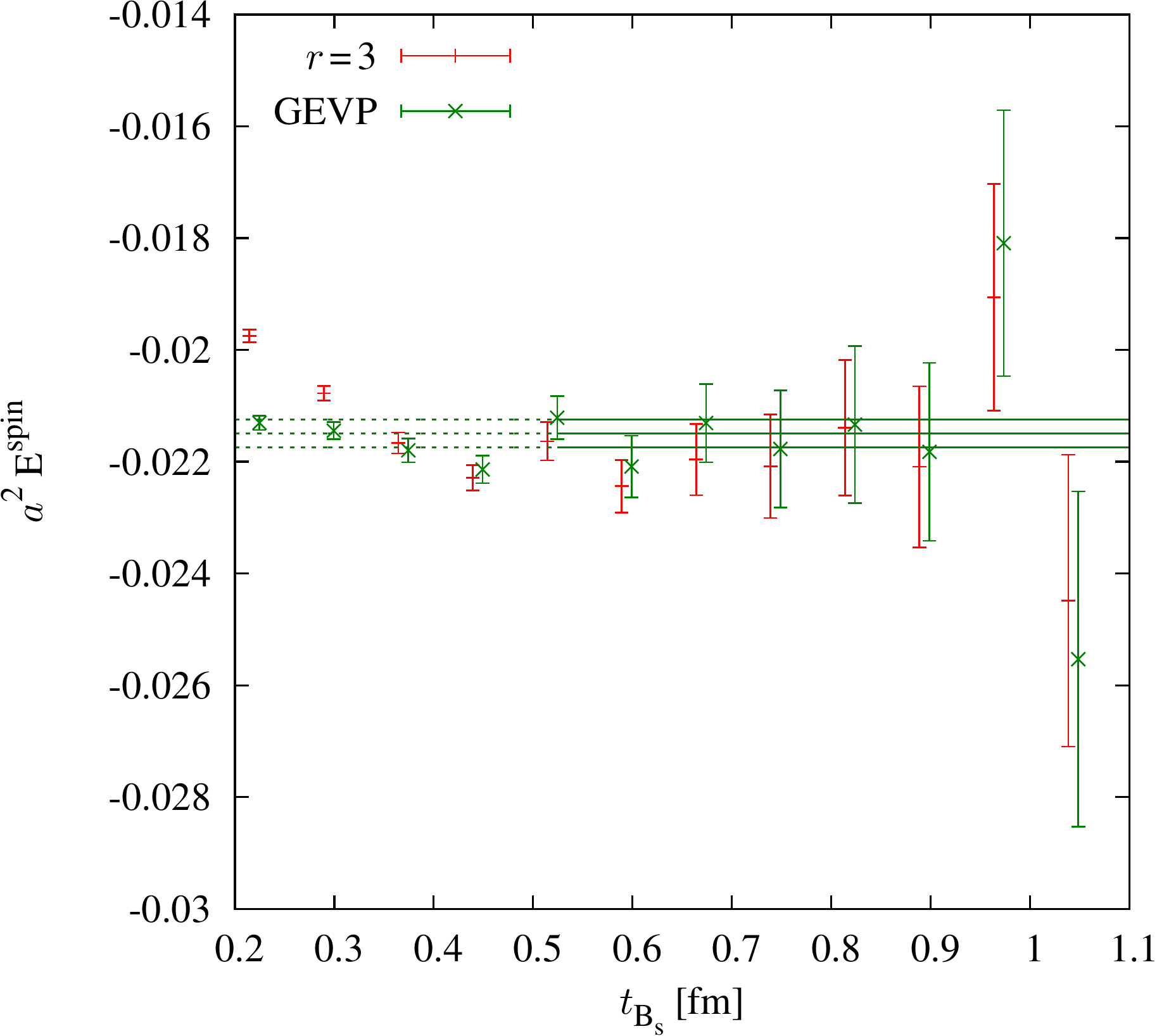}
\hspace{0.2cm}
\includegraphics[height=7cm]{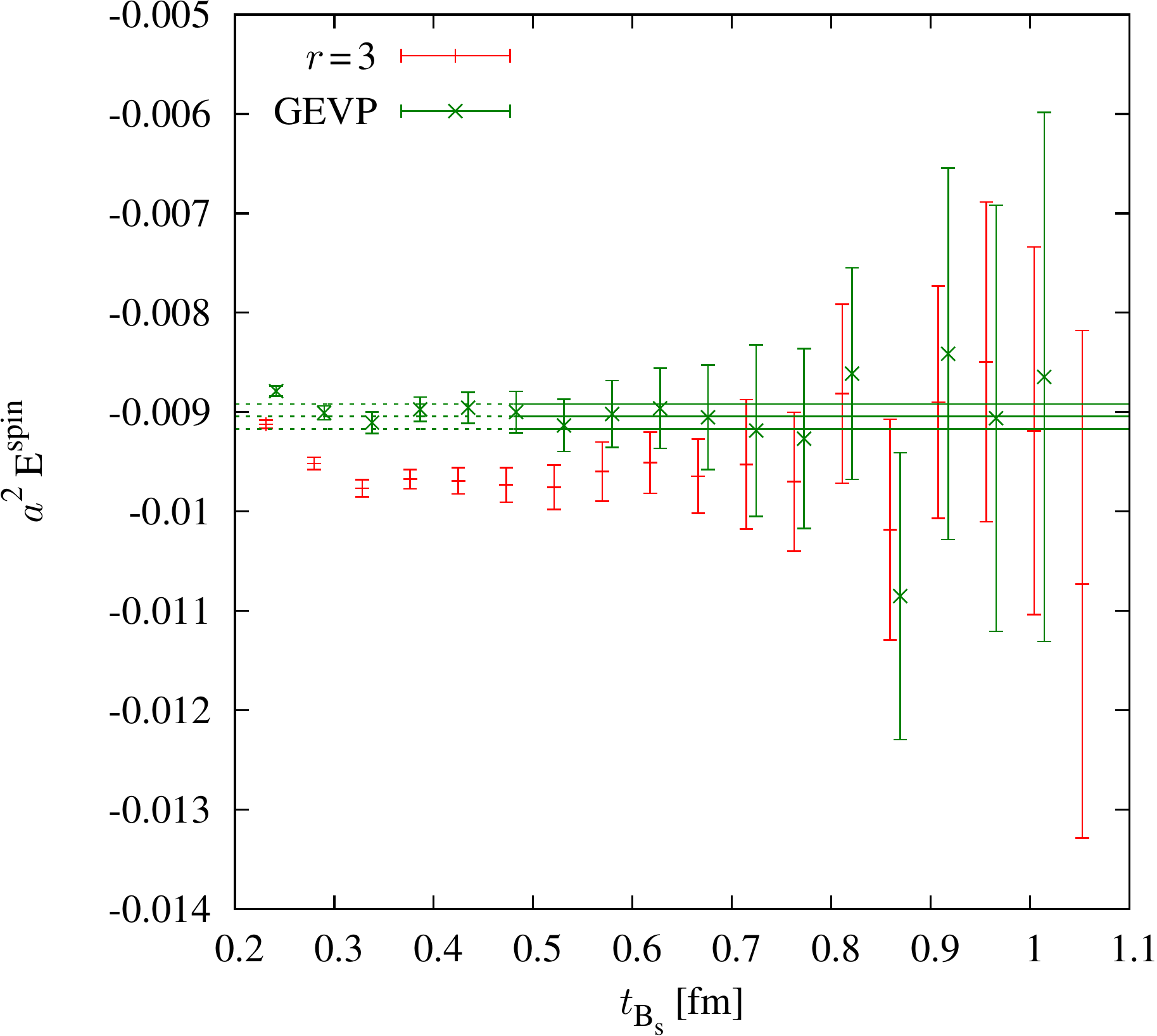}
}
\caption{$a^2 E_{\rm spin}$ for ensembles A5 (left) and O7 (right). The highest
smearing and the GEVP result are shown. The band shows the GEVP plateaux average.}
\label{fig:espin}
\end{center}
\end{figure}
 
\subsection{Numerical results for the kin and spin insertions at $1/m$ order}
\label{sec:kinspin}
As a first analysis we performed combined fits with parameters
$\Axxx{\B}{k}{r}$, $\Axxxx{\B\to\K}{k}{\mu}{r}$ and $\Exx{k}{}$
 to \eq{eq:A2pt} and \eq{eq:A3pt}
in the time region where a linear behavior in $t$ is observed. 
This was done for fixed insertion $k\in \{\kin,\spin\}$ and 
fixed smearing level, typically the largest smearing. The resulting 
errors on the matrix elements were rather large because the data 
did not constrain the fitted energies $\Exx{k}{}$ so well. An improvement 
could be achieved by determining $\Exx{k}{}$ from the GEVP at order 
$1/m$, exactly as described in
Ref.~\cite{Blossier:2009kd} and then using that as a constraint in  
 \eq{eq:A2pt} and \eq{eq:A3pt}. The GEVP takes into account information from all smearing levels of the two point functions. In a little more detail,
 we expand the ground state GEVP eigenvalues in $1/m$ (see Ref.~\cite{Blossier:2009kd} for explicit formulae) as $\lambda^{(0)}(t,t_0)
 = \lambda^{\stat}(t,t_0)+
 \omega_\kin \lambda^{\kin}(t,t_0)+
\omega_\spin \lambda^{\spin}(t,t_0)$ and then form
\begin{equation}
E_{\rm eff}^{k}(t,t_0) = -{\partial}_t \log\big(\lambda^{k}(t,t_0)\big)\,.
\end{equation}
We consider just $t_0\geq t/2$ as the {\em asymptotic} convergence 
is proven to be much better \cite{Blossier:2009kd} under that condition. For the present case, it turns out that there is rather little dependence on $t_0$ in practice but errors of course grow as it is increased. We then form a weighted average of the first
(up to) three values with $t_0$ at and above $t/2$. These 
averages are
shown as $E^k_\mathrm{eff}(t)$ in Fig.~\ref{fig:ekin} and \ref{fig:espin} together with the effective masses of the best smearing level. The GEVP estimates behave significantly 
better than just the best smearing which we show for
comparison. Our final numbers
come from  plateaux fits of the GEVP estimates starting at $t=0.5\fm$, and are shown as bands in Figs.~\ref{fig:ekin} and \ref{fig:espin}.

\begin{figure}[tbp]
\begin{center}
\makebox[\textwidth][c] {
\includegraphics[height=9.25cm]{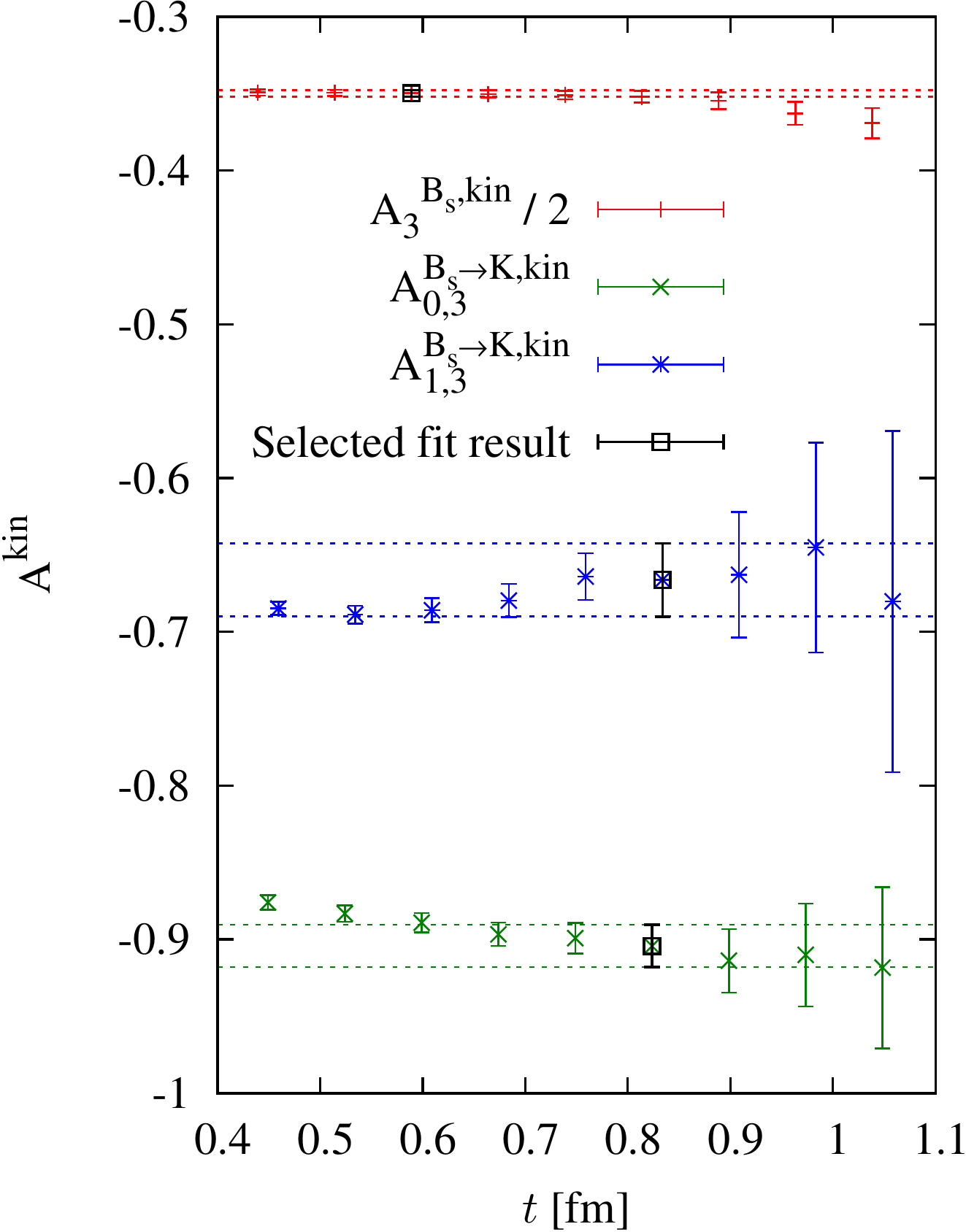}
\hspace{0.2cm}
\includegraphics[height=9.25cm]{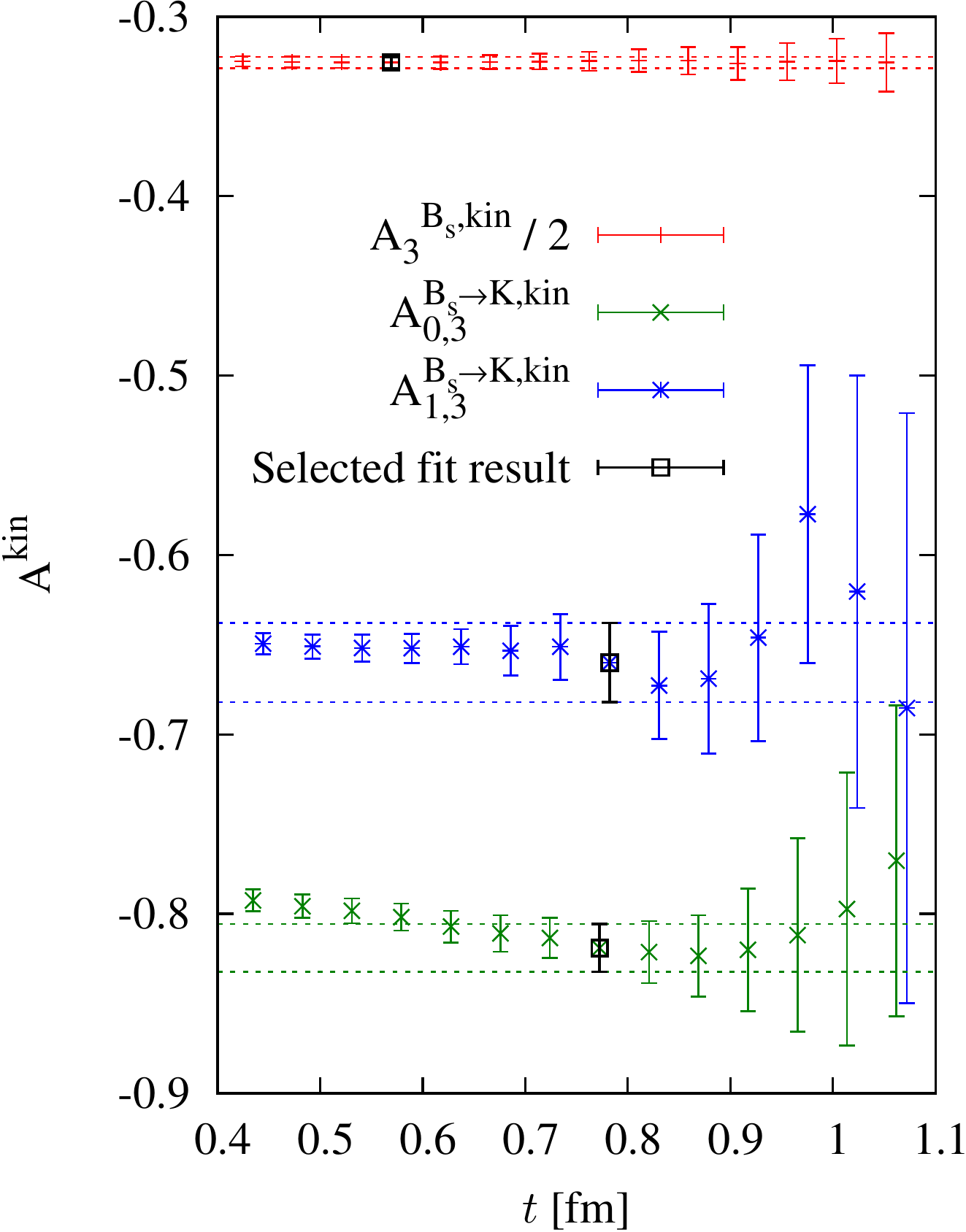}
}
\caption{Fitted $\Axxxx{\B\to\K}{\kin}{\mu}{3}$ and  $\Axxx{\B}{\kin}{3}$ for ensembles A5 (left) and O7
  (right). See text for explanations.}
\label{fig:kinplat}
\end{center}
\end{figure}

\begin{figure}[tbp]
\begin{center}
\makebox[\textwidth][c] {
\includegraphics[height=9.25cm]{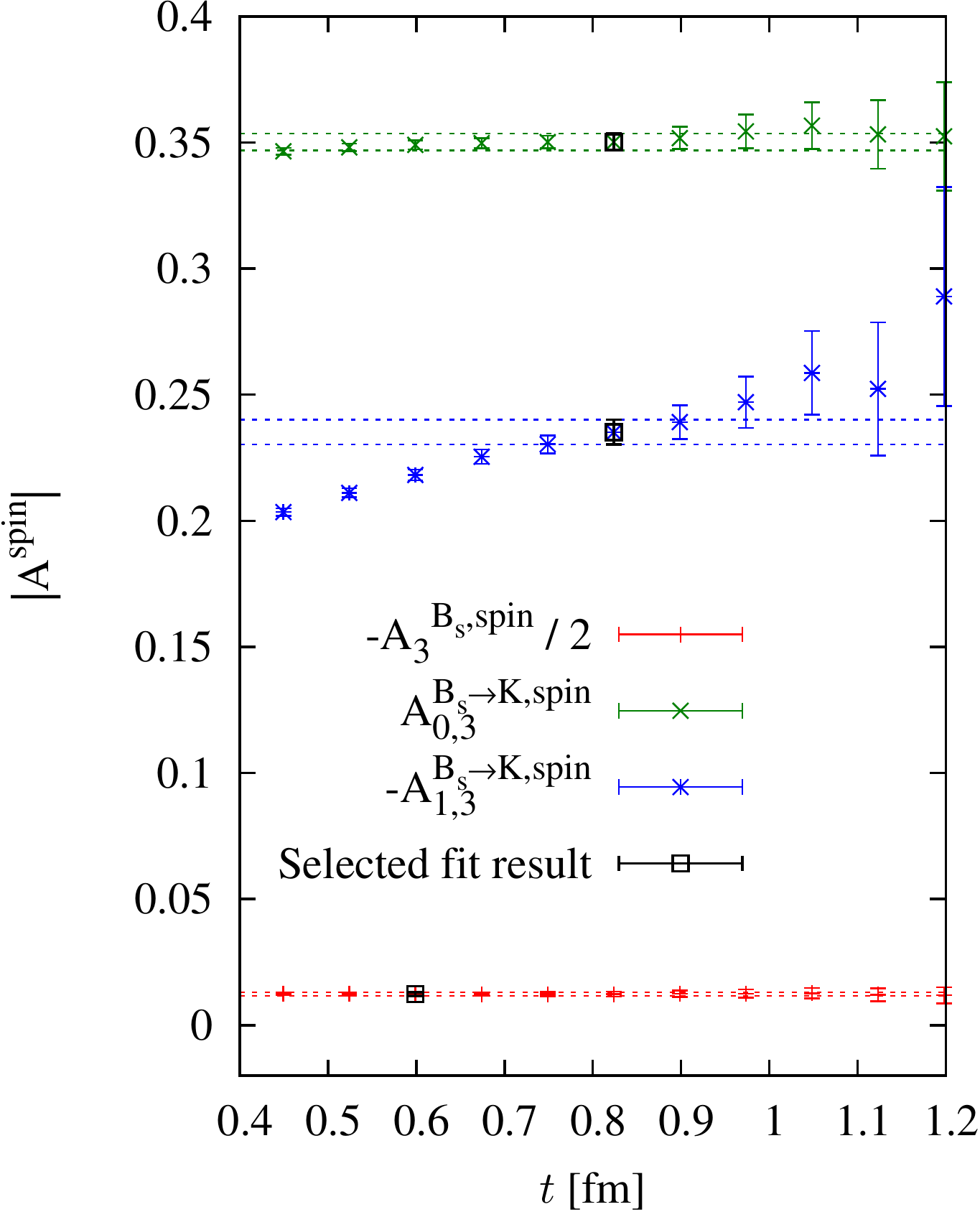}
\hspace{0.2cm}
\includegraphics[height=9.25cm]{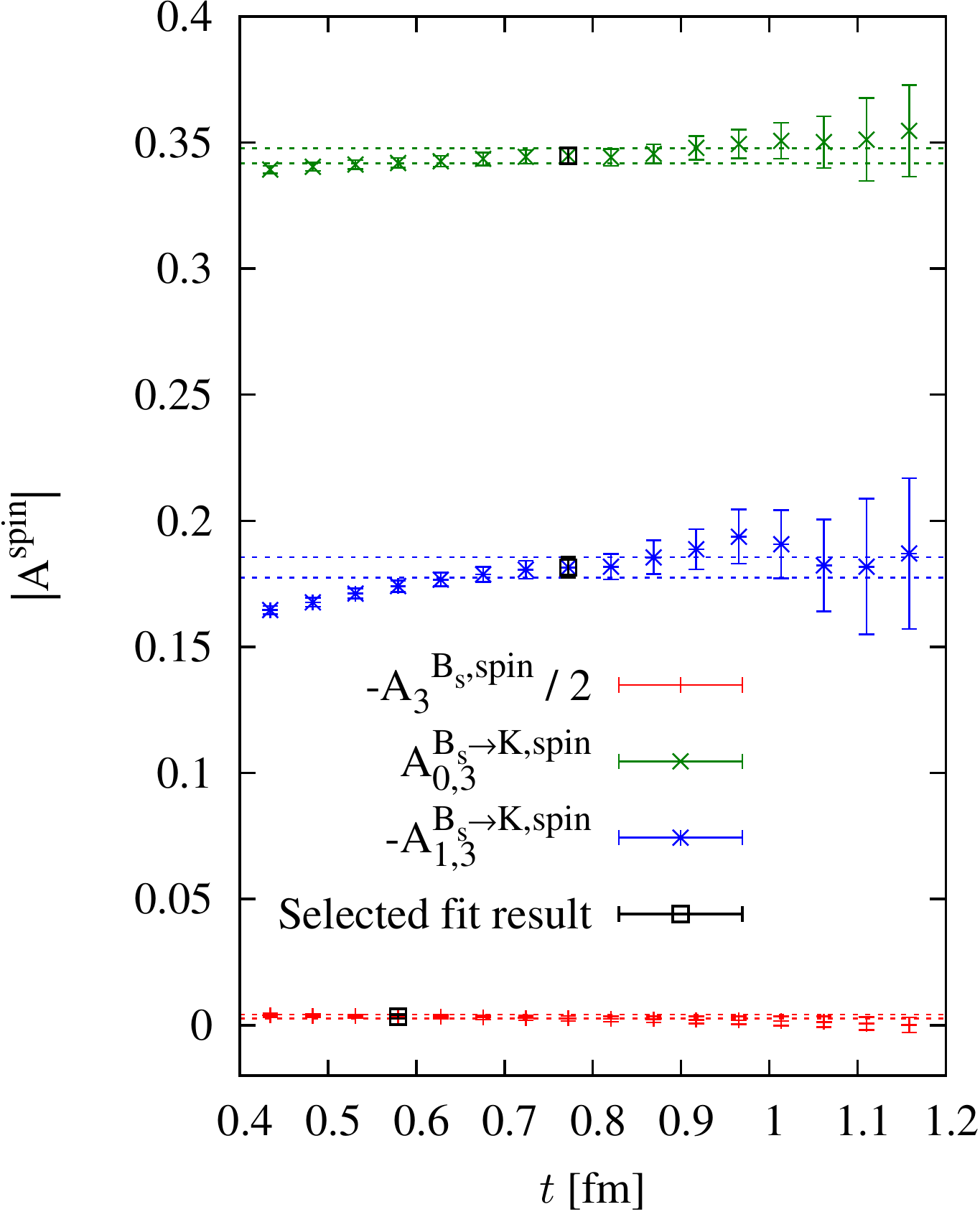}
}
\caption{Absolute values of fitted $\Axxxx{\B\to\K}{\spin}{\mu}{3}$ and  $\Axxx{\B}{\spin}{3}$ for ensembles A5
(left) and O7 (right). See text for explanations.}
\label{fig:spinplat}
\end{center}
\end{figure}

Effective  $\Axxx{\B}{k}{r}(t)$ are determined
by inserting the energies into
$ \Axxx{\B}{\kn}{r}(t) = \Cxxx{\B}{\kn}{rr}(t) / \Cxxx{\B}{\stat}{rr}(t) + \Exx{\kn}{} t $ and similarly for  $\Axxxx{\B\to\K}{k}{\mu}{r}$. 
These effective A-estimates are shown in Figs.~\ref{fig:kinplat}
and \ref{fig:spinplat}. Precise, early and long plateaux are present
for the two-point function $\Axxx{\B}{k}{r}$. In contrast,
for the three-point function $\Axxxx{\B\to\K}{k}{\mu}{r}$
an agreement with a plateau is only seen starting at 
$t=0.8\,\fm$ and in fact we would like the plateaux to be more convincing 
in one or two cases. Nevertheless, taking weighted averages starting at $t=0.8\,\fm$ 
is reasonable and we collect their results combined to $\rho^{k}_{\mu}$
in \tab{tab:kinspin}. The bands in Figs.~\ref{fig:kinplat} and \ref{fig:spinplat}
show the chosen fit values.
\begin{table}[htbp!]
\begin{center}
 \begin{tabular}{cccccc}
 \toprule
 $\mu$ & $k$  & $\rho_{\mu}^k$ A5 &  $\rho_{\mu}^k$ O7 \\
\midrule
 0 & kin & -0.554(13) & -0.493(12)\\
 1 & kin & -0.316(24) & -0.335(21)\\
\midrule
 0 & spin & \phantom{-}0.363(\phantom{0}3) & \phantom{-}0.348(\phantom{0}2)\\
 1 & spin & -0.223(\phantom{0}5) & -0.177(\phantom{0}3)\\
\bottomrule
\end{tabular}
\end{center}
\caption{Matrix elements at $\ord{1/m}$: kin and spin contributions.}
\label{tab:kinspin}
\end{table}

\begin{figure}[tbp]
\begin{center}
\makebox[\textwidth][c] {
\includegraphics[height=7cm]{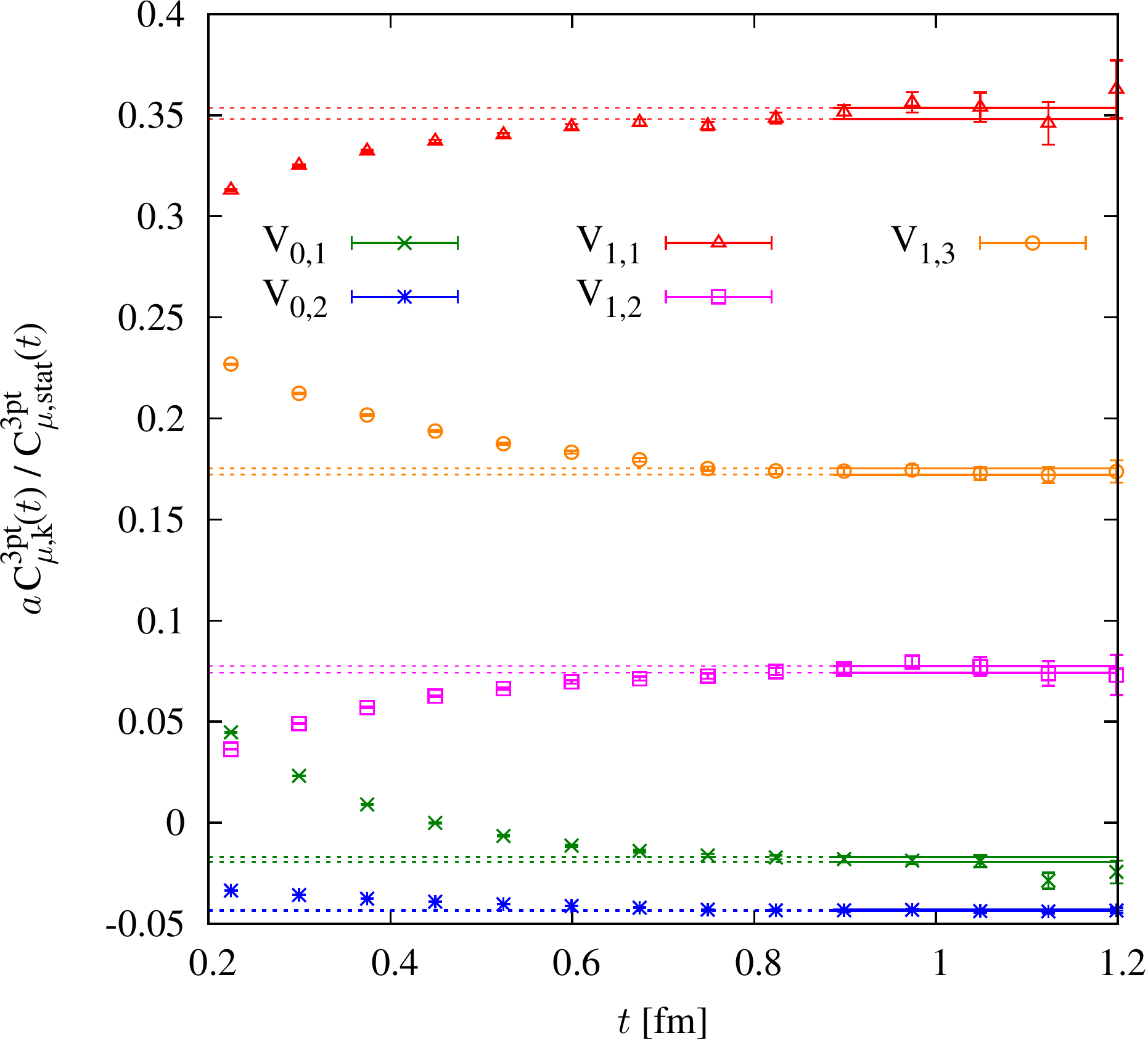}
\hspace{0.2cm}
\includegraphics[height=7cm]{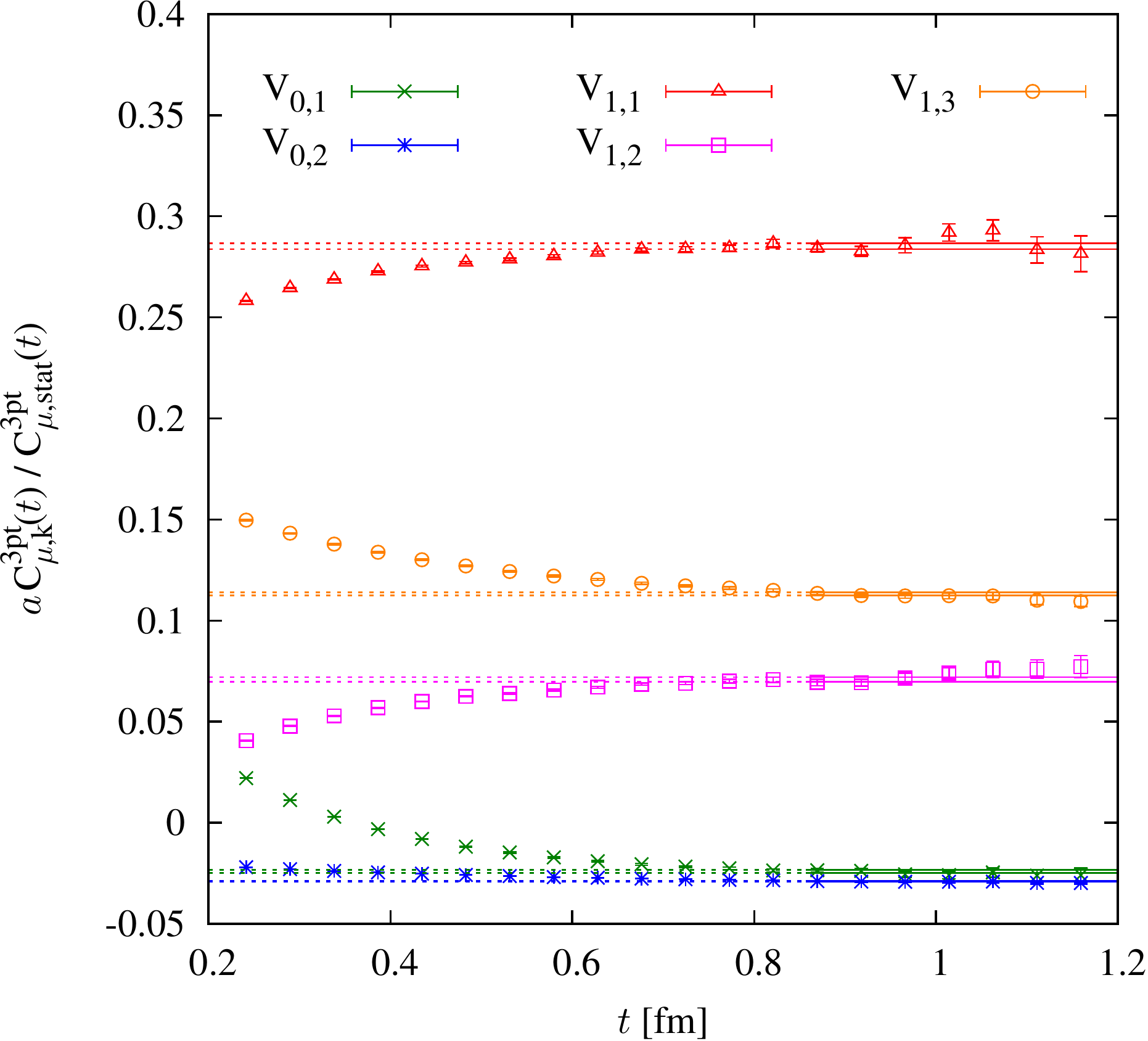}
}
\caption{Overview of the current insertions for ensembles A5 (left) and O7
(right). Fit bands are plateaux averages starting at 0.86 fm. For A5
the $V_{1,4}$ contribution is exactly equal to $V_{1,3}$, while on the O7 they
are not exactly identical but still indistinguishable on the plot, therefore
$V_{1,4}$ was not plotted.}
\label{fig:curr}
\end{center}
\end{figure}
\subsection{Current insertions at $1/m$ order}
\label{sec:currins}
 The $1/m$ vector current contributions are
\begin{align}
&V_0^{\mathrm{HQET}}(x) = Z_{V_0}^{\mathrm{HQET}}\big(V_0^{\mathrm{stat}}(x)+{\textstyle
\sum}_{j=1}^2 \omega_{0,j}V_{0,j}(x)\big),\\[3pt]
&V_i^{\mathrm{HQET}}(x) = Z_{V_i}^{\mathrm{HQET}}\big(V_i^{\mathrm{stat}}(x)+{\textstyle
\sum}_{j=1}^4 \omega_{i,j}V_{i,j}(x)\big)\,,\quad V_{\mu,j}(x)=\psibar_u\Gamma_{\mu,j}\heavy \,,
\end{align}
with the operators $\Gamma_{\mu,j}$ detailed in Table~\ref{tab:vec}. 
Note that with a momentum purely along the x-axis (A5 ensemble), only $i=1$ contributes
and $\rho_1^3 = \rho_1^4$ is exact. 

\begin{table}[htbp!]
\begin{center}
\begin{tabular}{ccccrr}
 \toprule
 $\mu$ & $j$ & $\Gamma_\mu^j$ & $\omega^\textnormal{tree}_{\mu,j}\cdot\mh$
 & $a\rho_{\mu}^j$ A5~~ &  $a\rho_{\mu}^j$ O7~~ 
 \\ \midrule
 0 & 1 & $\sum_l  \gamma_l
 \tfrac12(\nabla_l^\textnormal{S} -
 \overleftarrow\nabla_l^\textnormal{S})$  & 1/2 &
 -0.0182(12) & -0.0241(\phantom{0}7)\\
 0 & 2 & $\sum_l  \gamma_l
 \tfrac12(\nabla_l^\textnormal{S} +
 \overleftarrow\nabla_l^\textnormal{S})$ & 1/2 &
  -0.0434$(\phantom{0}4)$ & -0.0289(\phantom{0}2) 
 \\ \midrule
 $i$ & 1 & $\sum_l  \tfrac12 (\nabla_l^\textnormal{S} -
 \overleftarrow\nabla_l^\textnormal{S})\gamma_l\gamma_i$
 & 1/2 & 0.3508(27) & 0.2852(15) \\
 $i$ & 2 & $\tfrac12 (\nabla_i^\textnormal{S} -
 \overleftarrow\nabla_i^\textnormal{S})$ & -1 &
 0.0759(17) & 0.0709(11) \\
 $i$ & 3 & $\sum_l  \tfrac12 (\nabla_l^\textnormal{S} +
 \overleftarrow\nabla_l^\textnormal{S})\gamma_l\gamma_i$ &
 1/2 & 0.1737(16) & 0.1132(\phantom{0}8)\\
 $i$ & 4 & $\tfrac12 (\nabla_i^\textnormal{S} +
 \overleftarrow\nabla_i^\textnormal{S})$ & -1 &
 0.1737(16) & 0.1133(\phantom{0}8)
 \\ \bottomrule
\end{tabular}
\end{center}
\caption{Overview of the $1/m$ vector current insertions,
their tree-level matching coefficients, and the results extracted 
from the
highest light-quark smearing, $r=3$.  
We use symmetric covariant derivatives $\nabla^\textnormal{S}_i$.}
\label{tab:vec}
\end{table}
The results obtained for \eq{eq:rhojmut}
 are presented in Fig.~\ref{fig:curr}. We see plateaux
starting at roughly 0.8-0.9~fm and average from 
0.85 fm on.
The precision is better than that of the kin
and spin terms. A full quantitative error budget for the
$1/m$ contributions to the form factors has to wait until the
corresponding non-perturbative matching coefficients $\omega$ 
are available.

\section{Discussion} 
\label{sec:Conclion_outlook}

Flavor changing transitions are important channels for learning
about possible limitations of the standard model. Exclusive 
semi-leptonic decays of B or $\B$ mesons are particularly 
clean theoretically. However, lattice computations of the 
relevant form factors are needed and they are non-trivial
in practice. A major reason is the infamous signal-to-noise
problem. At large Euclidean time separations the noise in 
Monte Carlo evaluation of the correlation functions is too
large to determine the form factors (matrix elements) with
interesting precision.

While the issue is not new, we have exposed the problem in a
few graphs more clearly than often done, see Figs \ref{fig:E0_gevp}, \ref{fig:summed}.
For the three-point functions and derived (summed) ratios, 
this was possible because we have evaluated correlation 
functions at all time separations. We are only considering the 
pseudoscalar sector, where the signal-to-noise
problem is very mild for a relativistic formulation
and at zero momentum. However, we need finite momentum
(see Fig.~\ref{fig:kaon_meff} for the momentum dependence)
and for reasons explained in the beginning of the paper 
we use HQET
for the b-quark
(see Fig.~\ref{fig:E0_gevp} for the difference of Kaon and 
static $\B$ meson). In HQET, the signal-to-noise
problem becomes worse as one decreases the lattice spacing.
Nevertheless, we are determining the matrix elements 
at total time separations $\tau$ of the three-point functions
of around 2~fm and for $a\approx 0.05\,\fm$. Only for the summed ratio we use about 
half that time-separation (Fig.~\ref{fig:summed}), in agreement with the predicted better suppression of excited states after
summation. 

Despite our use of HQET, these separations are larger or similar to the ones typically used. E.g. most recently Ref.~\cite{Bazavov:2019aom} used a fixed $\tau \approx 2.2\,\fm$.

We have presented good evidence
that the chosen plateaux or fit-windows are reasonably safe, but nevertheless it would be better to have larger times accessible. 
Maybe multilevel strategies
~\cite{Ce:2016ajy,Ce:2017ndt} will help to reach those separations in the future. At present we derive our confidence
from the good agreement of 
\newcommand{\grtsim}{\raisebox{-.6ex}{$\stackrel{\textstyle{>}}{\sim}$}}

\begin{itemize}
\item
    fits with $ \tb \grtsim 0.3-0.6$ and 3 states in the $\B$ sector (and also with 2 $\B$ states and $ \tb \grtsim 0.5-0.7$, cf. Appendix 
    \ref{sec:fit12_stat})
\item ratios \eq{eqn:r2} with $\tau=\tb+\tk \,\grtsim\, 2\,\fm$,
\item and summed ratios \eq{eq:summed} at total separation
    $\tau \,\grtsim \,1\,\fm$ .
\end{itemize}
Note that also the precision of the different estimates 
is quite comparable. It is therefore preferable to use
the technically easier ratio methods.

At the lowest (static) order in $1/m$ the most relevant form factors
for $\mu=1$ have an accuracy around 2\%, which is good 
for precision physics. The first order corrections
in $1/m$ are actually more precise than that. The relative errors induced into the form factors are given by the absolute ones
of the quantities $\rho$ in Tables~\ref{tab:kinspin},\ref{tab:vec} multiplied with the appropriate $\omega$-coefficients.  
From the errors of $\rho^\mathrm{kin}$ and $\rho^\mathrm{spin}$, using the non-perturbative $\omega_\mathrm{kin}\,,\omega_\mathrm{spin}$
\cite{Blossier:2012qu}, we get error contributions of 1\% and 0.5\% to the form factors, respectively.
The uncertainties of $\rho^j$, translate into below 0.5\% errors
 assuming coefficients $\omega_j$
which do not exceed the tree-level values by more than a factor of two.
Of course, the uncertainties in the coefficients $\omega$ are to be added separately.

A positive result of our detailed analysis is thus that 
$1/m$ corrections can be 
determined precisely, when the coefficients (HQET parameters) are known
with reasonable accuracy. Once they are available it will be possible to provide further crosschecks on the existing analysis,
see e.g. Fig.~13 of \cite{Bazavov:2019aom}.

We would finally like to point out a possible danger in current and future
semi-leptonic form factor computations on the lattice. 
It consists in the contribution of 
multi-hadron states, such as $|\B^{(\ast)} ,\pi\rangle$ or $|\Bs^{(\ast)} ,\K\rangle$, to our Euclidean correlation functions. 
Formally, these contributions are just particular ones in the 
sums over excited states in Eqs.~(\ref{eq:c2ll}-\ref{eq:c3})
and are thus covered by our analysis. However, there are two properties
which make such states special. 
First, when the spatial volume becomes large 
and the light-quark masses small, there are several low-lying states
with small gaps $E^{(n)} - E^{(0)}$. All used methods may have difficulties in separating those. 
This has also been pointed out recently in \cite{Hashimoto:2019pgh}.
Second, normalized overlaps 
$\beta_r^{(n)}/\beta_r^{(0)}$
may be very similar for different smearing levels $r$ when $n$ corresponds
to a multi-hadron state.\footnote{See the discussion of the 
field in the chiral effective theory representing the action density at finite flow time, $E(x,t)$ in Ref.~\cite{Bar:2013ora}.}
As mentioned in Sect.~\ref{s:gevp} the GEVP-method is of little
help in such a situation.
For B-mesons these contaminations at finite $\tb$ have not been 
investigated at all, while there is considerable discussion in the analogous determination of nucleon matrix elements (see Refs.~\cite{Bar:2017gqh,Green:2018vxw} and references therein).  
It appears likely that form factors $\B\to K$, $\rm B \to \pi$
are in somewhat better shape than  nucleon matrix elements,
since larger Euclidean time separations are reached. Nevertheless 
systematic studies, especially theoretical ones such as the ones
carried out in
chiral perturbation theory \cite{Bar:2016uoj,Bar:2018xyi} for nucleon matrix elements are urgently needed to make quantitative statements.

\appendix

\section{Computation of the correlation functions}
\label{sec:cf}

Integration over the fermion fields yields
the correlation functions as 
\begin{equation}
  \C^{\Bs\to\K}_{\mu,r}(\tk,\tb;\veca{p}) = \langle \Chat^{\Bs\to\K}_{\mu,r} \rangle_U\,,
  \quad 
  \CK(t;\veca{p}) = \langle \ChatK \rangle_U\,,
  \quad 
  \CB_{rr'}(t) = \langle \ChatB_{rr'} \rangle_U\
  \label{e:gaugeav}
\end{equation}
averaged over the gauge fields, $U$, (with effective action including the 
log of the quark determinant). The dependence of the functions
\begin{align}
   \Chat^{\Bs\to\K}_{\mu,r} & =
  \frac{a^9}{L^3} \Tra\left[
    \gamma_5 P_\mathrm{f} F_- W_\K S_\mathrm{u} \gamma_\mu P_\mathrm{v} 
    F_+ S_\mathrm{h} P_\mathrm{i} \gamma_5 W_r S_{\mathrm{s}} W_\K 
  \right]
  \label{e:Chat3pt}
  \\
  \ChatK & = 
  \frac{a^6}{L^3} \Tra\left[
    \gamma_5 P_\mathrm{f} F_- W_\K S_\mathrm{u} W_\K \gamma_5 P_\mathrm{i} F_+ W_\K S_\mathrm{s} W_\K 
  \right]
  \label{e:ChatK}
  \\
  \ChatB_{rr'} & =
  \frac{a^6}{L^3} \Tra\left[
     \gamma_5 P_\mathrm{f} S_\mathrm{h} P_\mathrm{i} \gamma_5 W_{r'} S_\mathrm{s} W_r
  \right]
    \label{e:ChatB}
\end{align}
on the times $(x_0)_s$ ($s= \mathrm{f}, \mathrm{v}, \mathrm{i}$)
as well as the gauge fields 
is suppressed. In the above expressions  $S_q$ are the quark propagators, $W_\K = G^{\Nsm^\K},\; G=1 - \kappa_\mathrm{G}\, a^2 \Delta $ and $W_r=G^{\Nsm^\Bs_r}$ are the smearing 
operators, $P_s$ is the projection on time slice $(x_0)_s$ and $F_\pm = \exp(\pm i \vecx\vecp)$ is the multiplication with the Fourier phases on the corresponding time slice. Evaluation of the full 
traces which include the space coordinates becomes possible
by representing  the projector
\begin{equation}
 P_\mathrm{f} = \langle \eta \eta^\dagger \rangle_\eta
   \label{e:etaav}
\end{equation}
in terms of a random U(1) field $\eta$ with support on on time-slice 
$(x_0)_\mathrm{f}$. This  yields 
\begin{align}
   \Chat^{\Bs\to\K}_{\mu,r} & =
  \frac{a^9}{L^3} \left\langle \eta^\dagger
    F_- \gamma_5 W_\K S_\mathrm{u} P_\mathrm{v} \gamma_\mu  
    F_+ S_\mathrm{h} P_\mathrm{i} \gamma_5 W_r S_{\mathrm{s}} W_\K 
  \, \eta \right\rangle_\eta
  \\
  & = \left\langle \phi_\mathrm{u}^\dagger \gamma_5 P_\mathrm{v} \gamma_\mu  
    F_+ S_\mathrm{h} P_\mathrm{i} \gamma_5 W_r \phi_\mathrm{s} \right\rangle_\eta
    \label{e:Chat3pt1}
    \\
    & \phi_\mathrm{s} = S_{\mathrm{s}} W_\K \, \eta\,,
    \quad
    \phi_\mathrm{u} = S_{\mathrm{u}} W_\K F_-^\dagger\,\eta\,
\end{align}
and similar for the two-point functions. 
In this form we see that for each vector $\eta$ two solutions of the Dirac equation are needed in order to compute the fields $\phi_\mathrm{u}, \phi_\mathrm{s}$. The static propagator $S_\mathrm{h}$ is inserted by explicit forward 
propagation. Translation invariance in time is used by averaging over
all source time-slices $(x_0)_\mathrm{f}$, each one with a random U(1) field $\eta$.
Since the averages \eq{e:gaugeav} and \eq{e:etaav} are independent of each other, any number of $\eta$ fields per gauge field is correct; we 
use a single one per gauge field and time $(x_0)_\mathrm{f}$.

In practice, we compute propagators of periodic quark fields in 
a gauge field $e^{i\theta_\mu a/L} U(x,\mu)$, which includes 
the constant $U(1)$ background field $\theta^q_\mu a$ and only 
use the integer part of the momentum in the Fourier factors $F_\pm$.
This is equivalent to \eq{eqn:theta}, apart from a phase in the Gaussian
smearing, which we set to zero. Choosing a different phase in the smearing 
along the lines of \cite{Bali:2016lva} might be a further optimization.

\subsection{Improvement and $1/m$ terms}
\label{s:appNLO}
The $1/m$ terms are simple generalizations of the above.
First,
$[\C^{\Bs\to\K}_\mu]_k\,,k\in$ kin, spin and
$[\CB_{rr'}]_k$, are given by replacing $S_\mathrm{h} \to S_k$,
where the latter are 
\begin{align}
 S_\mathrm{kin} &= S_\mathrm{h} \nabla_i^*\nabla_i  S_\mathrm{h} 
 \\
 S_\mathrm{spin} &= \frac{i}{4} S_\mathrm{h} [\gamma_i,\gamma_j] \hat F_{ij}  S_\mathrm{h}
\end{align}
and the chromo-magnetic field strength tensor $\hat F_{ij}$ is discretized in terms
of the clover leaf, see e.g. \cite{Jansen:1998mx}. 

Second, the 
NLO three-point functions 
$[\C^{\Bs\to\K}_\mu]_j$ are given by the substitution  $\gamma_\mu \to \Gamma_\mu^j$ 
in \eq{e:Chat3pt} and \eq{e:Chat3pt1},
with $\Gamma_\mu^j$ listed in \tab{tab:vec}.
The O($a$) improvement corrections to the static three-point functions are just linear combinations thereof.

\section{Alternative static fit with $N_\K=1, N_\B=2$}
\label{sec:fit12_stat}

In this section we discuss a simpler version of the fit which includes only one
excited state in the $\B$ sector. We use only the two highest $\B$ smearings,
therefore we have a total of 12 fit parameters (as opposed to 20 in the fit
with $N_\B=3$).

Similar to Sec.~\ref{sec:fits_stat} we show the $\Bs$ energy states. Here, the
growth of the errors at larger $\tfit{B2/3}{min}$ is much milder than for the
$N_\B=3$ fit, but on the other hand the stability of the results wrt.\ changes
of $\tfit{B2/3}{min}$ is in some cases rather unsatisfying. It can be even more
 clearly seen for $\EB{0}$ (cf.~Fig.~\ref{fig:stab_1x2_eb0}) which, in
 particular for O7, shows a characteristic drift upwards with growing
 $\tfit{B3}{min}$.

\begin{figure}[tp!] \centering
\makebox[\textwidth][c]{
\includegraphics[height=7cm]{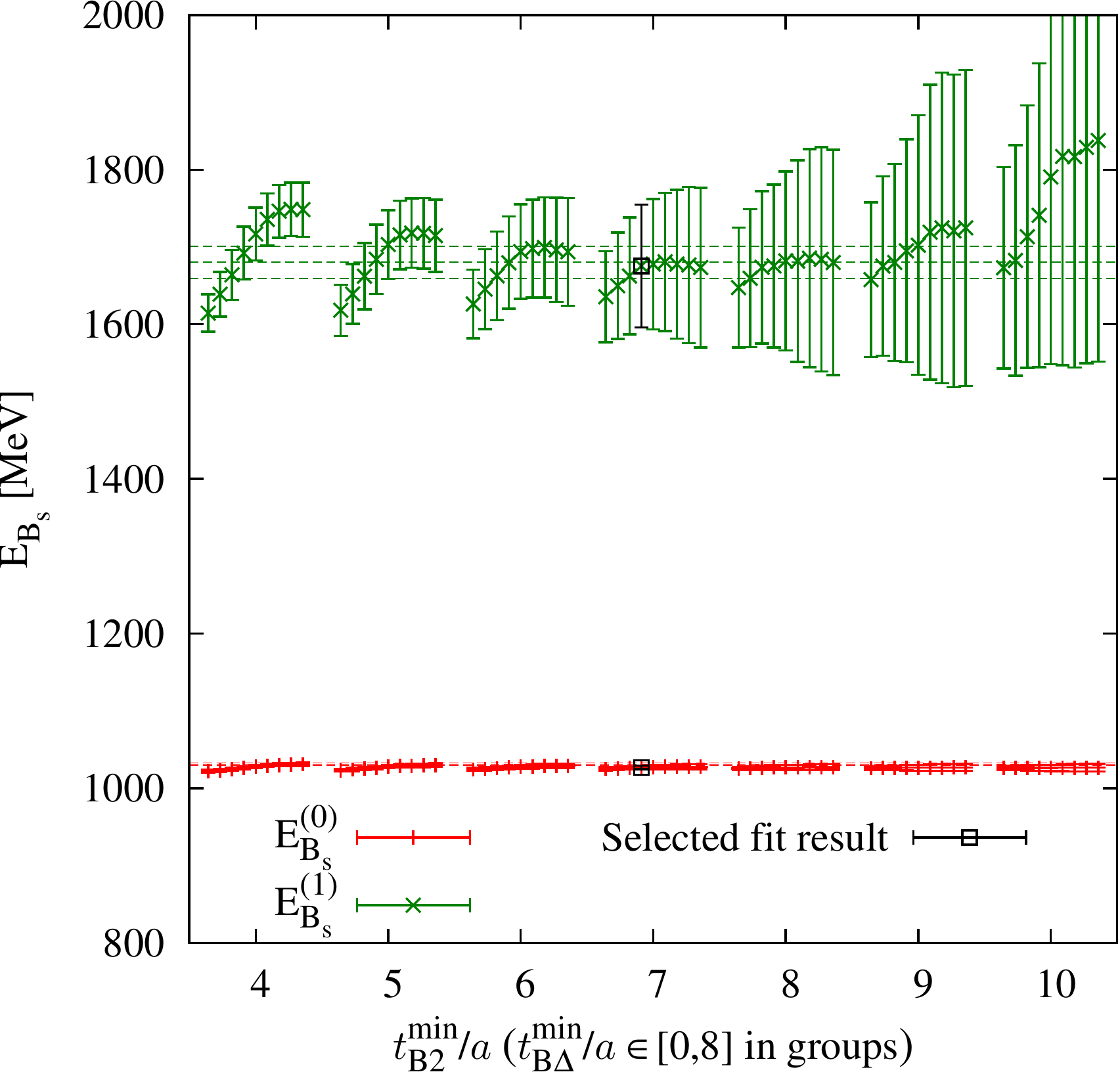}
\hspace{0.2cm}
\includegraphics[height=7cm]{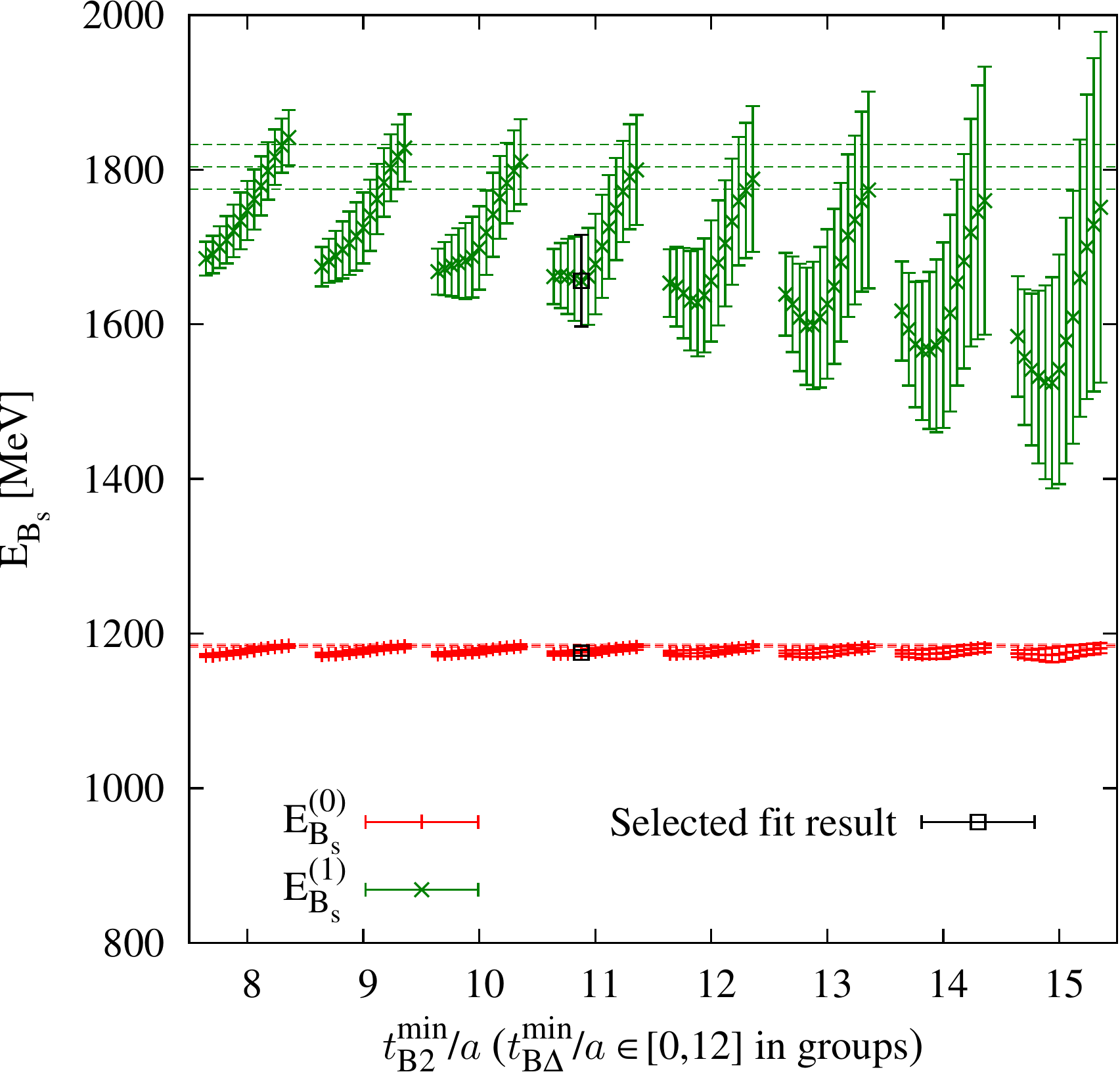} }
\caption{Stability plots for the $\B$ energies on ensemble A5
(left) and O7 (right). The selected $\tfit{B2/3}{min}$ values are highlighted
with a black color, while the dashed lines are values from the GEVP, cf.\
Sec.~\ref{sec:gevp}. A zoom into the behaviour of $\EB{0}$ is shown in
Fig.~\ref{fig:stab_1x2_eb0}.}
\label{fig:stab_1x2_all_eb}
\end{figure}

\begin{figure}[tp!] \centering
\makebox[\textwidth][c]{
\includegraphics[height=7cm]{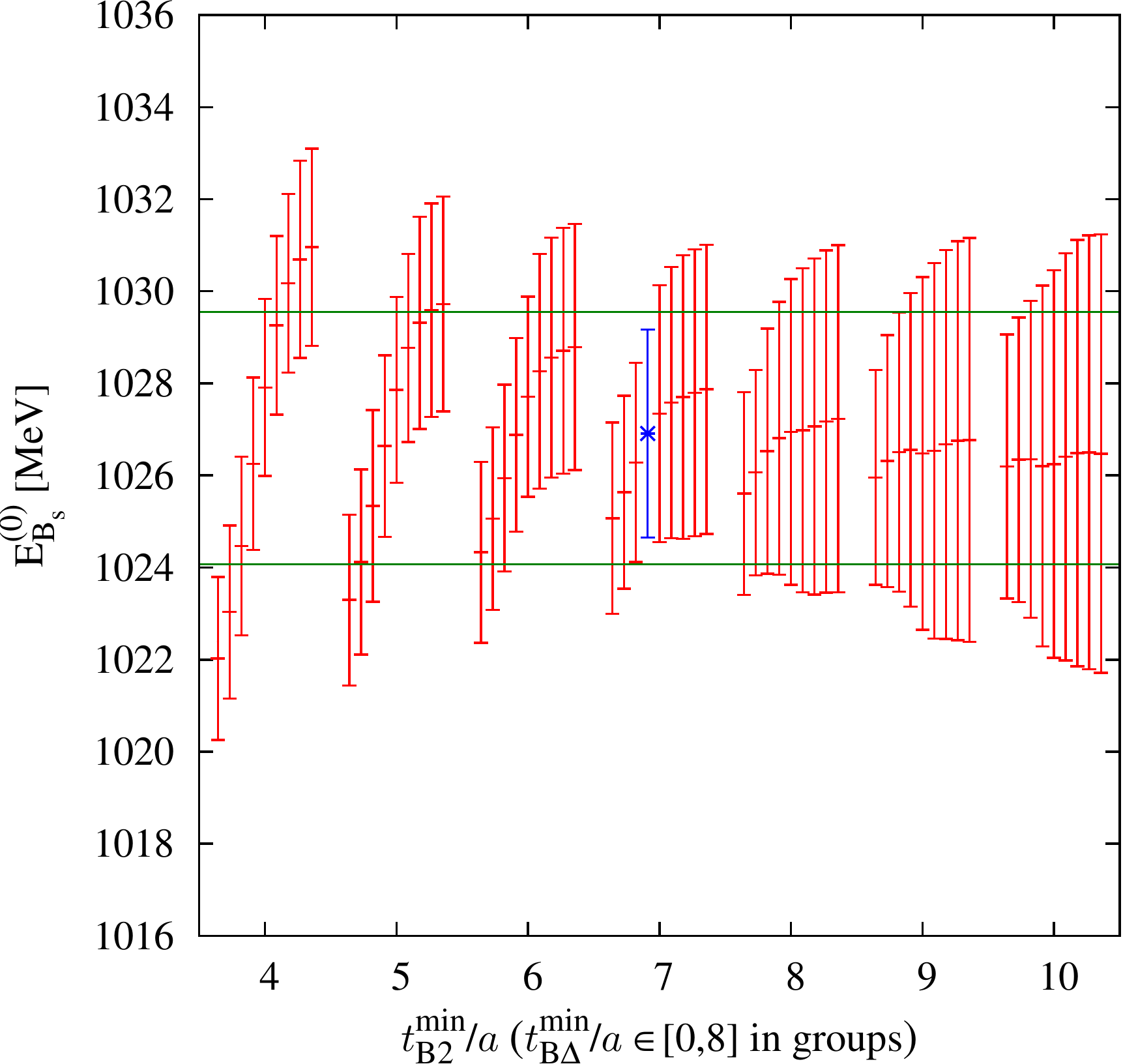}
\hspace{0.3cm}
\includegraphics[height=7cm]{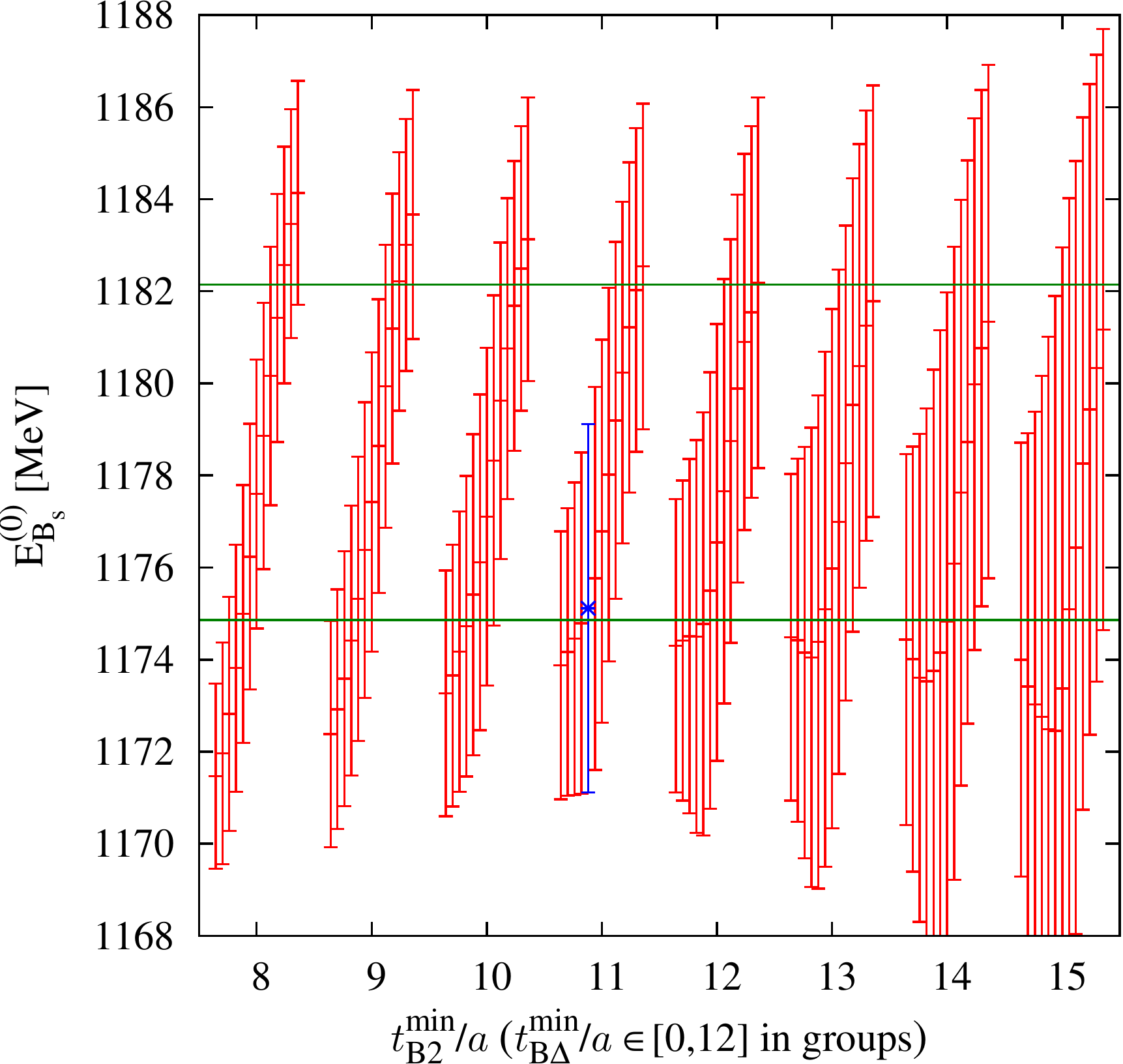}}\\[9pt]
\caption{Stability plots for $\EB{0}$ for ensemble A5 (left) and O7 (right).}
\label{fig:stab_1x2_eb0}
\end{figure}

\begin{figure}[p!] \centering
\makebox[\textwidth][c]{
\includegraphics[height=9.25cm]{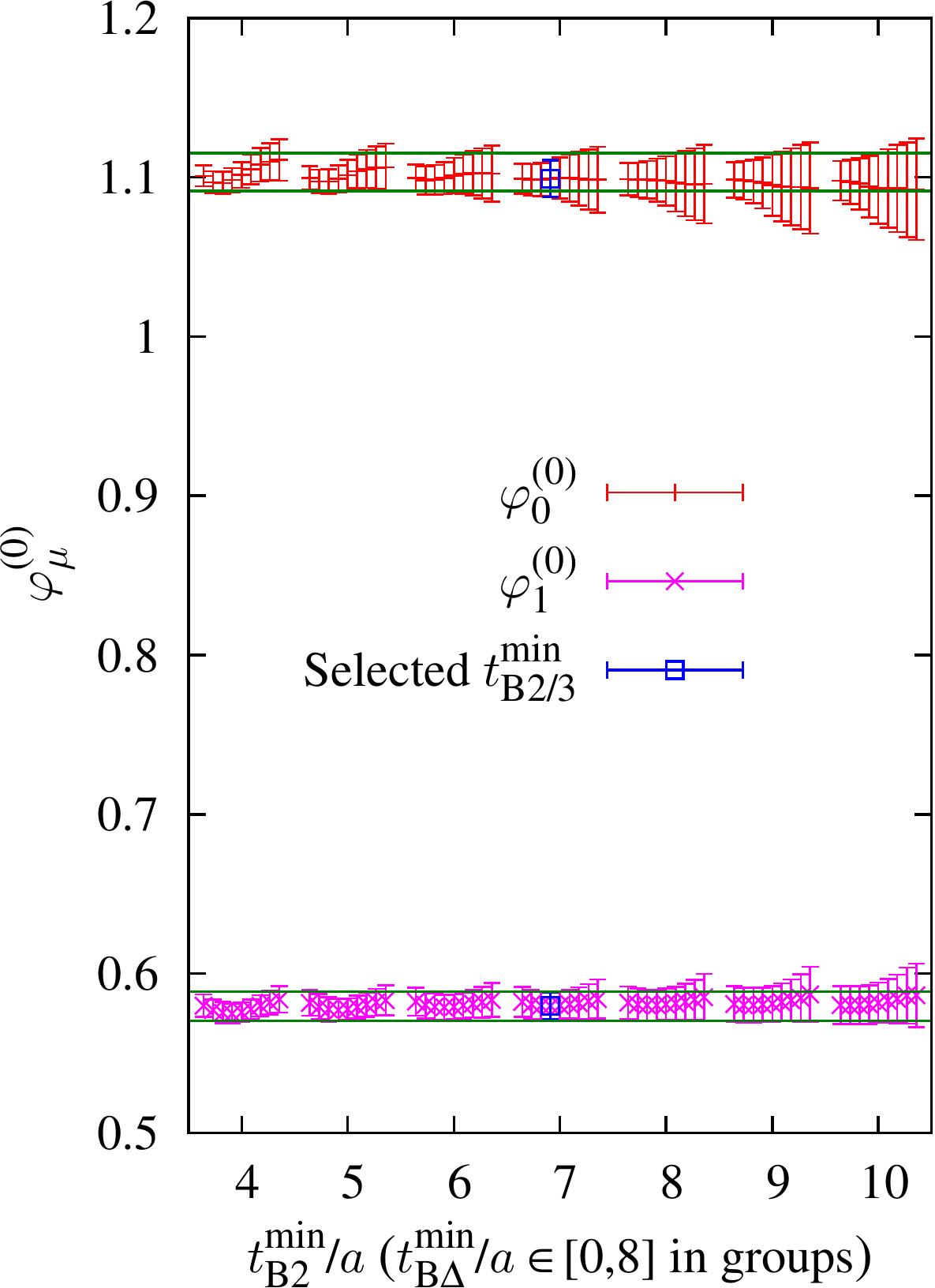}
\hspace{0.2cm}
\includegraphics[height=9.25cm]{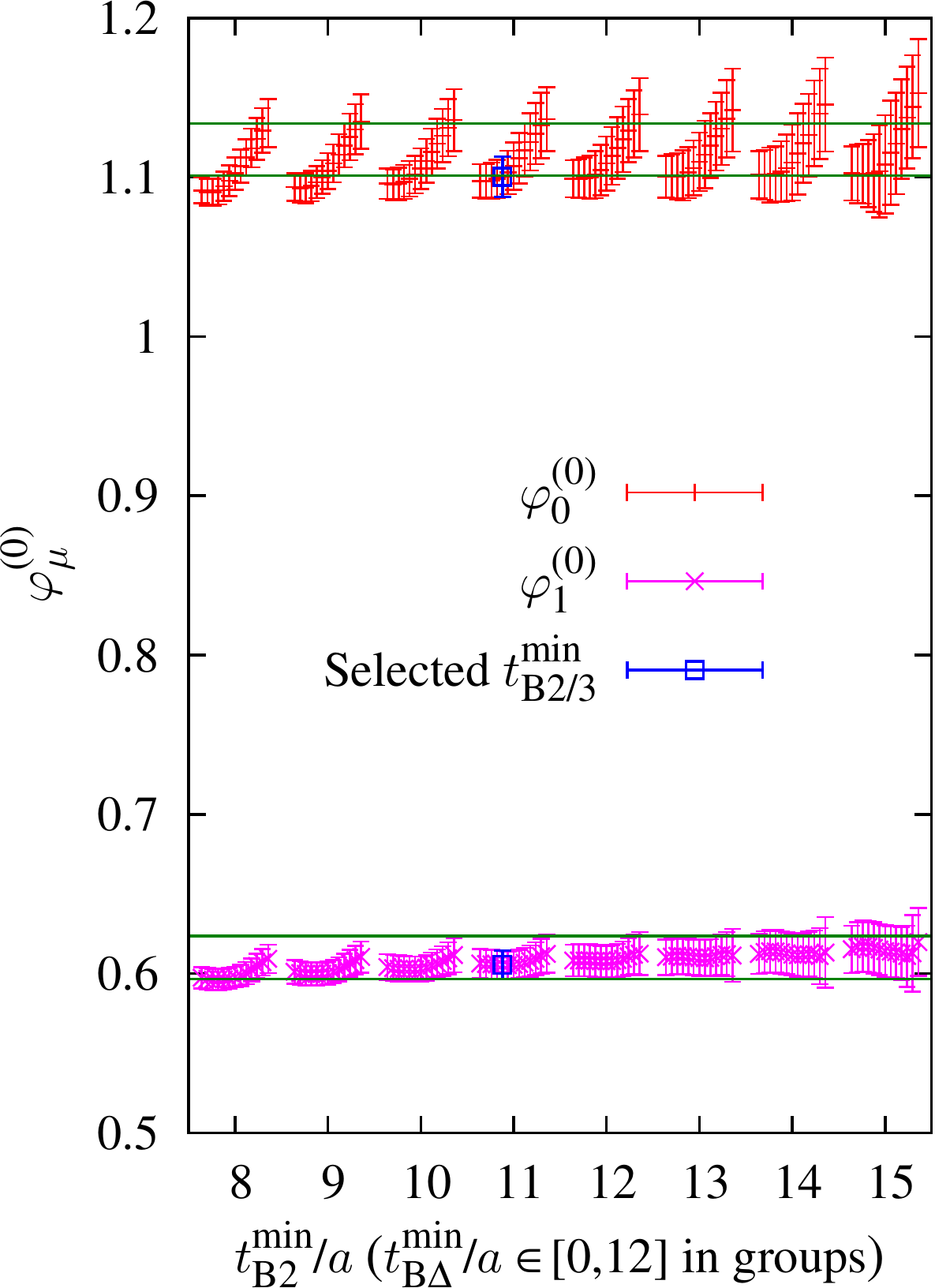}}
\caption{Stability plots of the ground-state form factors for ensemble A5 (left)
and O7 (right).}
\label{fig:stab_1x2_phi0}
\end{figure}

On the other hand, considering these drifts in energies (and amplitudes as
well), the ground-state form factors show reasonable stability and agreeement
with the $N_\B=3$ fit results, cf.~Fig.~\ref{fig:stab_1x2_phi0}. The case of
$\varphi_0^{(0)}$ for O7 is by far the worst, but most look more like the stable 
and consistent $\varphi_1^{(0)}$.

When choosing a final form factor value from the fit, one clearly has to keep
$\tfit{B2/3}{min}$ slightly higher than for the $N_\B=3$ fit. We use
$\tfit{B2}{min}\approx0.52$ fm and the same $\tfit{B\Delta}{min}$ as before. The
values summarized in Table \ref{tab:res_fit12} are consistent
within errors with other methods and give similar precision.

\begin{table}[h!]
\begin{center}
\begin{tabular}{ccc}
\toprule
id & $\mu$ & Fit value \\
\midrule
A5 & 0 & 1.099(12) \\
A5 & 1 & 0.580(9)  \\
O7 & 0 & 1.100(13) \\
O7 & 1 & 0.606(9) \\
\bottomrule
\end{tabular}
\caption{Results for the ground-state form factors $\varphi_\mu^{(0)}$ using
the fit with $N_\Bs=2$.}
\label{tab:res_fit12}
\end{center}
\end{table}

\begin{figure}[p!]
\begin{center}
\includegraphics[width=12cm]{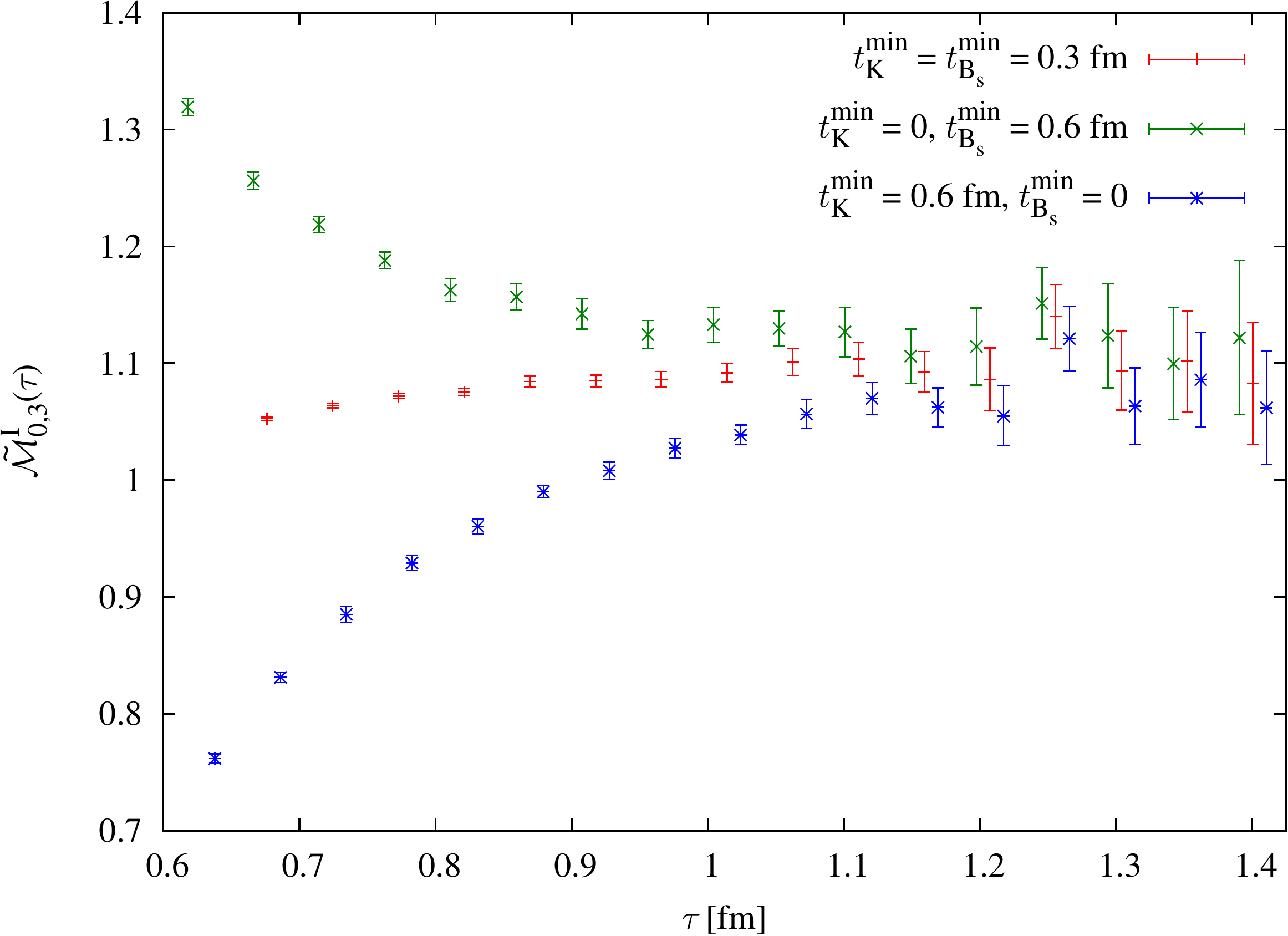}
\caption{The summed ratio for ensemble O7, $\mu=0$ with three versions: with symmetric
sum limits and with either only $t_\B^{\rm min}$ or $t_\K^{\rm min}$ larger than
0. Note that due to different (worse) convergence properties, these plots cannot
be directly compared with the ones in Fig.~\ref{fig:summed}.}
\label{fig:truncated}
\end{center}
\end{figure}

\section{Note on the convergence of the static summed ratio}
\label{sec:sumrat_conv}

In Section \ref{sec:sum_stat}, devoted to static summed ratios, we observe that
the plateaux for $\mu=0$ in Fig.~\ref{fig:summed} seem to start very early.
Taking the results for O7 at face value, one could start the plateau fit as
early as 0.4 fm. To see whether this is a genuine ground state dominance we investigate
truncated sums of the form:
\begin{equation}
\tilde\Mi_{\mu,r}(\tau)=\partial_\tau\,
a\sum_{\tb=t_\B^{\rm min}}^{\tau-t_\K^{\rm min}}\Ri_{\mu,r}(\tau-\tb,\tb).
\label{eq:sum_trunc}
\end{equation}
By putting $t_\B^{\rm min}$ or $t_\K^{\rm min}$ larger than 0 we can suppress
the excited states in the $\B$ and $\K$ sector respectively.\footnote{Note
however that this increases the excited state corrections
from \mbox{$\sim \tau \Delta \exp(-\tau\Delta)$} to
\mbox{$\sim \tau' \Delta \exp(-\tau'\Delta)$} with 
$\tau'=\tau- t_\B^{\rm min} - t_\K^{\rm min}$.} 
The
results of this procedure are shown in Fig.~\ref{fig:truncated}. We see that the
excited states in the $\K$ sector push the result upwards, while the excited
states in the $\Bs$ sector push the results downwards. Their cancellation
results in a flat ``fake'' plateau which can start at very early times. A
similar but much less pronounced effect is at work for $\mu=1$. One should
therefore not rely solely on the flatness of the plots but also devise an
independent criterion for the beginning of the plateaux, as was done in Section
\ref{sec:sum_stat}.

\begin{acknowledgement}%
 We would like to acknowledge useful discussions with Oliver B\"ar, 
 Stefan Schaefer, Michele Della Morte, Brian Colquhoun.

We gratefully acknowledge the Gauss Centre for Supercomputing (GCS)
for providing computing time through the John von Neumann Institute for
Computing (NIC) on the GCS share of the supercomputer JUQUEEN at J\"ulich
Supercomputing Centre (JSC). GCS is the alliance of the three national
supercomputing centres HLRS (Universit\"at Stuttgart), JSC (Forschungszentrum
J\"ulich), and LRZ (Bayerische Akademie der Wissenschaften), funded by the
German Federal Ministry of Education and Research (BMBF) and the German State
Ministries for Research of Baden-W\"urttemberg (MWK), Bayern (StMWFK) and
Nordrhein-Westfalen (MIWF).
We acknowledge PRACE for awarding us access to resource JUQUEEN in Germany at J\"ulich and to resource SuperMUC in Germany at M\"unchen.
We also thank the LRZ for a CPU time grant on SuperMUC, project pr85ju, and
DESY for access to the PAX cluster in Zeuthen.

\end{acknowledgement}
\providecommand{\href}[2]{#2}\begingroup\raggedright\endgroup

\end{document}